\documentclass[twocolumn]{aastex631}
\usepackage{amsmath}
\usepackage{graphicx}
\usepackage{hyperref}

\defcitealias{Cehula+24}{C24}

\newcommand{\be}{\begin{equation}}
\newcommand{\ee}{\end{equation}}

\shorttitle{{\textit{r}-process nucleosynthesis in magnetar giant flares}}

\shortauthors{A.~Patel et al.}

\begin{document}

\title{\textbf{\textit{r}}-Process Nucleosynthesis and Radioactively Powered Transients from Magnetar Giant Flares}

\author[0009-0000-1335-4412]{Anirudh Patel}
\affil{Department of Physics and Columbia Astrophysics Laboratory, Columbia University, New York, NY 10027, USA}

\author[0000-0002-4670-7509]{Brian D.~Metzger}
\affil{Department of Physics and Columbia Astrophysics Laboratory, Columbia University, New York, NY 10027, USA}
\affil{Center for Computational Astrophysics, Flatiron Institute, 162 5th Ave, New York, NY 10010, USA} 

\author[0000-0003-1012-3031]{Jared A.~Goldberg}
\affil{Center for Computational Astrophysics, Flatiron Institute, 162 5th Ave, New York, NY 10010, USA} 

\author[0000-0002-4914-6479]{Jakub Cehula}
\affil{Institute of Theoretical Physics, Faculty of Mathematics and Physics, Charles University, V Hole\v{s}ovi\v{c}k\'{a}ch 2, Prague, 180 00, Czech Republic}

\author[0000-0002-1730-1016]{Todd A.~Thompson}
\affil{Department of Astronomy, Ohio State University, 140 West 18th Avenue, Columbus, OH 43210, USA}
\affil{Center for Cosmology \& Astro-Particle Physics, Ohio State University, 191 West Woodruff Ave., Columbus, OH 43210, USA}
\affil{Department of Physics, Ohio State University, 191 West Woodruff Ave., Columbus, OH 43210, USA}

\author[0000-0002-6718-9472]{Mathieu Renzo}
\affil{Steward Observatory, University of Arizona, 933 N. Cherry Avenue, Tucson, AZ 85721, USA}

\correspondingauthor{Anirudh Patel}
\email{anirudh.p@columbia.edu}

\begin{abstract} We present nucleosynthesis and light-curve predictions for a new site of the rapid neutron capture process ($r$-process) from magnetar giant flares (GFs). Motivated by observations indicating baryon ejecta from GFs, \citet{Cehula+24} proposed mass ejection occurs after a shock is driven into the magnetar crust during the GF. We confirm using nuclear reaction network calculations that these ejecta synthesize moderate yields of third-peak $r$-process nuclei and more substantial yields of lighter $r$-nuclei, while leaving a sizable abundance of free neutrons in the outermost fastest expanding ejecta layers. The final $r$-process mass fraction and distribution are sensitive to the relative efficiencies of $\alpha$-capture and $n$-capture freeze-outs. We use our nucleosynthesis output in a semi-analytic model to predict the light-curves of {\it novae breves}, the transients following GFs powered by radioactive decay. For a baryonic ejecta mass similar to that inferred of the 2004 Galactic GF from SGR 1806-20, we predict a peak UV/optical luminosity of $\sim 10^{39}$--$10^{40}\,\rm erg\,s^{-1}$ at $\sim 10$--$15$ minutes, rendering such events potentially detectable to several Mpc following a gamma-ray trigger by wide-field transient monitors such as {\it ULTRASAT/UVEX}. The peak luminosity and timescale of the transient increase with the GF strength due to the larger ejecta mass. Although GFs likely contribute $1$--$10 \%$ of the total Galactic $r$-process budget, their short delay-times relative to star-formation make them an attractive source to enrich the earliest generations of stars.

\end{abstract}

\keywords{}

\section{Introduction} \label{sec:intro}

Magnetars are a class of highly magnetized and slowly rotating neutron stars (NS), with typical surface dipole magnetic field strengths $B\sim 10^{14}$--$10^{15} \,\rm G$ and spin periods of several seconds \citep{Duncan&Thompson92,Kouveliotou+98}. They are further distinguished by a range of transient outbursts, powered by the sudden dissipation of their ultra-strong magnetic fields and categorized according to the total energy scale of their X-ray and/or soft $\gamma$-ray emission (\citealt{Mereghetti+2015,Turolla+2015, Kaspi2017}). At the top of this hierarchy lie giant flares (GFs) rare and noncatastrophic events releasing $E \sim 10^{44}\text{--}10^{46}\,\rm ergs$ in $\gamma$-rays, in under 1 second. Seven extragalactic GF candidates have been identified \citep{Burns+21, Beniamini+2024, Rodi+2025} and three GFs have been observed within our Galaxy or the nearby Large Magellanic Cloud over the last 50 years: in 1979 \citep{Mazets+1979, Evans+1980}, 1998 \citep{Hurley+1999}, and 2004 \citep{Palmer+2005, Hurley+05}. Despite originating from different magnetars, all three events comprised a brief ($\lesssim 0.5$ s) $\gamma$-ray burst followed by a $\sim$ minutes long pulsating hard X-ray tail modulated by the magnetar spin period. The two Galactic GFs (1998 and 2004) were also followed by a bright synchrotron radio afterglow \citep{Frail+1999, Hurley+1999, Cameron+2005, Gaensler+2005}. The remarkable similarities in the emission properties across the three events imply a common underlying mechanism behind GFs. Recently, it has become apparent that flaring magnetars are also sources of another type of transient emission: ms duration coherent bursts known as ``fast radio bursts'' (FRBs; \citealt{Lorimer+07,Bochenek+2020,CHIME+20}).

The prodigious December 2004 GF from SGR 1806-20 is amongst the most luminous transients observed in our Galaxy \citep{Hurley+05}. Its radio afterglow was found to be initially decaying on a timescale $\gtrsim 10$ days, followed by a brief rebrightening epoch at around 25 days. The initial decay phase was interpreted to arise from an expanding, baryon-loaded shell released from the magnetar surface during the flare, which rebrightens as the ejecta sweeps up and shocks matter in the ambient medium \citep{Gelfand+2005, Taylor+2005, Granot+2006}. Insofar as the initial $\gamma$-ray burst must arise from an $e^{-/+}$ pair--photon outflow with negligible baryon contamination (which would otherwise have softened the emission due to scattering), the baryonic ejecta powering the GF radio afterglows were likely released after this initial emission phase, as a byproduct of the flare operating mechanism \citep{Turolla+2015}.

 Magnetar GFs are believed to arise from sudden magnetic field reconfiguration leading to magnetic energy deposition and charged particle acceleration in the magnetosphere, for which several possible triggers have been invoked (see, e.g., \citealt{Turolla+2015, Kaspi2017} for reviews). An unstable magnetic field configuration evolving from perturbations to the strongly wound up internal field, arising due to mild crustal deformation or turbulent motion in the superconducting liquid core, can be relaxed through magnetic reconnection in the magnetosphere \citep{Thompson&Duncan1995, Thompson&Duncan2001, Ioka2001, Gill+Heyl2010}. Strongly twisted fields may accumulate in the magnetosphere due to such internal stresses, also resulting in a sudden energy release through reconnection instabilities similar to those driving solar flares \citep{Lyutikov2003, Lyutikov2006, Gill+Heyl2010, Parfrey+2013}. While the exact operating mechanism(s) remains uncertain, most models agree that the initial $\gamma$-ray spike is driven by an efficient conversion of magnetic energy through reconnection. 

\begin{figure*} 
    \centering
    \includegraphics[width=1.\textwidth]{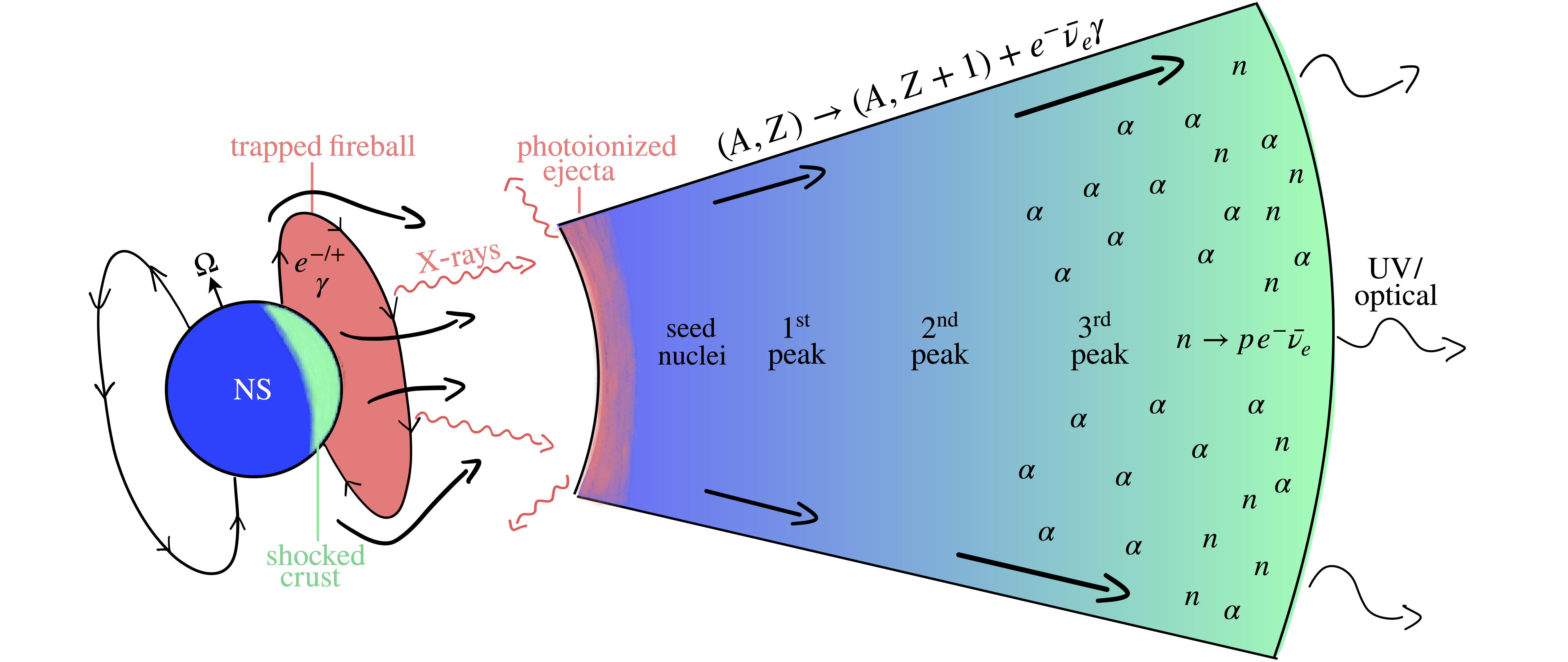}
\caption{Schematic illustration of the envisioned scenario of mass-ejection, $r$-process production, and {\it nova brevis} emission following a magnetar GF. Matter unbound from the shocked NS crust flows outwards along temporarily open field lines, ultimately reaching a homologous state characterized by radial stratification of the expansion velocity and entropy. While the innermost layers primarily synthesize light seed-nuclei ($A \approx 80-100$), faster expanding outer mass-layers undergo an $\alpha$--rich freeze-out, enabling the production of progressively heavier nuclei, including 2nd and 3rd $r$-process nuclei.  The outermost fastest layers expand so quickly that the $r$-process itself freezes-out prematurely, leaving a high abundance of free-neutrons.  Radioactive decay of $r$-process nuclei and neutrons powers the optical/UV {\it nova brevis} emission lasting up to several minutes.  Also potentially contributing to the ejecta heating is reprocessing by X-rays from the trapped $e^{\pm}$ fireball on the magnetar surface (Sec.~\ref{sec:tail}).}
    \label{fig:cartoon}
\end{figure*}

A generic consequence of such local magnetic energy dissipation is the generation of a high--pressure pair--photon outflow capable of temporarily forcing open the field lines (i.e., by exceeding the magnetic-field pressure); the field lines subsequently re-close, confining any lagging hot plasma near the surface (see Fig.~\ref{fig:cartoon} for an illustration). Radiative cooling of this trapped fireball naturally accounts for the prolonged X-ray tail of GFs \citep{Thompson&Duncan1995}. While the resultant radiative flux can exceed the magnetic Eddington limit and ablate some baryons from the magnetar surface (\citealt{Demidov2023}, see their Eq.~4), the predicted ejecta mass falls several magnitudes short of the $\gtrsim 10^{24.5}\,{\rm g}$ inferred for the 2004 GF radio afterglow \citep{Gelfand+2005, Granot+2006}; this implies that the bulk of the ejecta mass must have been released earlier, likely immediately after the initial outflow when the field lines were still open. \citet[hereafter \citetalias{Cehula+24}]{Cehula+24} developed a model for such prompt baryon ejection, where the primary magnetic field dissipation acts to generate a high pressure region just above the magnetar surface, which drives a shockwave into the crust capable of unbinding enough crustal material to satisfy the ejecta mass constraints. \citetalias{Cehula+24} further posit that the hot baryon ejecta may undergo heavy-element nucleosynthesis, as described below.

Roughly half the elements heavier than iron are produced through rapid neutron capture process ($r$-process) nucleosynthesis \citep{Burbidge+57,Cameron57}. The notion that decompressing NS crustal material could undergo an $r$-process was first considered by \citet{Lattimer&Schramm74, Lattimer+Schramm76, Symbalisty&Schramm82} who showed that a small fraction of the stellar mass can be stripped and dynamically ejected by tidal forces during the coalescence of a NS and another compact object. \citet{Lattimer+1977, Meyer1989} conducted more detailed studies of the nucleosynthesis in initially cold ($T \approx 0\,\rm K$), dense ($\rho \sim \rho_{\rm nuclear} \approx 3 \times 10^{14}\,\rm g\,cm^{-3}$), and neutron-rich matter (electron fraction $Y_e \lesssim 0.2$), which they posit characterizes the ejecta from such systems. They showed that under these conditions, an initial ``cold" $r$-process proceeds during decompression, whereby existing nuclei and free neutrons in the ejected crustal material are reprocessed into ``superheavy nuclei" through neutron drip, $(n, \gamma)$ reactions, $\beta$-decays, and fission cycling. The energy released from $\beta$-decays and fission would heat the material sufficiently (to temperatures $T_9 \equiv T/(10^{9}\rm \, K) \gtrsim 1$) such that neutron capture and photodisintegration equilibrium, $(n, \gamma) \rightleftharpoons (\gamma, n)$, is established and $\beta$-decays build up successively heavier nuclei in this subsequent ``hot" $r$-process (\citealt{Sato1974, Norman+Schramm1979}).

Modern studies of mass ejection in NS mergers have altered the picture laid out in these pioneering works (e.g., \citealt{Shibata&Hotokezaka19}).  Although early $r$-process scenarios envisioned material that remains cold (on nuclear energy scales) during its initial phases of decompression, most ejecta from neutron star-mergers likely experiences some form of dynamical heating (e.g., due to shocks, \citealt{Evans&Mathews1988}) to high temperatures ($T_9 \gtrsim 1)$, resulting in the dissociation of any preexisting nuclei in the crust. The resultant pool of free nucleons is transmuted into $\alpha$-particles and successively heavier seed-nuclei during expansion, and a hot $r$-process ensues once the ejecta cools to allow $(n, \gamma) \rightleftharpoons (\gamma, n)$ equilibrium. Whether a limited or full $r$-process is successful in this scenario (i.e., able to reach the second or third abundance peaks, respectively) depends on the electron fraction, initial thermodynamic state, and the hydrodynamics of the expansion (e.g., \citealt{Hoffman+97,Meyer&Brown97}).

LIGO's first detection of a binary NS merger and its observed electromagnetic counterpart \citep[e.g.][]{Abbott+17, Abbot+2017_EM, Arcavi+2017, Coulter+2017} confirmed that an $r$-process occurs in the ejecta from compact object mergers. In such events, the large ensemble of freshly synthesized radioactive nuclei with various decay lifetimes provides a continuous energy source powering the observed {\it kilonova} afterglow \citep{Li&Paczynski98, Metzger+10}, and the ejecta mass, opacity, and velocity play a key role in setting the timescales and dominant wavelengths of the emission. 

While NS mergers are a major if not dominant source of $r$-process elements in the Universe, both Galactic chemical evolution studies and isotopic analyses of meteorites indicate mergers are likely not the only source \citep{Qian&Wasserburg07, Ott+Kratz2008,Lugaro+2014, Cote+19, Tsujimoto+2021,Simon+2023}. Observations of $r$-process enhanced stars in ultrafaint dwarf galaxies implicate rare events with large $r$-process yields early in Galactic history (i.e., at low metallicity) corresponding to short delay times after star formation (e.g., \citealt{Cote+19,Zevin+2019,vandeVoort+20}), such as collapsars (e.g., \citealt{Siegel19,Miller+20,Issa+24}), magnetorotational SNe (e.g., \citealt{Thompson+04,Winteler+12}), or other sources involving rapidly accreting compact objects \citep{Grichener+22}. However, more frequent and lower yield $r$-process sources at low metallicity may also be required by observations \citep{Sneden+08,Qian&Wasserburg07,Thielemann+2020,Ou+2024}. The proto-NS winds following core collapse supernovae could in principle satisfy such a constraint; however, these sites likely do not achieve the requisite entropy to support a robust $r$-process\footnote{Possible exceptions include core-collapse SNe driven by a hadron-quark phase transition \citep{Fischer+2020} or PNS winds from particularly massive proto-NS \citep{Wanajo13}, with inclusion of convective wave heating \citep{Nevins&Roberts23}, and/or very strong magnetic fields and/or rapid rotation \citep{Thompson+03,Thompson+04,Metzger+07, Winteler+12,Mosta+14,Vlasov+17, Thompson&udDoula18,Desai+2022,Desai+2023,Prasanna+2023,Prasanna+2024}. However, the rate of such explosions remains uncertain and thus their ability to solve the missing $r$-process problem remains poorly constrained.}\citep{Qian&Woosley96}, and the production of modest quantities of $r$-process nuclei in ordinary explosions are challenging to test observationally, e.g. through their effects on the supernova light-curve \citep{Patel+24}. Alternate sites that operate promptly after star formation with rates comparable to supernovae are thus likely required to provide a complete description of the origin of heavy elements in the Universe.

Using 1D hydrodynamical simulations, \citetalias{Cehula+24} demonstrated that shock-heated ejecta from magnetar GFs can in principle possess the requisite combination of entropy $s$, $Y_e$, and hydrodynamical evolution to support a hot $r$-process. Despite the outer NS crust possessing only a moderate neutron excess ($Y_e \approx 0.39-0.46$), an $r$-process up to and/or surpassing the third peak is still possible within the decompressing ejecta via the ``$\alpha$-rich freeze-out'' mechanism (e.g., \citealt{Meyer+92}). In this scenario, a high ratio of neutrons to the ``seed nuclei'' onto which neutrons capture, can be achieved by suppressing seed formation given sufficiently high entropy (low density) and/or rapid expansion timescale (e.g., \citealt{Hoffman+97}).

If magnetar GFs eject radioactive nuclei, they should be accompanied by kilonova-like electromagnetic counterparts. However, given the very low ejecta masses $\lesssim 10^{-6}\,M_\odot$ (and associated short photon diffusion timescale) relative to the debris of NS-mergers, such emission is expected to peak on a timescale of $\sim$minutes after the GF, much faster than in kilonovae. \citetalias{Cehula+24} aptly denominate such rapidly-evolving radioactively powered GF transients as {\it novae breves}, events they estimate will emit at high effective temperatures $T_{\rm eff} \gtrsim 20,000\,\rm K$ corresponding to the ultraviolet (UV) band. 

Because the detection prospects of radioactively-powered emission following magnetar GFs are sensitive to its duration, luminosity, and spectral colors, better predictions for their multi-band light-curves are essential to motivate and design potential search strategies. These considerations are timely given that several wide-field UV survey missions are planned over the next few years: Ultraviolet Transient Astronomy Satellite ({\it ULTRASAT}, launch date 2027; \citealt{Sagiv+14}), Ultraviolet Explorer ({\it UVEX}; \citealt{Kulkarni+21}), and Quick Ultra-Violet Kilonovae surveyor ({\it QUVIK}; \citealt{Werner+23}). Aside from confirming and characterizing a new source of $r$-process nucleosythesis in the universe, {\it novae breves} in principle offer direct probes of the uncertain baryon ejection process in magnetar GFs. They may also provide insights into the magnetar-FRB connection, particularly the compact magneto-ionic synchrotron nebulae observed to surround repeating FRB sources (e.g., \citealt{Michilli+18}), which require enormous baryon injection rates, possibly from the accumulation of many magnetar flares over decades or longer \citep{Margalit&Metzger19}.

In this paper we expand upon the work of \citetalias{Cehula+24} by conducting full nuclear reaction network calculations of nucleosynthesis in magnetar GF ejecta, and then use our results to construct a semi-analytic model of {\it nova brevis} lightcurves. In Sec.~\ref{sec:model} we summarize the key dynamical and thermodynamic quantities from \citetalias{Cehula+24} and describe our nucleosynthesis calculations and lightcurve model. Results for a fiducial model, as well as weaker or more powerful GFs, all producing total ejecta mass within the range required to explain the baryon ejection from the 2004 flare from SGR 1806-20 are presented in Sec.~\ref{sec:results}. Sec.~\ref{sec:discussion} discusses implications of our results, including the contribution from magnetar GFs to the Galactic $r$-process inventory, detection prospects of {\it novae breves} with optical/UV telescopes, and related topics. We summarize our conclusions in Sec.~\ref{sec:conclusion}.

\section{Model Description} \label{sec:model}

\subsection{Crust Ablation in Giant Flares and Unbound Ejecta Properties} 

\label{sec:ejecta_estimates}

Here we overview the hypothesized mechanism of mass ejection from magnetar GFs, as supported by the hydrodynamic simulations of \citetalias{Cehula+24}. We also summarize the unbound ejecta properties used in our nucleosynthesis calculations. A schematic picture of the mass ejection and the site of its EM counterpart is given in Fig.~\ref{fig:cartoon}.  For additional details and derivations of the quoted results, see Appendix \ref{sec:ejecta}.

We assume that energy released during a magnetar GF acts to generate a pair-photon fireball above the NS surface, which exerts an external pressure, $P_{\rm ext}$, on the crust higher than in the pre-flare state. Because the flare's energy derives from the magnetic energy of the magnetosphere, the latter's magnetic energy density defines a characteristic scale for the external pressure:
\be
\label{eq:Pext}
P_{\rm ext} \sim P_{\rm mag} = \frac{B^{2}}{8\pi} \approx 4\times 10^{28}\,{\rm erg\,cm^{-3}}\,\left(\frac{B}{10^{15}\,{\rm G}}\right)^{2},
\ee
where $B$ is the surface magnetic field strength. However, note that $P_{\rm ext}$ produced by the GF may locally differ from the average $P_{\rm mag}$ across the whole magnetosphere, so this mapping should be taken only as a rough estimate (e.g., $B$ need not correspond to the global dipole field). Moreover, in some GF models the flare is powered by the much larger internal magnetic field of the NS \citep{Thompson&Duncan1995, Thompson&Duncan2001, Ioka2001}. 

The external pressure $P_{\rm ext} \equiv P_{\rm ext,30} \times 10^{30}$ erg cm$^{-3}$ drives a shock into the surface layers of the NS (of mass $1.4 M_{1.4}\,M_{\odot}$ and radius $12R_{12}\,\rm km$), heating the outer crust to a high thermal pressure. Once the flare concludes, the pair-photon fireball is free to expand to infinity along regions of the magnetosphere unconfined by the remaining magnetic field, typically on the light-crossing time, thus relieving the external pressure. Then the hot, shock-compressed crustal material finds itself over-pressurized with respect to its surroundings and begins to decompress and accelerate off the surface, akin to a spring uncoiling, with a portion of the shocked crust being ejected into space. Specifically, those outer layers of the crust which are shocked to a specific enthalpy exceeding the NS gravitational binding energy can be ejected. This critical condition defines the maximum mass-depth within the crust contributing to the unbound ejecta, and hence the total ejecta mass (Eq.~\eqref{eq:Mej_app}),
\be
    M_{\rm ej}\,[M_{\odot}] \approx 6.4\times 10^{-8}\,P_{\rm ext,30}^{1.43}R_{12}^{5.43}M_{1.4}^{-2.43}.
    \label{eq:Mej}
\ee
Given the applied external pressure, the Rankine–Hugoniot jump conditions define the immediate post-shock density, $\rho_{\rm sh}$, and thermal energy density, $e_{\rm sh}$, of each unbound layer. Labeled by Lagrangian mass-coordinate $m \leq M_{\rm ej}$ where $m=0$ corresponds to the outermost layer of the ejecta originating from magnetar surface, these follow profiles of the form (Eq.~\eqref{eq:rhocr_app}, \eqref{eq:e_sh_app}),
\be
    \rho_{\rm sh}(m)\,[\mathrm{g\, cm^{-3}}] \approx 3.5 \times 10^{10} M_{1.4}^{0.7} R_{12}^{-2.8}\bigg( \frac{m}{10^{-7} M_\odot}\bigg)^{0.7};
    \label{eq:rho_sh}
\ee
\be
    e_{\rm sh}(m)\,[\mathrm{erg\, g^{-1}}] \approx 8.6 \times 10^{19} P_{\rm ext, 30} M_{1.4}^{-0.7}R_{12}^{2.8}\bigg( \frac{m}{10^{-7} M_\odot} \bigg)^{-0.7}.
    \label{eq:e_sh}
\ee
Given an assumed equation of state of the post-shock ejecta, $\{ \rho_{\rm sh}(m), \, e_{\rm sh}(m) \}$ can be mapped to the associated specific entropy profile $s(m)$ (e.g., Eq.~\eqref{eq:s_app}). The latter is assumed to be conserved during the decompression phase until significant heating from nucleosynthesis begins. 

Asymptotically, the velocity profile of the unbound ejecta should approach that of approximately spherical homologous expansion. Motivated by the hydrodynamical simulations of \citetalias{Cehula+24}, we adopt the approximate density profile,
\be
\rho(v) = \frac{3}{4\pi}\frac{M_{\rm ej}}{(\bar{v}t)^{3}}\left(\frac{v}{\bar{v}}\right)^{-6}.
\label{eq:rhoprofile} 
\ee
which corresponds to a velocity profile,
\begin{eqnarray} \frac{v(m)}{c} &=& \frac{\bar{v}}{c}\left(\frac{m}{M_{\rm ej}}\right)^{-1/3} \nonumber \\
&\approx& 0.2 P_{\rm ext, 30}^{1/2} M_{1.4}^{-0.8}R_{12}^{1.8}\left(\frac{m}{10^{-7} M_\odot} \right)^{-1/3},
\label{eq:v}
\end{eqnarray}
where we take $\bar{v} = 0.18c$ (Eq.~\eqref{eq:dMdv_app}).
Thus, given an assumed equation of state and an initial electron fraction $Y_e$ at the depths in the outer NS crust of interest, the profiles $\{s(m), \rho(m), v(m)\}$ uniquely specify the initial thermodynamic state and approximate expansion profile of each ejecta layer, as described further in the next section. Analytic estimates of other key quantities to the nucleosynthesis outcome of each layer, such as the expansion time through the region where alpha-particles and seed nuclei form, are also provided in Appendix \ref{sec:ejecta} (though we do not make use of those estimates in our nucleosynthesis calculations). 

\subsection{Nucleosynthesis Calculations} 
\label{sec:sky_calcs}

We employ the nuclear reaction network {\it SkyNet} \citep{Lippuner&Roberts15,Lippuner&Roberts17} to calculate the nucleosynthetic yields of the magnetar flare ejecta. {\it SkyNet} evolves the abundances of 7843 species from free nucleons to $^{337}{\rm Cn}$ ($Z =112$), with over 140,000 nuclear reactions; it includes strong reaction rates from the JINA REACLIB database \citep{cyburt:10}, weak reaction rates from \citet{fuller:82, oda:94, langanke:00}, and REACLIB, and spontaneous and neutron-induced fission rates from \citet{frankel:47, mamdouh:01, wahl:02, panov:10}. The code uses a modified version of the Helmholtz equation of state (EOS) of \citet{Timmes&Swesty00} to calculate the entropy and include internal partition functions separately for each nuclear species (hereafter, {\it Helmholtz EOS} refers to this modified version). Nuclear masses and partition functions are included from the WebNucleo XML file distributed with REACLIB (i.e., with experimental data where available and otherwise using the FRDM mass model of \citealt{Moller+2016}).

We model the total ejecta $M_{\rm ej}$ with a one dimensional grid of $N$ discrete zones. The $n$'th ejecta layer contains a mass $\Delta m_{n}$ where $n = 1, 2,..., N$ (therefore, $\sum^N_{\mathrm{n = 1}} \Delta m_{n} = M_{\rm ej})$. We use the radial mass coordinate $m \in [0, M_{\rm ej}]$ to be ejected mass, such that $m = 0 (M_{\rm ej})$ corresponds to outermost(innermost) ejeca layer furthest(closest) from the magnetar. The ejecta layers are logarithmically spaced with finer resolution near the stellar surface (e.g. $\Delta m_1 \ll \Delta m_{N}$) and with the characteristic mass coordinate for each layer taken to be at its center. We follow a unique nucleosynthesis evolution for each mass layer as the matter decompresses away from the NS surface to lower densities and temperatures. The initial state of each layer is determined by the jump conditions across the shock (Appendix \ref{sec:ejecta}; Sec.~\ref{sec:ejecta_estimates}). The temperature at this post-shock stage is sufficiently high $k_{\rm B}T \gg 1$ MeV that all nucleons are free and the electron/positrons are relativistic. Alpha particles form as a layer decompresses and its temperature drops to $\approx k_{\rm B}T_{\alpha} \lesssim 1$ MeV (Eq.~\eqref{eq:kTalpha}). Because significant heating from nucleosynthesis only really begins at alpha particle formation, we have the freedom to start our {\it SkyNet} calculations around or before this time. For the same reason, our final results for the abundance distribution are not overly sensitive to the details of the earliest phases of expansion prior to alpha formation (as shown in Appendix~\ref{sec:tD_test} for variations to the $t_{\rm D}$ parameter entering the density profile described below).  

We adopt Lagrangian density trajectories of the general form:
\begin{equation} \label{eq:rho_LR}
    \rho(t)  = \rho_{\rm D}
	\begin{cases}
		\left(\frac{t}{t_{\rm D}}\right)^{-1},\,\,\,\, t_0 \le t \le t_{\rm D} \\
		\left(\frac{t}{t_{\rm D}}\right)^{-3},\,\,\,\,\, t \ge t_{\rm D},
	\end{cases} 
\end{equation}
where $t_0$ is the start time of the network calculation (after the shocked ejecta has already begun to decompress off the surface; see below) and $t_{\rm D} \equiv 2R_{\rm NS}/v$ is the time for the layer to expand a distance comparable to the NS diameter. This profile is motivated by the desire to smoothly interpolate between an early phase of approximate planar one-dimensional constant-velocity expansion ($t < t_{\rm D}$) and a late-time three-dimensional expansion phase. Matching Eq.~\eqref{eq:rho_LR} to the density profile achieved in the homologous phase (\citetalias{Cehula+24}; Eq.~\eqref{eq:rhoprofile}), provides the density normalization for each layer:
\begin{eqnarray}
\rho_{\rm D} &=& \frac{3}{4\pi}\frac{M_{\rm ej}}{(t_{\rm D}\bar{v})^{3}}\left(\frac{v}{\bar{v}}\right)^{-6} \nonumber \\
&\approx& 3.4\times 10^{6}\,{\rm g\,cm^{-3}} R^{-3}_{12} \left(\frac{M_{\rm ej}}{10^{-7}M_{\odot}}\right)\left(\frac{v}{\bar{v}}\right)^{-3}.
\label{eq:rho0}
\end{eqnarray}

The initial temperature of a given layer, $T_0 = T(t_0)$, at the start-time of the network calculation ($t = t_0 > 0$), is determined from the density $\rho_0 = \rho(t_0)$ (Eq.~\eqref{eq:rho_LR}) and the specific entropy $s$ of the layer. The latter is calculated from the {\it Helmholtz} EOS using the immediate post-shock density $\rho_{\rm sh} \approx 7 \rho_{\rm cr}$ and specific internal energy $e_{\rm sh} \approx 3P_{\rm ext}/\rho_{\rm sh}$, as determined from the jump conditions (see discussion after Eqs.~\eqref{eq:rho_sh}, \eqref{eq:e_sh}). We assume a radiation-dominated $\gamma = 4/3$ shock, as justified given the high entropy $s \gtrsim 10 k_{\rm B}$ baryon$^{-1}$ (\citealt{Qian&Woosley96}, their Eq.~34). The initial time $t_0$ is chosen such that the density and temperature start sufficiently high (e.g., $T_0 > few$ MeV) to justify nuclear statistical equilibrium in the initial state. We have verified that our results for the final abundances and nuclear heating rate are not sensitive to adopting earlier starting times.

We fiducially assume the same initial electron fraction $Y_{e}(t_0) = 0.44$ for all ejecta layers. This choice is informed by \citetalias{Cehula+24} (their Fig.~$2$), who find that $Y_{e}$ in the NS crust spans a modest range $0.39$--$0.46$ across mass-depths $M_{\rm ej} \lesssim 10^{-6}\,M_\odot$ corresponding to the ejecta masses of our models. However, we consider models with lower and greater $Y_e$ given the possibility of aspherical mass ejection and/or weak interactions which change $Y_e$ during the ejection process (Sec.~\ref{sec:Ye_param_study}).

Each layer is evolved self-consistently out of NSE to capture the full nucleosynthesis chain and resultant radioactive heating. A resolution of $N = 30$ ejecta layers is adopted for all models presented in this work, which we have checked is sufficient for convergence on the final abundance distribution and {\it nova brevis} light-curve properties (see Appendix~\ref{sec:tests}).

As an example, Fig.~\ref{fig:fid_inputs} shows the initial properties of each mass layer $m \leq M_{\rm ej}$ for the fiducial model ($B = 6 \times 10^{15}\,{\rm G} \Rightarrow P_{\rm ext} \approx 1.4 \times 10^{30}\,\rm erg\,cm^{-3} \Rightarrow M_{\rm ej} \approx 10^{-7} M_\odot$; Sec.~\ref{sec:fiducial}).  The middle and bottom panels compare various critical densities and temperatures, which in descending order include: the immediate post-shock state; the start of the reaction network calculations; and the time of alpha-particle formation (defined as when the alpha-particle mass fraction $X_\alpha$ first reaches $Y_e$, half its maximum value of $\simeq 2Y_e$). The top panel in Fig.~\ref{fig:fid_inputs} also shows the timescale of alpha-formation, $t_{\alpha}$. For the power-law density trajectory of the form \eqref{eq:rho_LR}, $t_{\alpha}$ also equals the expansion time of the fluid element at $\alpha$ formation $\{ \rho_\alpha, T_\alpha \}$, which is one of the critical quantities (along with the entropy and electron fraction) that enters the analytic criterion for achieving an $\alpha$-rich freeze-out (Eq.~\eqref{eq:zeta_app}). This value, shown with open circles in Fig.~\ref{fig:fid_inputs}, is comparable (within a factor of $\lesssim 3$) to an independent analytic estimate of $t_{\rm \alpha}$ derived in Appendix \ref{sec:ejecta} under different assumptions (Eq.~\eqref{eq:app_t_alpha}), which we show for comparison with solid circles.  

\begin{figure} 
    \centering
    \includegraphics[width=0.5\textwidth]{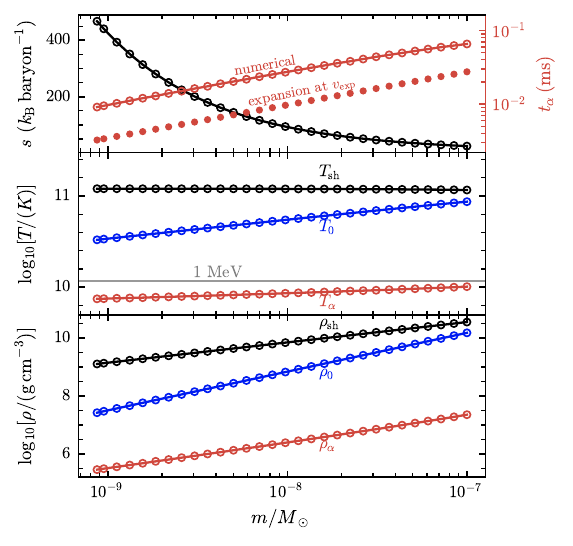}
\caption{Ejecta properties for the fiducial model ($P_{\rm ext} \approx 1.4 \times 10^{30}\,\rm erg\,cm^{-3}$; $B = 6\times 10^{15}$ G) with $N$ = 30 grid zones represented by circles. The bottom two panels show the temperature and density at three different times (in chronological order): immediately post-shock (black), at the start of nucleosynthesis calculations (blue), and at $\alpha$-formation (red). The entropy shown in the top panel (black; left axis) is calculated from the {\it Helmholtz} EOS using the immediate post-shock properties and is assumed to be conserved during the decompression phase from $\{\rho_{\rm sh},\, T_{\rm sh}\} \rightarrow \{\rho_{\alpha},\, T_\alpha\}$ and is thus used to describe the ejecta state at the start of the network calculations $\{\rho_{0},\, T_{0}\}$. 
The top panel also shows the expansion timescale through the $\alpha$-particle formation region $t_{\alpha} = (\rho_{\rm D}/\rho_{\alpha})t_{\rm D}$ based on our adopted Lagrangian density trajectory (red circles; right axis) in comparison to an independent analytic estimate based on the initial expansion speed of the layer (solid red dots; Eq.~\eqref{eq:app_t_alpha}).}
    \label{fig:fid_inputs}
\end{figure}
\subsection{Light-Curve Model}

We follow a semi-analytic layer-by-layer light-curve model for the {\it nova brevis}, similar to that presented in \citet{Metzger19} in the context of kilonovae. The optical depth external to a given layer of mass $m$ and velocity $v$ (Eq.~\eqref{eq:v}) is
\be \label{eq:tau_numerical}
\tau(v, t) = \int_v^\infty \kappa(v', t) \rho(v', t) dv',
\ee
where the density profile $\rho \propto v^{-6}$ is given by Eq.~\eqref{eq:rho0} and
\begin{eqnarray}
&& \kappa(t) = \nonumber \\
&& X_{p}(t)\kappa_{p} + X_{\alpha}\kappa_{\alpha} + X_{\rm seed}\kappa_{\rm seed} + X_{\rm 2nd}\kappa_{\rm 2nd} + X_{\rm 3rd}\kappa_{\rm 3rd}, \nonumber \\
\end{eqnarray}
is the opacity of the layer in question. Here, we define separate contributions to the opacity from protons ($\kappa_{p} \approx 0.38$ cm$^{2}$ g$^{-1}$), alpha-particles ($\kappa_{\alpha} \approx \sigma_{T}/2m_p \approx 0.2$ cm$^{2}$ g$^{-1}$ assuming full ionization, a reasonable approximation given the high temperatures of interest), light seed nuclei near or just above the Fe-peak ($\kappa_{\rm seed} \approx 0.5$ cm$^{2}$ g$^{-1}$), second-peak/light $r$-process nuclei ($\kappa_{\rm 2nd} \approx 3$ cm$^{2}$ g$^{-1}$), and third-peak $r$-process nuclei ($\kappa_{\rm 3rd} \approx 20$ cm$^{2}$ g$^{-1}$). The fiducial $r$-process opacity values adopted here are rough estimates based on atomic structure calculations (e.g., \citealt{Kasen+13,Tanaka+20}), but our results are not overly sensitive to these assumptions. We account for the changing proton fraction $X_p(t)$ as a result of free neutron decay over the duration of the transient (see Sec.~\ref{sec:results} for further discussion). The abundances of most nuclei and the $\alpha$-particles are approximately constant over timescales relevant to the light-curve, $t \gtrsim 10\, \rm s$.

Numerically integrating Eq.~\eqref{eq:tau_numerical}, we obtain the time-dependent optical depth $\tau(t)$ for each layer. The homologous condition $r =vt$ connects the velocity to the radius of each layer, from which the radial optical depth profile $\tau(r)$ at any time follows. Since our simulation grid consists of a discrete number of layers, we linearly interpolate between the optical depth at these radii to achieve a smooth radial profile. 

Now it is straightforward to determine the evolution of two locations in the ejecta critical to the emission.  First we consider the diffusion surface, located at $\tau(r) = c/v$, and corresponding to mass coordinate $M_{\rm diff}(t)$ above which the photon diffusion time is shorter than the dynamical expansion timescale of the ejecta, thus enabling radiation to escape without significant adiabatic losses. By contrast, radiation released below the diffusion surface, $m > M_{\rm diff}(t)$, experiences sizable adiabatic losses (i.e., is ``trapped'') and hence its contribution to the light-curve can be neglected (we check this approximation at the end of this section). The diffusion surface passes through a given mass shell on a timescale which can be roughly approximated as (e.g., \citealt{Khatami&Kasen19})
\be
t_{\rm pk}(m) \approx \left(\frac{\kappa(t_{\rm pk}) m}{4\pi v c}\right)^{1/2}.
\label{eq:tpk}
\ee

Ejecta layers above the diffusion surface can further be divided into two regions, separated by the photosphere $M_{\rm ph}(t)$ located at $\tau(r) \approx 2/3$. Layers below the photosphere are optically thick, while those above the photosphere are optically-thin and will radiate with a different spectrum, as discussed below. 

Each ejecta layer experiences a unique nucleosynthesis evolution (Sec.~\ref{sec:results}), and hence will experience a different radioactive heating rate. The specific heating rate of a given layer $m$ can be written as
\be
\dot{q}_{\rm heat}(m,t) = f_{\rm th}(m,t)\dot{q}(m,t),
\ee
where $\dot{q}(m,t)$ is the specific radioactive power of layer $m$ (the {\it total} power, $m\dot{q}(m,t)$, is shown in Fig.~\ref{fig:trajectories} for the fiducial model) and $f_{\rm th} \le 1$ is the thermalization efficiency; the latter accounts for the fraction of the radioactive decay power that shares its energy with the plasma and is available to be radiated, instead of being lost to neutrinos or escaping gamma-rays (e.g., \citealt{Metzger+10,Barnes+16,Hotokezaka+16}). Given the high ejecta densities at times of interest, we assume that 100\% of the kinetic energy of $\beta$--decay electrons is thermalized with the plasma (e.g., \citealt{Metzger+10}). A conservative estimate of the thermalization timescale accounting only for bremsstrahlung losses \citep{Metzger+10} shows this to be a good assumption for all ejecta layers at early times $t < 10^3\,\rm s$ corresponding to the peak of the light-curve. Thus, we have $f_{{\rm th},n} \approx 0.38$ for heating due to free-neutron $\beta$--decay because 38\% of the decay energy goes into the $\beta$--decay electron while the remaining 62\% goes into the escaping neutrino (e.g., \citealt{Kulkarni05}). The radioactive heating from light $r$-process elements is also dominated by $\beta$--decays, but in this case typically only $\sim 25\%$ of the decay energy goes into the fast-electron while $\sim 40\%$ goes into gamma-rays (e.g., \citealt{Barnes+16}, their Fig.~4). A gamma-ray will deposit its energy in the ejecta with a probability given by $\tau (\kappa_{\gamma}/\kappa)$, where $\kappa_{\gamma} \approx 0.2$ cm$^{2}$ g$^{-1}$ is an estimate of the opacity to $\beta$--decay gamma-rays of typical energy $\approx 0.5$ MeV (e.g., \citealt{Metzger+10}). The combined efficiency of neutron and $r$-process heating can thus be written:
\begin{eqnarray}
f_{\rm th}(t)  &=& 0.38\frac{\dot{q}_{n}(t)}{\dot{q}(t)} \nonumber \\
&+& \frac {\dot{q}_{r}(t)}{\dot{q}(t)}\bigg[0.25 + 0.4\left\{1-\exp\left(-\tau(t) \kappa_{\gamma}/\kappa(t)\right)\right\}\bigg], \nonumber \\
\label{eq:fth}
\end{eqnarray}
where $\dot{q}_{n}(t)$ and $\dot{q}_{r}(t) = \dot{q}(t)-\dot{q}_{n}(t)$ are the contributions to the radioactive power from free neutrons and $r$-process elements, respectively (we show these as separate dashed and dotted lines for the fiducial model in Fig.~\ref{fig:fid_LC}; see Appendix~\ref{sec:n_decay} for derivation of free neutron decay power).  

Arnett's law dictates that the radiated luminosity at any time can be estimated as the total radioactive heating above the diffusion shell \citep{Arnett80}. This emission can be divided into that emitted in optically-thick and optically-thin regions of the ejecta. The luminosity of optically-thick emission corresponds to the radioactive heating of those shells between the diffusion shell and the photosphere, i.e., $L_{\rm ph} = \int_{M_{\rm ph}}^{M_{\rm diff}} \dot{q}_{\rm heat}dm$. This emission component is assumed to radiate as a blackbody spectrum at a temperature
\be
T_{\rm ph} = \left(\frac{L_{\rm ph}}{4\pi \sigma R_{\rm ph}^{2}}\right)^{1/4},
\ee
where $R_{\rm ph}$ is the photosphere radius. 

The luminosity of optically-thin or ``nebular" emission corresponds to the total radioactive heating rate released by ejecta layers above the photosphere, i.e., $L_{\rm neb} = \int_{0}^{M_{\rm ph}} \dot{q}_{\rm heat}dm$. Although a crude approximation, we also adopt a Maxwellian spectrum for the optically-thin emission, with an assumed effective temperature $T_{\rm eff} = 6000\,\rm K$ motivated by supernova nebular emission, particularly its quasi-thermal spectral shape and peak (see discussion in \citealt{Barnes&Metzger22} for justification). While most supernovae do not contain sizable $r$-process elements, we may expect the nebular line emission from light seed nuclei to be similar to those of the Fe-rich ejecta of e.g. Type Ia supernovae.  When assessing the detectability of the nova brevis emission we shall consider the uncertainty associated with this choice by exploring how our results change if we neglect the optically-thin component altogether (i.e., if we assume $L_{\rm neb} = 0)$.

Together, the radiation from the optically-thick and optically-thin regions with their respective spectral energy distributions, contribute to the photometric band evolution (see Fig.~\ref{fig:fid_LC} for the fiducial model). For a source at distance $d$, the AB magnitude of emission at observing frequency $\nu$ is calculated by summing the optically-thick and optically-thin contributions to the flux according to:
\begin{eqnarray}
F_\nu &=& F_{\nu, \, \rm ph} + F_{\nu,\, \rm neb} \nonumber \\
&=& \frac{1}{4 d^2\sigma_{\rm SB}} \left[ \frac{L_{\rm ph}}{T_{\rm ph}^4} B_\nu(T_{\rm ph}) + \frac{L_{\rm neb}}{T_{\rm eff}^4} B_\nu(T_{\rm eff}) \right]
\label{eq:Fnu}
\end{eqnarray}
Finally, we note that our assumption that the transient luminosity at time $t$ follows the instantaneous radioactive heating of all layers with $t_{\rm diff} < t$ relies on the implicit assumption that thermal energy deposited at smaller radii/earlier times ($t \ll t_{\rm diff}$), correcting for adiabatic losses until $t_{\rm diff}$, is negligible relative to the thermal energy deposited over the recent expansion time.  This assumption is valid provided that the radioactive heating rate decays sufficiently shallowly, e.g. $\delta \le 2$ if $\dot{Q} \propto t^{-\delta}$ since $\delta \approx 2$ corresponds to the energy loss rate via adiabatic expansion of radiation dominated matter. This condition is satisfied for our radioactive heating law, as shown in the top panel of Fig.~\ref{fig:fid_LC} with a dashed line to guide the eye.

\section{Results} \label{sec:results}
\subsection{Fiducial Model}
\label{sec:fiducial}

\subsubsection{Nucleosynthesis Yields}

\begin{figure*} 
    \centering
    \includegraphics[width=1.0\textwidth]{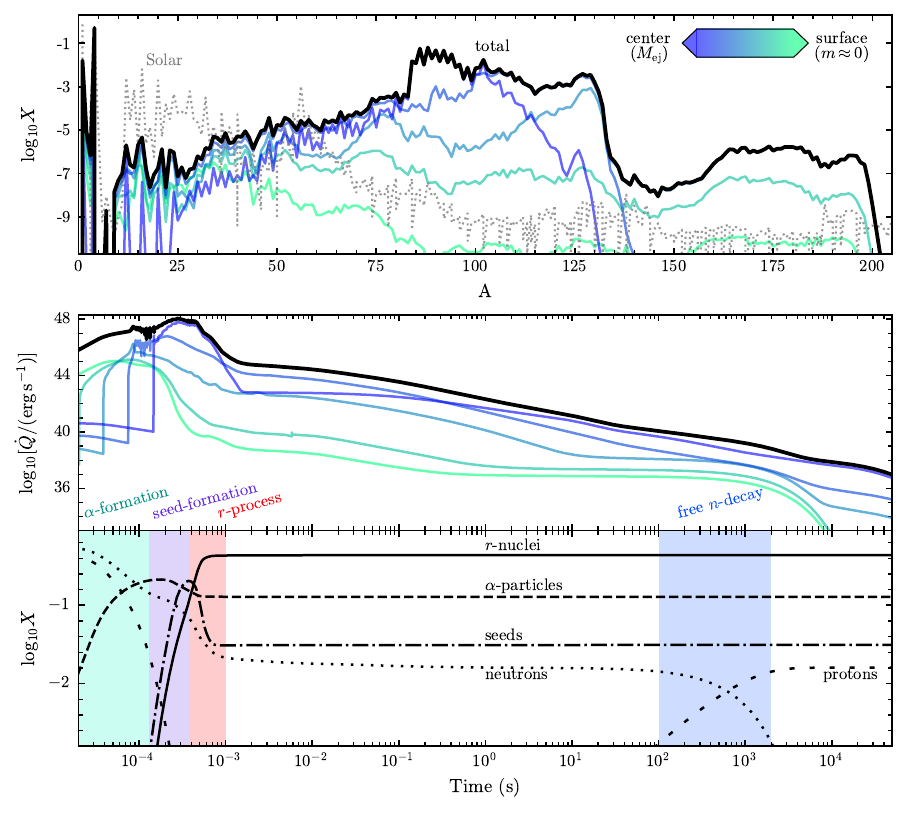}
\caption{Results of nucleosynthesis calculations for the fiducial model. The 30 ejecta mass layers are binned in 5 groups for visual clarity, and their mass weighted nucleosynthetic yields (top panel) and radioactive heating (middle panel) are shown. The lightest green line shows the first bin containing the six ejecta layers closest to the ejecta surface, while successively bluer lines show the deeper bins through to the bin with the six deepest ejecta layers in darkest blue. The black lines show mass weighted average quantities for the total ejecta. The Solar System abundances \citep{Lodders+2020} are included in gray dotted lines for comparison (top panel). The compositional evolution of the total ejecta is shown in the bottom panel.}
\label{fig:trajectories}
\end{figure*}

Detailed results from the nuclear reaction network calculations are shown in Fig.~\ref{fig:trajectories} for the fiducial model: final nucleosynthesis yields (top panel), radioactive power (middle panel), and time evolution of the total ejecta composition highlighting key nucleosynthesis phases. The 30 ejecta layers are binned into 5 groups of 6 layers each; the mass-weighted contributions to the total abundances and radioactive power of each bin are illustrated by a qualitative color scheme, with greener lines representing bins containing layers closer to the ejecta surface ($m = 0$) and bluer lines indicating bins with deeper ejecta layers ($m = M_{\rm ej}$). We describe below the differing nucleosynthesis trends from the outer ejecta to the inner ejecta, showing how they together produce the mass-weighted total ejecta properties represented by black lines.

The outermost ejecta layers experience extremely rapid expansion $t_\alpha \sim 10^{-2}\,\rm ms$ with high initial entropy $s \gtrsim 200\,\rm k_{\rm B}\,baryon^{-1}$ (Fig.~\ref{fig:fid_inputs}). These are the requisite conditions for a successful hot $r$-process via the $\alpha$-rich freeze-out mechanism \citep{Meyer+92}, as described qualitatively in Sec.~\ref{sec:intro}. In NSE with high entropy and $Y_e < 0.5$, protons efficiently convert into $\alpha$-particles via (the effective reaction) $2n + 2p \rightarrow \alpha + \gamma$ at $t \sim t_{\alpha}$ corresponding to $T_9 \approx 10$ (Fig.~\ref{fig:fid_inputs}), with the leftover neutrons remaining free. When the ejecta cools to $T_9 \approx 5$, carbon is produced by combination of ``normal" and neutron-aided $3\alpha$-reactions (i.e., $^4{\rm He}(\alpha,\gamma)^8{\rm Be}(\alpha, \gamma)^{12}\rm C$ and $^4{\rm He}(\alpha n, \gamma)^9{\rm Be}(\alpha, n)^{12}\rm C$, respectively), enabling chain production of successively heavier nuclei up to the Fe-group ($A \approx 60$) via $(\alpha , \gamma)$ reactions (e.g., \citealt{Woosley&Hoffman92}). However, this $\alpha$-ladder freezes out by $T_9 \approx 2.5$, which, for high entropy and/or rapid expansion time-scales, occurs before $\alpha$-particles are efficiently converted to seed nuclei. Thus, fewer seed nuclei are produced and a large fraction of the mass is instead trapped in inert $\alpha$-particles, thereby providing a greater neutron to seed ratio and enabling the subsequent $r$-process to proceed to heavier nuclei, potentially up to the third peak ($A \approx 190)$ or beyond. Indeed, we find substantial $\alpha$-particles and a moderate fraction of third-peak nuclei at the end of the evolution, as shown in the top panel of Fig.~\ref{fig:trajectories}. 

In the outermost layers with the shortest expansion timescales, the density drops so rapidly that even the $n$-capture reactions themselves freeze-out (e.g., \citealt{Mumpower+12}) despite the free neutron mass-fraction remaining sizable (e.g., \citealt{Goriely+14,Metzger+15_n_precursor,Lippuner&Roberts15}). This is evident by the plateau in the free neutron fraction at $X_n \approx 0.02$ at $t \gtrsim 10^{-2}$ s, indicating neutron captures have ceased and the $r$-process has frozen out. Thus, it is not the extremal outer-most ejecta layers ($m/M_{\rm ej} \lesssim 0.05$), but rather those slightly further in ($ 0.05 \lesssim m/M_{\rm ej} \lesssim 0.1$) that undergo a strong $\alpha$-rich freeze-out but {\it not} a significant $n$-capture freeze-out, which dominantly contribute to the third $r$-process peak. The residual free neutrons in the outermost ejecta layers decay to protons with a lifetime $\tau_n \approx 15$ minutes, as indicated by the blue shaded region. The energy released from neutron decay is visible at $t \sim \tau_n$ in the radioactive heating of the outer ejecta layers, ultimately producing a subtle ``bump" in the total mass-averaged heating rate. 

In deeper ejecta layers, a robust third peak $r$-process is not achieved for the fiducial model. The low entropy and slower expansion timescales are insufficient to suppress seed formation enough to enable a high ratio of neutrons to seed nuclei. However, a weaker $\alpha$ freeze-out still occurs, resulting in production up to the second peak ($A \approx 130$) in significant quantities, with some such nuclei produced in mass fractions comparable to that of seed and light $r$-process nuclei ($A \approx 90$). The deepest ejecta layers, with the lowest fraction of left-over $\alpha$-particles, contribute much of the prominent seed/light $r$-process peak in the final abundance distribution.

In summary, the abundance pattern of the entire ejecta receives contributions from different ejecta layers characterized by distinct nucleosynthetic paths. The fastest outermost layers primarily contribute free-neutrons and $\alpha$-particles due to a strong $\alpha$-rich freeze-out and a premature $n$-capture freeze-out preventing significant $r$-nuclei production. At greater depths but still well within the outer ejecta, a strong $\alpha$-rich freeze-out occurs with a weaker $n$-capture freeze-out, thus producing most of the 3rd peak $r$-process in the ejecta. The remainder of the ejecta at yet greater depths exhibits a natural continuum of this behavior as the dominant nuclei production shifts down to second-peak $r$-process and light $r$-process/Fe-group in the moderately deep layers, and exclusively to light-$r$-process/seed nuclei in the deepest layers. These layer-by-layer contributions are illustrated schematically in Fig.~\ref{fig:cartoon}.  

The condition to achieve an $\alpha$-rich freeze-out can be framed in terms of an approximate analytic criterion based on the ratio $\zeta \equiv s^{3}/t_{\alpha}Y_e^3$ (Eq.~\eqref{eq:zeta_app}; \citealt{Hoffman+97}), where higher values of $\zeta$ result in a higher neutron-to-seed ratio and hence the synthesis of heavier elements (see extensive discussion in Appendix \ref{sec:ejecta}). In the top panel of Fig.~\ref{fig:params}, we show the range of $\zeta$ achieved by the ejecta layers of the fiducial model ($B_{15} = 6)$, illustrating that the approximate threshold value of $\zeta$ required to reach the 3rd $r$-process peak (gray shaded region) is indeed achieved most robustly by the faster outer ejecta layers.

\subsubsection{Light-Curves}

\begin{figure*} 
    \centering
    \includegraphics[width=1.\textwidth]{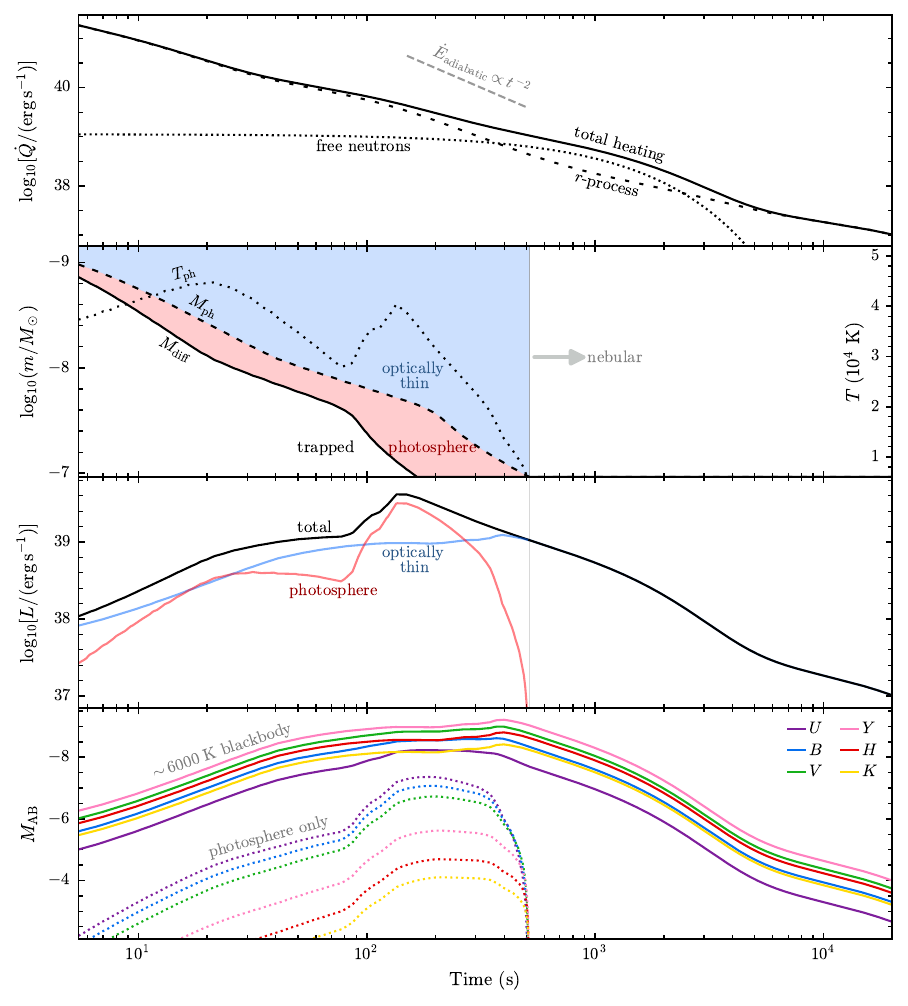}
\caption{Top panel: Total radioactive heating rate including thermalization corrections, broken down into separate contributions from decaying free neutrons and $r$-process nuclei. A dashed line shows the slope of the adiabatic energy loss rate $\propto t^{-2}$; this being steeper than the radioactive heating rate reveals that at a given time $t$, stored thermal energy from heating at times $\ll t$ has a negligible impact on the light-curve. Second panel: Mass coordinate of the diffusion shell, $M_{\rm diff}$, and photosphere, $M_{\rm ph}$, showing those regions of the ejecta where radiation is trapped ($m > M_{\rm diff}$), radiated from optically-thick regions below the photosphere ($M_{\rm diff} < m < M_{\rm ph}$; red), and radiated above the photosphere as optically thin emission ($m < M_{\rm ph}$; blue). The dotted line also shows the effective temperature of the photosphere (right vertical axis). The photosphere recedes all the way through the ejecta by $t \approx 500\,\rm s$ (light gray vertical line), after which point all emission is optically thin. Third panel: The total luminosity (black line), accounting for both photospheric (red line) and optically thin (blue line) contributions. Bottom panel: Absolute magnitudes in UV, optical, and infrared bands. The solid lines include radiation from both the photospheric and optically thin regions, while the dotted lines include only the photospheric emission, representing a lower bound on the emission, given uncertainties in the SED of the optically thin emission.} 
\label{fig:fid_LC}
\end{figure*}

Here we present the {\it nova brevis} light-curve for the fiducial model, calculated as described in Sec.~\ref{sec:model}. We decompose the total radioactive heating into the contributions from free neutrons and $r$-process nuclei, accounting for the different thermalization efficiencies of each (Eq.~\ref{eq:fth}), as shown in the top panel of Fig.~\ref{fig:fid_LC}. Decay of $r$-process nuclei dominates the heating rate for most of the evolution, but free neutron decay dominates at times $t \sim \tau_n \approx 900\,\rm s$. As shown in the second panel, the heating is shared among three regions of the ejecta: (1) the ``trapped'' region below the diffusion shell from which radiation is not able to diffuse outwards over a dynamical timescale (below the solid black line); (2) the optically-thick ``photosphere'' emission region above the diffusion shell but below the photosphere (shaded red); and (2) the optically-thin emission region above the photosphere (shaded blue). By $t \approx 180\,\rm s$, the diffusion shell has receded through all mass layers to the center of the ejecta, after which all the thermalized radioactive heating contributes to the escaping radiation. The nebular phase begins at $t \approx 500\,\rm s$ when the photosphere has also receded to the ejecta center, such that all of the emission now occurs from optically-thin regions.

The evolution of the bolometric luminosity shown in panel three results from the evolving distribution of mass between the three regions described above. The latter in turn arises from the non-uniform opacity structure of the ejecta corresponding to the radial distribution of different nuclei through the ejecta (Fig.~\ref{fig:cartoon}). At early times $t \lesssim 20\,\rm s$, the diffusion shell recedes through the ejecta at a slightly faster rate than the photosphere and thus the mass contained between the two surfaces (which contributes to the observed radiation) increases at a faster rate than the mass contained above the photosphere (note the logarithmic scale); this results in the photospheric emission increasing faster than the optically thin emission. This behavior is also reflected as an increase in the photosphere emission temperature (dotted line in panel 2, right y-axis). The opposite is true for $20~{\rm s} \lesssim t \lesssim 100~\rm s$, where the mass contained between the photosphere and diffusion shell is decreasing, resulting in a decline in the photosphere luminosity and temperature.

The diffusion shell exhibits a notable ``knee''-like feature around $t \approx 100\,\rm s$. This corresponds to when the diffusion shell has receded to the inner ejecta layers containing primarily seeds and light $r$-process nuclei, which contribute significantly less opacity compared to the larger abundances of second-peak and heavier nuclei in the exterior layers. The diffusion shell thus recedes through these low-opacity layers more quickly, while the photosphere is still receding slowly through the high opacity layers above it. During this time, mass in the previously-trapped region quickly accumulates in the photospheric region, generating a steeply rising photospheric temperature and luminosity. This continues until the diffusion shell reaches the innermost edge of the ejecta at $t \approx 180\,\rm s$, corresponding to the peak luminosity $L_{\rm pk} \approx 3\times 10^{39}$ erg s$^{-1}$. After this point, the light-curve monotonically decreases since there is no remaining trapped ejecta yet to become visible and the total heating rate in the ejecta is declining.

Finally, absolute magnitudes in the AB system, calculated from the weighted combination of photosphere and optically thin emission (Eq.~\ref{eq:Fnu}), are shown in the bottom panel with solid colored lines. The total spectral energy distribution peaks at all times in the near-infrared bands, reflecting the dominant contributions from the lower temperature ($T_{\rm eff} = 6000 \,\rm K$) of the assumed optically thin emission. The bump in the $U$-band light-curve arises from the spike in the bolometric luminosity that occurs near peak ($t \approx 200\,\rm s$) since around this time the emission becomes dominated by the photosphere component, with $T_{\rm ph} \approx (3\text{-}4)\times 10^{4}{\rm K} \gg T_{\rm eff}$, which enhances the contribution to the bluer bands.

Due to the large uncertainties associated with assigning a $6000 \, \rm K$ blackbody spectrum to the nebular-phase emission, we also show with dotted lines light-curves calculated based only on the photospheric emission. As expected due to the high photosphere temperatures $T_{\rm ph} \gtrsim 2\times 10^{4}$ K, the spectrum in this case peaks at ultraviolet frequencies. These ``photosphere only" light-curves provide a conservative lower bound in the predicted emission since they neglect any optically thin emission contribution altogether.

\subsection{Impact of Varying the Flare Strength}

\label{sec:param_study}
Now we explore models of various flare strengths, chosen to sample a range of total ejecta mass $\sim10^{-8}\text{--}10^{-6} M_\odot$ similar to the ejecta mass range inferred based on modeling the 2004 GF's radio afterglow \citep{Gelfand+2005,Granot+2006}. Our models are parameterized by $B_{15} \equiv B/(10^{15}\,{\rm G}) =  \{3, 6, 9, 13 \}$ (Eq.~\ref{eq:Pext}), where $B_{15} = 6$ corresponds to the fiducial model already described. We reiterate that the magnetic field here is just a convenient way to parameterize the external pressure generated by a given GF, which need not represent the magnetar's large-scale dipole field (see Sec.~\ref{sec:ejecta_estimates}). All other parameters remain unchanged from the fiducial case.

\begin{figure} 
    \centering
    \includegraphics[width=0.5\textwidth]{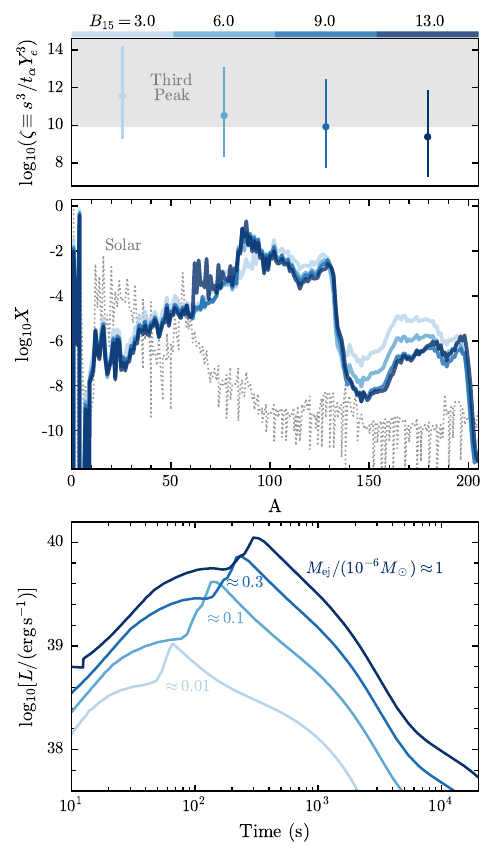}
\caption{Magnetar GF properties varying the external pressure generated by the magnetar GF (equivalently, surface magnetic field strength $B$; Eq.~\eqref{eq:Pext}) compared to the fiducial model ($P_{\rm ext} = 1.4\times 10^{30}$ erg cm$^{-3}$; $B_{15}=6$).  The middle panel shows the total final abundance patterns, while the bottom panel shows the bolometric {\it nova brevis} light-curves. The top panel shows, for each model as marked along the top, the spread across the ejecta layers $m \in [0,M_{\rm ej}]$ of the critical parameter $\zeta \equiv s^{3}/t_{\alpha}Y_e^3$ (Eq.~\eqref{eq:zeta_app}) relative to the analytic threshold necessary to achieve third-peak $r$-process via the $\alpha$-rich freeze-out. The mass-weighted average $\zeta$ across all layers for each model represented by a dot lies towards the lower end of the spread in $\zeta$ since the outer ejecta layers characterized by relatively large $\zeta$ make up only a small fraction of the total ejecta mass (due to the logarithmically spaced mass grid; Sec.~\ref{sec:model})}
\label{fig:params}
\end{figure}

Nucleosynthesis yields and bolometric light-curves for the model suite are presented in Fig.~\ref{fig:params}. One key feature of our results is that lower-energy flares produce a larger mass-fraction of heavy $r$-process nuclei. This can be understood by considering how the shock energy $\propto B^2$ (Eq.~\ref{eq:Pext}) is distributed over the total ejecta mass $M_{\rm ej} \propto B^3$ (Eq.~\ref{eq:Mej}). The shock deposits more energy per gram of ejecta for lower values of $B$, enhancing the specific entropy of a given ejecta layer. The combination of greater specific entropy and slower expansion time-scale through $\alpha$-formation $t_\alpha$, more readily satisfy the \citet{Hoffman+97} criterion for a robust $r$-process as shown in the top panel of ~\ref{fig:params}. Indeed, comparison of the the nucleosynthesis yields for different models (middle panel) makes evident that a stronger $\alpha$-rich freeze-out is achieved as we reduce $B$. 

However, the strongest flare $B_{15} =13$ achieves a more robust third peak $A\approx 190$ than the weaker flares, despite possessing a lower \citet{Hoffman+97} metric, $\zeta$. While this metric can delineate which ejecta achieves a robust $\alpha$-rich freeze-out, it does not directly probe the efficiency of the neutron capture freeze-out. The $B_{15} = 13$ model does achieve a weaker $\alpha$-rich freeze-out than the less energetic flares, consistent with its lower $\zeta$ value; however, in the outer $\sim 10\%$ of ejecta layers (those that contribute to the total third peak abundance) the freeze-out of $n$-captures is suppressed sufficiently that the ratio of (capturable) neutrons to seed nuclei is actually enhanced compared to the $B_{15} = 6,\,9$ models, enabling synthesis of heavier nuclei. Although not shown here, we have confirmed that this trend continues with greater flare strengths. This implies that for sufficiently rapid expansion time-scales, one must also consider the strength of $n$-capture freeze-out, in addition to the $\zeta$ metric, to make more detailed predictions of the mass distribution between neutrons, $\alpha$-particles, and heavy nuclei during the $r$-process.

Irrespective of how mass is distributed between the $r$-process abundance peaks, the total $r$-process mass fraction, $X_r$, increases with flare strength. We find $X_r\approx 0.27, \,0.44,\, 0.49,$ and $0.57$ for $B_{15} = 3,\,6,\,9,\, 13$, respectively. This can be understood with same arguments as above, however, instead of considering only the ejecta layers which contribute to the third $r$-process peak, we consider the total ejecta, since all layers achieve at least a first peak $r$-process. Then it is evident that, averaged over the total ejecta, the $n$-capture freeze-out grows disproportionately weaker compared to the $\alpha$-freeze-out, enabling more mass to accumulate in $r$-process nuclei. In other words, the ratio between frozen $\alpha$-particles and frozen neutrons (and therefore, the neutron to seed ratio) is increasing monotonically with flare strength due to slower expansion time-scales.

The enhanced opacity from the greater $r$-process production in stronger flares could in principle play a role in the light-curve evolution of the {\it nova brevis}, particularly during the optically thick phase; however, in practice the order $\sim 10\%$ increase in the $r$-process mass fraction has only a minor effect on the primary transient properties compared to the trends which follow simply from the overall larger ejecta mass. As expected, stronger flares with greater ejecta masses generate brighter and longer duration transients (Fig.~\ref{fig:params}, bottom panel).

\subsection{Impact of Varying the Electron Fraction}

\label{sec:Ye_param_study}

\begin{figure} 
    \centering
    \includegraphics[width=0.5\textwidth]{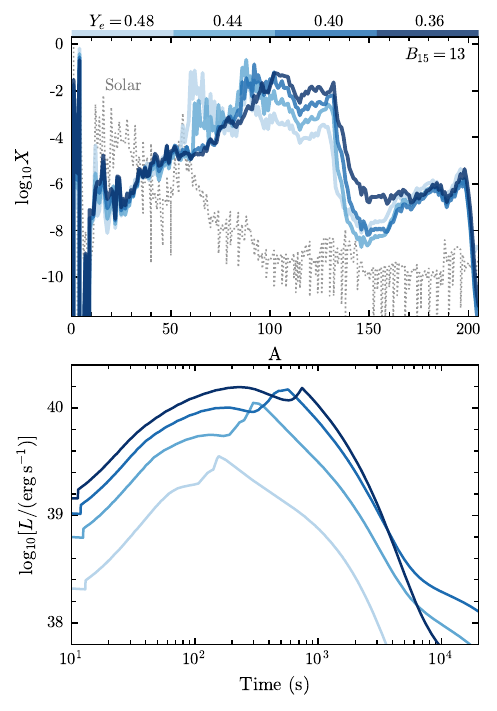}
\caption{Nucleosynthesis yields and {\it nova brevis} light-curves varying $Y_e$ of the unbound ejecta of a high flare strength model ($P_{\rm ext} = 6.7\times 10^{30}$ erg cm$^{-3}$; $B_{15} = 13$) achieving $M_{\rm ej} \approx 10^{-6}M_\odot$. The top panel shows the total final abundance patterns (mass-weighted across each layer of the ejecta), while the bottom panel shows the bolometric {\it nova brevis} light-curves.}
\label{fig:Ye_params}
\end{figure}

We have thus far considered models with $Y_e = 0.44$. This estimate assumes the ejecta mass is excavated uniformly across the entire neutron star surface. The fact that the pulsating X-ray tail emission was not blocked by the ejecta in either of the two Galactic giant flares supports that a modest fraction of the magnetar was shrouded by the ejecta \citep{Granot+2006}, consistent with the one-sided outflow inferred from VLBI imaging \citep{Taylor+2005} for the 2004 GF. This implies that the mass ejected by the shock may have been excavated from some fraction $<1$ of the NS surface area, and thus from greater radial depths within the NS crust possessing lower $Y_{\rm e}$ than assumed in spherically symmetric ejection. We show two lower values of $Y_e$ for a high flare strength ({$B_{15} = 13$}) capable of excavating $M_{\rm ej} \approx 10^{-6}M_\odot$ in Fig.~\ref{fig:Ye_params}. We consider only the high flare strength in this $Y_e$ parameter exploration as it is more likely to unbind deeper low $Y_e$ crustal material. We also show a model with greater $Y_e$, to account for the possibility that high energy flares can shock heat the crust sufficiently to enable rapid $e^{-/ +}$ absorption (\citetalias{Cehula+24}), the effects of which may be to increase or decrease $Y_e$ from its initial (pre-shock) value. Self consistent treatment of these weak interactions to determine the resultant $Y_e$ is beyond the scope of the current work. 

As expected, we find that models with greater neutron excess produce a greater total mass fraction of $r$-process nuclei. For example, the $Y_e = 0.36$ model yields approximately $80\%$ $r$-process nuclei by mass. The distribution of mass between the abundance peaks is also sensitive to the electron fraction, as shown in the top panel of Fig.~\ref{fig:Ye_params}, where lower $Y_e$ models produce fewer seeds/light $r$-process nuclei, and more second peak nuclei. The greater mass fraction of $r$-process nuclei in low $Y_e$ models leads to the generation of more radioactive decay energy, thus enhancing the luminosity of the {\it nova brevis} light-curve (bottom panel, Fig.~\ref{fig:Ye_params}). The variations in mass distribution across the abundance peaks significantly affects the opacity and therefore the light-curve shape and duration, as evident by the double peaked feature and longer (second) peak time-scale in the $Y_e = 0.36$ light-curve due to the large opacity contribution from the substantial abundances of second peak nuclei.

\section{Discussion}
\label{sec:discussion}

\subsection{Implications for Galactic Chemical Evolution}

\begin{figure*}[htbp]
    \centering
    \includegraphics[width=1.0\textwidth]{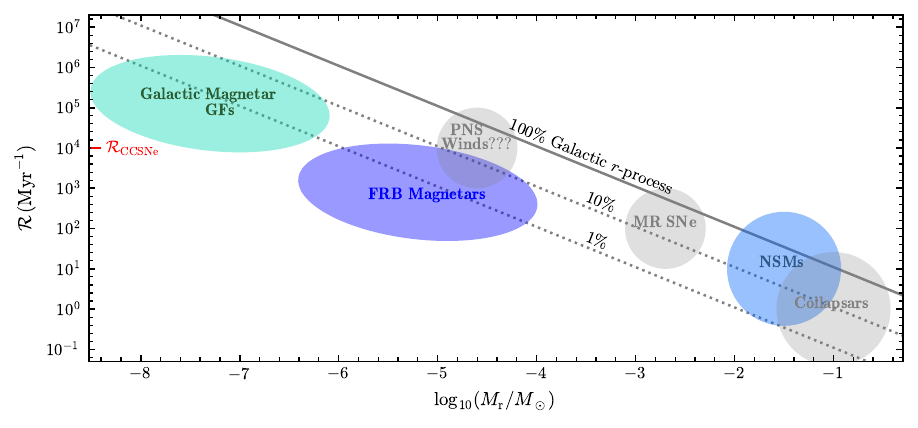}
\caption{Approximate rates and $r$-process mass yields of Galactic magnetar GFs (green) and FRB magnetars (purple), as compared to neutron star mergers (blue) and other proposed but not-yet-observed $r$-process sources (gray) . The diagonal lines are contours denoting $100\%$, $10\%$, and $1\%$ contribution to the total Galactic $r$-process mass, as labeled.} 
\label{fig:rates}
\end{figure*}

To what extent can magnetar GFs contribute to the heavy-element enrichment of the Galaxy? Given the total Galactic abundances of $r$-process elements at the present day, the contribution from any potential $r$-process site can be expressed as the product of the $r$-process mass yield per event, $M_r$, and the event rate, $\mathcal{R}$ (e.g., \citealt{Hotokezaka+2018}), as illustrated schematically in Figure \ref{fig:rates}. For example, if the neutrino-driven winds of all proto-NSs were to successfully create $r$-process nuclei,\footnote{Existing studies suggest that such winds generally {\it do not} achieve the requisite conditions for alpha-rich freeze-out (e.g., \citealt{Thompson+01}; however see \citealt{Nevins&Roberts23}).} then their estimated per-explosion yield of $\sim 10^{-5}$--$10^{-4}M_{\odot}$ (e.g., \citealt{Meyer+92}) times the Galactic rate of core-collapse SNe $\sim 10^{4}$ Myr$^{-1}$ (e.g., \citealt{Adams+13}) suggests a potential contribution corresponding to $\sim 10$--$100\%$ of the Galactic $r$-process abundances.

\citetalias{Cehula+24} found that across a range of flare energies (pressures) magnetar GFs can eject up to $\sim 10^{-6}M_{\odot}$ of baryonic material, in the case of a particularly strong flare. Our simulations reveal that a fraction $\lesssim 80\%$ is synthesized into $r$-process nuclei (Sec.~\ref{sec:results}), corresponding to an $r$-process mass yield $M_{\rm r} \lesssim 8\times 10^{-7}M_{\odot}$ per Galactic GF.  Given an estimated present day Galactic GF rate of roughly once per decade to once per century \citep{Beniamini2019, Burns+21} and given the star-formation rate (and hence magnetar-formation rate) was higher early in the history of the Milky Way than today, this suggests that Galactic Magnetar GF likely contribute at least $\sim 1$--$10\%$ of the Galactic $r$-process, as illustrated by the light green region in Fig.~\ref{fig:rates}. 

The observed properties of cosmological FRB sources, particularly those which are seen to repeat over timescales as long as decades (e.g., \citealt{Spitler+16}), point to much stronger flaring activity than that exhibited by magnetars in our own Galaxy. If the most powerful FRBs are accompanied by baryon ejection similar to Galactic magnetar GFs, then given the higher rate of bursting activity, even a single such ``hyper-active'' magnetar could, over its lifetime, eject a much larger quantity of NS crustal material $\gtrsim 10^{-5}$--$10^{-4}M_{\odot}$ and corresponding $r$-process mass, $M_{\rm r} \gtrsim 10^{-6}$--$10^{-5}M_{\odot}$. In particular, the enormous rotation measure $\gtrsim 10^{5}$ rad m$^{-2}$ of the repeating FRB source 121102 \citep{Michilli+18} was found to require $\sim 10^{-5}$--$10^{-4}M_{\odot}$ of total baryons injected into a magnetized nebula surrounding the FRB source over a timescale of several decades or longer \citep{Margalit&Metzger18}.  Despite their comparatively large $r$-process output, the birth-rates of the hyper-active magnetar population needed to explain the cosmological FRB population are likely $2$--$4$ orders of magnitude less than the core-collapse supernova rate \citep{Margalit+2020}; thus, even the FRB-powering magnetar population is likely to contribute $\sim 1$--$10\%$ of the total $r$-process production rate, indicated by the purple region in Fig.~\ref{fig:rates}.

Observations of metal poor stars in the Galactic halo or nearby dwarf galaxies do not exhibit a correlation between $r$-process elements and Fe \citep{Farouqi+2022}, indicating that $r$-process production sites should be spatially distinct from core-collapse supernovae, the dominant sites of Fe production in the early Galaxy. Since magnetars are born in core-collapse supernovae, one might expect their GF ejecta to mix with the surrounding Fe-rich supernova remnant, resulting in a strong correlation between $r$-process elements and Fe. However, we note that magnetars are likely formed in only a fraction of core-collapse explosions (e.g., \citealt{Gill&Heyl2007, Sautron+2025}). Furthermore, it is unclear whether all magnetars experience the same persistent flaring behavior; some may be docile, experiencing few if any GFs, while others are hyper-active (like the ``FRB magnetars" described above). Therefore, the wide range of $r$-process enrichment observed in metal poor stars likely does not preclude a magnetar GF origin. 

In summary, magnetar GFs likely contribute sub-dominantly to total $r$-process production in the Galaxy. However, unlike NS mergers, they represent an $r$-process source that will usually promptly follow star-formation, similar to core-collapse supernovae (except for magnetars produced through rare alternative channels such as accretion-induced collapse; e.g., \citealt{Margalit+19}). As such, absent other sources of the $r$-process at low metallicity, magnetar GFs may contribute an important source of enrichment for the most metal-poor stars found in the Galactic halo and nearby dwarf galaxies.

\subsection{Nova Brevis Detection Prospects}

\begin{figure} 
    \centering
    \includegraphics[width=0.5\textwidth]{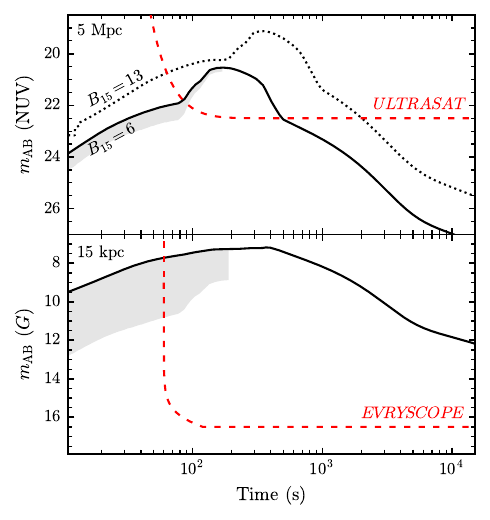}
\caption{Top panel: NUV-band (centered around $2500 {\rm \AA} $) light-curves for the fiducial ($B_{15} = 6$; solid lines) and a more energetic $(B_{15} = 9$; dotted lines) GF model, for an extragalactic source of distance $d = 5\,\rm Mpc$.  Bottom panel: $G$-band light-curve for a Galactic source at a distance of $d = 15 \,\rm kpc$. Estimated sensitivity ranges for {\it ULTRASAT} and {\it EVRYSCOPE} (both assuming a slew time of 60 s) are shown in the top and bottom panels, respectively.} 
\label{fig:detectability}
\end{figure}

At first glance, the likelihood to detect a {\it nova brevis} transient blindly with UV/optical surveys would appear to be exceptionally low, given their brief peak durations $\lesssim$ 15 minutes. However, we highlight two potential detection strategies. One are GFs from Galactic or Local Group magnetars, for which, given their favorable proximity, the {\it nova brevis} signal is sufficiently bright to observe even with a relatively modest-sized telescope provided it either has an enormous field-of-view covering a significant fraction of the sky, or is otherwise continuously monitoring the locations of known Galactic magnetars (one can imagine targeting magnetars showing recent X-ray flaring activity, possibly presaging a GF; e.g., \citealt{Hurley+05, Boggs+2007, Gill+Heyl2010}). As shown in the bottom panel of Fig.~\ref{fig:detectability}, our fiducial light-curve model exhibits a peak $G$-band magnitude of $m_{\rm AB} \approx 8$ for a source distance of 15 kpc, well above the detection sensitivity of extant very wide-field monitors such as {\it EVRYSCOPE} which covers approximately 38\% of the night sky with a 2-minute cadence \citep{Corbett+23}. This high detection probability remains true, even considering the large uncertainties associated with the emission from optically-thin portions of the flare ejecta (as indicated by the gray shaded region in Fig.~\ref{fig:detectability}).

Another potential avenue to detect even extragalactic {\it nova brevis} is by rapid slewing of an (narrower field) optical/UV telescope, following a trigger from a gamma-ray satellite. Magnetar GFs exhibit similar duration and hardness properties to ``short GRBs'' from NS mergers (e.g., \citealt{Hai-Ming+20,Burns+21}), and there is already a great interest in getting onto the source as early as possible to detect the earliest phases of the merger kilonova (e.g., \citealt{Metzger+18,Arcavi18}).  Several wide-field UV satellites with time-domain capabilities are planned over the next decade, including the Ultraviolet Transient Astronomy Satellite ({\it ULTRASAT}, launch date 2027; \citealt{Sagiv+14}), Ultraviolet Explorer ({\it UVEX}; \citealt{Kulkarni+21}), and the Czech Quick Ultra-Violet Kilonovae surveyor ({\it QUVIK}; \citealt{Werner+23}). We focus on {\it ULTRASAT}, which will reach a $5\sigma$ sensitivity of 22.4 AB magnitude in the NUV bands ($h \nu \approx $ 5 eV) across an instantaneous field of view of $\approx 200$ deg$^{2}$ for a 900 second (15 minute) integration \citep{Sagiv+14}.\footnote{
{\it UVEX} will possess sensitivity extending also into the FUV and reaching an AB magnitude depth of 24.5 for a 900 s integration \citep{Kulkarni+21}, roughly a factor of 6 deeper than {\it ULTRASAT}. However, {\it UVEX}'s smaller instantaneous field of view $\approx 12$ deg$^{2}$, results in a comparable survey speed time to {\it ULTRASAT}.} The top panel of Fig.~\ref{fig:detectability} shows the fiducial NUV light-curve falls just within {\it ULTRASAT}'s sensitivity range for a source at 5 Mpc (5 Mpc is approximately the distance out to which an extragalactic GF of energy $\gtrsim 10^{45}$ erg should be observed every decade; e.g., \citealt{Burns+21}) and assuming (optimistically) $\approx 60\,\rm s$ delay before slewing; a stronger flare at the same distance (corresponding to our $B_{15} = 13$ model) more easily surpasses detection constraints due to its slower evolution and brighter peak. Extragalactic {\it novae breves} from lower energy flares (e.g., Fig.~\ref{fig:params}) will be more challenging to detect, due to their more rapid evolution and lower peak luminosities.

\subsection{Impact of X-ray Heating}
\label{sec:tail}

The brief bursts of $\gamma$-ray emission from the 1998 flare of SGR1900+14 and the 2004 flare of SGR1806-20 were followed by luminous tails of X-ray emission lasting several minutes, modulated on the rotational period of the magnetar \citep{Feroci+01,Hurley+05}. This emission arises from the hot, slowly evaporating $e^{\pm}$ plasma fireball trapped within the NS magnetosphere (e.g., \citealt{Feroci+01}). The luminosity of the X-ray tail could be fit to the following functional form (e.g., \citealt{Hurley+05}):
\be
L_{\rm X}(t) = L_{0}\left(1-\frac{t}{t_{\rm evap}}\right)^{p}.
\label{eq:LX}
\ee
For the SGR 1806-20 flare, $t_{\rm evap} = 382$ s, $p \approx 1.5$, $L_{0} \approx E_{\rm tail}/t_{\rm evap} \approx 3\times 10^{41}d_{15}^{2}$ erg s$^{-1}$, where $E_{\rm tail} \approx 1.2\times 10^{44}d_{15}^{2}$ erg, and $d = 15d_{15}$ kpc is the source distance \citep{Svirski+11}.  Here we consider what impact irradiation of the ejecta by this luminous X-ray emission might have on the {\it nova brevis} signal.

A fraction of $L_{\rm X}$ can act to heat the baryonic ejecta, depending on the solid angle $\Delta \Omega$ the latter subtends as seen from the NS surface and the ejecta albedo (fraction of the incident X-rays scattered instead of being absorbed). The covering fraction is likely small $\Delta \Omega \ll 4\pi$, because the ejecta apparently did not block most of the X-ray emission from either SGR1900+14 or SGR1806-20. The ejecta albedo depends on the ratio of its scattering opacity $\kappa_{\rm es}$ to its absorptive (bound-free) opacity $\kappa_{\rm bf}$. Most of the X-ray tail  is emitted at photon energies $\sim 10$--$20$ keV (e.g., \citealt{Hurley+05}). Although the photo-ionization cross section of neutral $r$-process elements at these energies is a factor of $\sim 100$ times higher than the Thomson cross-section, the opacity to electron scattering is still likely to compete or dominate due to the very high ionization level of the ejecta. In particular, the ejecta will be highly photo-ionized by the X-rays emission, as indicated by the high ionization parameter: 
\begin{eqnarray}
&& \xi = \frac{L_{\rm X}}{n_{\rm ej}R_{\rm ej}^{2}} \approx 6\times 10^{4}\, {\rm erg\, cm\,s^{-1}}\left(\frac{L_{\rm X}}{3\times 10^{41}\,\rm erg\,s^{-1}}\right) \nonumber \\ &&\times \left(\frac{t}{300{\rm s}}\right)\left(\frac{\Delta \Omega}{4\pi}\right)\left(\frac{v}{0.2c}\right)\left(\frac{M_{\rm ej}}{10^{-7}M_{\odot}}\right)^{-1},
\end{eqnarray}
where $n_{\rm ej} \equiv (M_{\rm ej}/\Delta \Omega R_{\rm ej}^{3} m_p)$ and $R_{\rm ej} \approx v t$ are the ejecta density and characteristic radius on timescales $t \lesssim t_{\rm evap}$ relevant to the X-ray tail phase.  

Given such high ionization fractions $\xi \gtrsim 10^{4}-10^{5}$, most of the valence shell electrons will likley be stripped (e.g., \citealt{Kallman+21}, their Fig.~2), giving rise to a small ratio $\kappa_{\rm bf}/\kappa_{\rm es} \ll 1$ and hence high albedo.

\begin{figure} 
    \centering
    \includegraphics[width=0.5\textwidth]{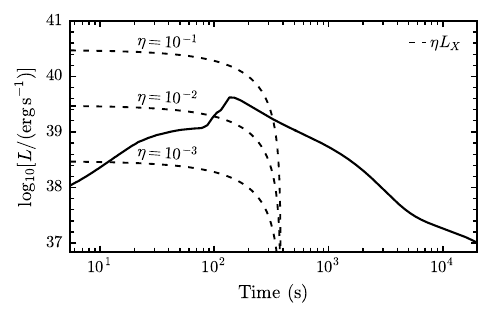}
\caption{Bolometric luminosity of the radioactively-powered {\it nova brevis} emission for the fiducial model (Fig.~\ref{fig:fid_LC}) in comparison to the heating rate $\eta L_{\rm X}(t)$ due to irradiation of the ejecta shell by the X-ray tail following the magnetar GF (dashed lines; see Eq.~\eqref{eq:LX}) for different assumed heating efficiencies $\eta$ (see text for details).} 
\label{fig:Lx}
\end{figure}

A more detailed photo-ionization calculation beyond the scope of this paper would be required to determine the fraction $\eta \approx (\Delta \Omega/4\pi)\kappa_{\rm bf}/\kappa_{\rm es}$ of $L_{\rm X}(t)$ absorbed by and used to heat the ejecta. Nevertheless, even for modest values of $\eta$, the resulting contribution to the transient's luminosity might compete with that from radioactivity alone.  Fig.~\ref{fig:Lx} shows $\eta L_{\rm X}(t)$ (using $L_{\rm X}$ from Eq.~\eqref{eq:LX}, for parameters fit to the 2004 flare from SGR1806-20) in comparison to the bolometric {\it nova brevis} light-curve of the fiducial model (Fig.~\ref{fig:fid_LC}). We see that for $\eta\gtrsim 10^{-2}$, X-ray heating could in principle compete with radioactively-powered emission, potentially enhancing the transient luminosity and detection prospects relative to the baseline model. However, because most of the X-ray heating occurs prior to peak luminosity when the ejecta is still optically thick, a radiative transfer calculation (accounting, e.g., for the depth into the ejecta where X-ray heating occurs) would be required to more precisely quantify this effect on the light-curve. We leave such a detailed calculation to future work.

\subsection{Gamma-Ray Line Diagnostics}

Gamma-rays released from the decay of $r$-process nuclei that are absorbed by the ejecta act as a source of heating and were taken into account in our light-curve calculations (Eq.~\ref{eq:fth}). However, a fraction of the gamma-rays are not absorbed and will begin to escape the ejecta once the latter becomes optically-thin to gamma-rays on a timescale $\gtrsim 100\,{\rm s}$. While the detection of a {\it nova brevis} optical/UV transient, as predicted in this paper, would establish $r$-process production in magnetar GFs, the detection of nuclear $\gamma$-rays would provide a more direct and powerful probe. In principle such a signal would enable probing the yields of individual isotopes, though the substantial Doppler broadening of the lines due to the high ejecta velocities $v/c \gtrsim 0.2$ would make this challenging in practice.

\section{Conclusions}
\label{sec:conclusion}
We have calculated nucleosynthesis and corresponding {\it nova brevis} light-curves resulting from baryon ejection in magnetar GFs. Adopting a one dimensional multi-zone ejecta model motivated by hydrodynamical simulations (\citetalias{Cehula+24}), we use a nuclear reaction network to conduct parameterized nucleosynthesis calculations unique to each ejecta layer. These determine the radioactive heating rate and approximate time-dependent opacity of each layer, which we combine in a semi-analytic framework to predict bolometric and band-specific light-curves, accounting for thermalization efficiency and adiabatic losses. We assess the detectability of such transient events and their role in Galactic $r$-process enrichment.

Our conclusions can be summarized as follows:
\begin{enumerate}
    \item The baryon ejecta from magnetar GFs support a moderate third peak $r$-process through the $\alpha$-rich freeze-out mechanism. The nucleosynthesis yields vary with radial mass coordinate through the ejecta: the extremal outer ejecta layers yield alpha particles and a significant fraction of free neutrons (which $\beta$--decay to protons over roughly 15 minutes); the remaining outer ejecta layers contribute significant quantities of alpha particles and fewer free neutrons, while the innermost layers dominantly contribute second peak, first peak, and light seed nuclei, respectively with increasing depth (schematic Fig.~\ref{fig:cartoon}). The total $r$-process mass fraction increases with flare strength; $M_r \approx 0.3 \text{--}0.8M_{\rm ej}$ across the parameter space explored. Weaker flares produce a larger relative fraction of third-peak elements because of the greater specific ejecta energy (larger entropy); however, the strongest flares also show an enhanced production of third-peak nuclei because neutron capture freeze-out is comparatively less important than for weaker flares. These trends demonstrate how the total $r$-process yield, and its division across the three abundance peaks, are sensitive to the efficiencies of $\alpha$-rich and neutron capture freeze-outs in the rapidly expanding ejecta.
    
    \item Neutron-rich $r$-process nuclei dominate the radioactive heating at most times, except around $t\sim15$ minutes when free neutrons briefly dominate.  Similar $n$-decay also occurs in the outermost ejecta layers of NS mergers, where it can power ``precursor'' emission to the main $r$-process-powered kilonova \citep{Metzger+15_n_precursor,Gottlieb&Loeb20,Combi&Siegel23}. Due to their large abundances, light/2nd-peak $r$-process nuclei ($A\lesssim 130)$ and $\alpha$-particles generally contribute most of the ejecta opacity.
    
    \item The photometric evolution is primarily controlled by the ejecta mass, which increases with flare strength. Larger ejecta yields from more powerful flares produce brighter and longer duration emission. Opacity also play a role in setting the peak time-scale and light-curve shape for ejecta with lower electron fraction due to the greater lanthanide/actinide abundance. Ejecta masses spanning the range $\approx 10^{-8}$--$10^{-6}\, M_\odot$ inferred from giant flare radio afterglows produce light-curve peaks at $L \lesssim 10^{40} \,\rm erg\,s^{-1}$ and $t\lesssim 15\,\rm mins$. Such transients may be detectable in the Milky Way or nearby galaxies out to several Mpc by rapid-slewing telescopes such as {\it ULTRASAT}.

    \item Considering the approximate rate and $r$-process ejecta yields of Galactic magnetar GFs, we estimate these events contribute $\sim 1$--$10\%$ to the Galactic $r$-process inventory. While significantly less than NS mergers (e.g., \citealt{Shibata&Hotokezaka19}) or other rare high-yield $r$-process sources, the relatively high event-rate and short delay-time of magnetar GFs relative to star-formation may favor them as dominant $r$-process sources at low metallicities. This warrants future investigation with Galactic chemical evolution modeling, to test these predictions and better constrain the role of GFs.
\end{enumerate}

The ejecta model employed in this work was motivated by one-dimensional hydrodynamical simulations (\citetalias{Cehula+24}). However, the mechanism of baryon ejection during magnetar GFs remains an area of active study warranting additional exploration in future work, e.g. with multi-dimensional magnetohydrodynamic simulations. To the extent that such future simulations will predict different conditions for the entropy and expansion rate of the ejected NS crust material, and the density profile achieved in the homologous state, we would expect quantitative differences in the resulting nucleosynthesis yields and {\it nova brevis} light-curves.  Nevertheless, because an $r$-process is likely a robust consequence of the sudden decompression of neutron star crust material (e.g., \citealt{Lattimer+1977}), the qualitative features of the {\it nova brevis} emission as predicted here are likely to remain intact. For a given ejecta mass and kinetic energy, such as that measured from the radio afterglow in the case of the 2004 GF of SGR 1806-20, the rise-time and peak luminosity of the light-curve are likely to be accurate to within factor of a few.  

Updated nucleosynthesis calculations based on future multidimensional simulations could in turn be complemented with full multi-group radiative transport, yielding improved light-curve estimates. New modeling may also offer an opportunity to include additional physics such as neutrino cooling and leptonization, which may affect both the dynamics and nucleosynthesis in the most powerful GFs (\citetalias{Cehula+24}).

During the final stages of completing this manuscript, we became aware of a delayed component of gamma-ray emission in the 2004 giant flare from magnetar SGR 1806-20 \citep{Mereghetti+05, Boggs+2007, Frederiks+07}, whose light-curve and spectrum match closely those due to radioactive decay of $r$-process nuclei.  In a separate letter \citep{Patel+2025}, we interpret this previously unexplained emission component as the first gamma-ray signature of $r$-process nuclei, freshly synthesized within $\sim 10^{-6}M_\odot$ of ejecta (with parameters similar to our high ejecta mass models in Sec~\ref{sec:Ye_param_study}; see Fig.~\ref{fig:Ye_params}). We outline several implications of this discovery for Galactic chemical evolution and the origin of heavy cosmic rays. This compelling evidence for magnetar giant flares as just the second confirmed site of the $r$-process motivates further efforts to study their baryon ejection and its radioactively powered emission.



\vspace{0.5cm}

A.~P.~thanks Yossef Zenati, Jens Mahlmann, and Jeremy Heyl for valuable discussions. B.~D.~M thanks Eric Burns and Dan Kasen for helpful discussions.  A.~P.~and B.~D.~M.~were supported in part by the National Science Foundation (grant No. AST-2009255) and by the NASA Fermi Guest Investigator Program (grant No.~80NSSC22K1574). The Flatiron Institute is supported by the Simons Foundation. This research was supported in part by grant no. NSF PHY-2309135 to the Kavli Institute for Theoretical Physics (KITP). J.~C. was supported by the Charles University Grant Agency (GA UK) project No. 81224.

\appendix

\section{Mass Ejecta Properties}
\label{sec:ejecta}

Here we estimate the layer-by-layer properties of the ejecta from a magnetar GF, as input for our nucleosynthesis calculations described in Sec.~\ref{sec:ejecta_estimates}. Much of our analysis follows that presented in \citetalias{Cehula+24}, to which we refer the reader for additional details and calibration based on hydrodynamical simulations. 

We approximate the pressure in the cold outer crust of the NS (densities $\rho_{\rm cr} \lesssim 10^{11.5}$ g cm$^{-3}$) as a polytrope:
\be
P_{\rm cr} = P_{0}\left(\frac{\rho_{\rm cr}}{\rho_{0}}\right)^{\Gamma_{0}},
\ee
where $P_{0} = 10^{19}$ erg cm$^{-3}$, $\rho_{0} = 10^{4}$ g cm$^{-3}$, and $\Gamma_{0} \approx 1.43$ based on a fit to standard pressure-density relationship (see \citetalias{Cehula+24}, their Fig.~1). Neglecting general relativistic effects, hydrostatic equilibrium reads:
\be
\frac{1}{\rho_{\rm cr}}\frac{dP_{\rm cr}}{dr} = -g_{\rm ns} \approx -\frac{GM_{\rm ns}}{R_{\rm ns}^{2}},
\ee
where $M_{\rm ns}$ and $R_{\rm ns}$ are the mass and radius, respectively of the NS, and we have taken $g_{\rm ns}$ to be constant because we are considering radii $r \approx R_{\rm ns}$. The mass above a given radius $r$ can thus be written:
\be
M_{\rm cr}(>r) \equiv \int_{r}^{R_{\rm ns}}4\pi r^{2}\rho_{\rm cr}dr \approx \frac{4\pi R_{\rm ns}^{4}}{GM_{\rm ns}}P_{\rm cr}(r) = M_{\rm ns}\frac{P_{\rm cr}(r)}{P_{\rm ns}},
\ee
where $P_{\rm ns} \equiv GM_{\rm ns}^{2}/4\pi R_{\rm ns}^{4}$ is a characteristic internal pressure. Inverting, the density-mass relationship in the crust reads:
\be
\rho_{\rm cr}(r) \approx \rho_{0}\left(\frac{M_{\rm cr}}{M_{\rm ns}}\frac{P_{\rm ns}}{P_{0}}\right)^{1/\Gamma_{0}} \propto M_{\rm cr}^{0.7}.
\ee

The sudden application of a external pressure $P_{\rm ext} \gg P_{\rm cr}$ above the NS surface from the magnetar GF drives a strong radiation-dominated ($\Gamma \simeq 4/3$) shock into the magnetar crust.  Each layer of the crust is compressed to a density $\rho_{\rm sh} \approx 7\rho_{\rm cr}$ and is heated to a specific internal energy density $e_{\rm sh} \approx 3P_{\rm sh}/\rho_{\rm sh} \approx 3P_{\rm ext}/\rho_{\rm sh} \approx 3c_{\rm s}^{2}$, where $c_{\rm s} \approx (P_{\rm ext}/\rho_{\rm sh})^{1/2}$ is the sound speed. After the external pressure $P_{\rm ext}$ is relieved, the shocked layers are now over-pressured and will re-expand, converting this thermal energy back into kinetic energy. A rough condition for a given shocked layer to ultimately escape from the star, is that its enthalpy $h_{\rm sh} \approx e_{\rm sh} + P_{\rm sh}/\rho_{\rm sh} \approx 4c_{\rm s}^{2}$ exceeds the gravitational binding energy, i.e.,
\be
4c_{\rm s}^{2} \gtrsim \frac{v_{\rm esc}^{2}}{2} \approx \frac{GM_{\rm ns}}{R_{\rm ns}} .
\ee
This condition defines a maximum density, corresponding to the innermost escaping layer:
\be
\rho_{\rm cr} \lesssim \rho_{\rm cr,max} = \frac{4}{7}\frac{P_{\rm ext}R_{\rm ns}}{GM_{\rm ns}} \approx  3.7\times 10^{9}\,{\rm g\,cm^{-3}}\,P_{\rm ext,30}R_{12}M_{1.4}^{-1}.
\label{eq:rhocr_app}
\ee
The total mass of the unbound ejecta is given by the mass-depth of this critical layer in the NS crust,
\be \Rightarrow m \lesssim M_{\rm ej} = M_{\rm ns}\frac{P_0}{P_{\rm ns}}\left(\frac{8}{7}\frac{P_{\rm ext}}{\rho_{0} v_{\rm esc}^{2}}\right)^{\Gamma_0} \approx 6.4\times 10^{-8}M_{\odot}\,P_{\rm ext,30}^{1.43}R_{12}^{5.43}M_{1.4}^{-2.43}.
\label{eq:Mej_app}
\ee
where we have adopted the notation for mass coordinate $m \equiv M(>r)$. This estimate somewhat exceeds the ejecta mass found by \citetalias{Cehula+24} by hydrodynamical simulations, pointing to limitations of our analytic estimates, such as our assumption that each layer of the star evolves independently after being shocked.  We nevertheless keep this estimate, keeping in mind that the true value of $P_{\rm ext}$ required to achieve a given $M_{\rm ej}$ may be somewhat higher than we have estimated. The internal energy density of  the shock crust can also be written in terms of mass coordinate,
\be
    e_{\rm sh}(m) \approx e_{\rm ej}\bigg(\frac{m}{M_{\rm ej}}\bigg)^{-0.7},
    \label{eq:e_sh_app}
\ee
where
\be
e_{\rm ej} \approx 1.2 \times 10^{20} {\rm erg\,g^{-1}}\, M_{1.4}R_{12}^{-1}.
\ee

\citetalias{Cehula+24} show that the unbound matter with $v \gtrsim 0.3c \approx v_{\rm esc}/2$ attains a distribution of asymptotic velocities which can be approximately described as a power-law of the form:
\be
\frac{dm}{dv} \propto \frac{v}{(v^{2}+v_{\rm esc}^{2})^{\Gamma_0+1}} \underset{v \gtrsim \bar{v}}\propto \left(\frac{v}{\bar{v}}\right)^{-\alpha},\,\,\, v > \bar{v},
\label{eq:dMdv_app}
\ee
where $\alpha = 2\Gamma_{0}+1 \approx 3.86$ and we have neglected special relativistic corrections. In this distribution, $M_{\rm ej} = m(v = \bar{v})$. \citetalias{Cehula+24} find the bulk of the ejecta has velocity $\bar{v}\sim 0.1$--$0.3c$; for simplicity we take $\bar{v} \approx 0.2c$ and $\alpha = 4$ and focus on those layers with $v \geq \bar{v}$, for which a homologous density profile is achieved of the form,
\be
\rho(v) \propto M_{\rm ej}/r^{3} \propto M_{\rm ej}/v^{3} \propto v^{-6}, v \gtrsim \bar{v},
\label{eq:rhov_app}
\ee
or, properly normalized,
\be
\rho = \frac{3}{4\pi}\frac{M_{\rm ej}}{(\bar{v}t)^{3}}\left(\frac{v}{\bar{v}}\right)^{-6}, t \gg t_{\alpha} .
\label{eq:rhoprofile_app} 
\ee

Given the relatively high electron fraction $Y_{\rm e} \gtrsim 0.4$ of the ejected layers of the NS crust at the depths of interest, the potential for generating $r$-process elements hinges on achieving an $\alpha$-rich freeze-out (e.g., \citealt{Meyer+92}). The latter depends on the entropy and expansion timescale of the outflow around the time $\alpha$-particles form, just prior to seed nucleus formation (e.g., \citealt{Hoffman+97}).

Insofar that the pressure of the shocked gas roughly matches that of the externally-applied pressure, the temperature and specific entropy of the shocked layers obey:
\be
k_{\rm B}T_{\rm sh} \approx k_{\rm B}\left(\frac{12P_{\rm sh}}{11a}\right)^{1/4} \approx 9.5\,{\rm MeV}\,P_{\rm ext,30}^{1/4}
\ee
\be
s_{\rm sh,min}[k_{\rm B}\,{\rm baryon^{-1}}] \gtrsim 5.21 \frac{(T_{\rm sh}/{\rm MeV})^{3}}{(\rho_{\rm sh,max}/\rm 10^{8} g\, cm^{-3})} \approx 18P_{\rm ext,30}^{-1/4}M_{1.4}R_{12}^{-1},
\label{eq:Sshmin}
\ee
where $\rho_{\rm sh,max} = 7\rho_{\rm cr,max}$ is the post-shock density.  These estimates assume an ideal gas of photons and $e^{-}/e^{+}$ pairs in the ultra-relativistic limit that $k_{\rm B}T_{\rm sh} \gg 2m_e c^{2} \approx 1$ MeV, which is marginally-satisfied at the time of $\alpha$-particle formation.  
This is the minimum entropy, attained for the densest layer. However, because $s_{\rm sh} \propto \rho_{\rm sh}^{-1} \propto \rho_{\rm cr}^{-1} \propto m^{-0.7}$, the entropy profile of the ejecta obeys,
\be
s \approx s_{\rm sh,min}\left(\frac{m}{M_{\rm ej}}\right)^{-0.7},
\label{eq:s_app}
\ee
for shallower layers $m < M_{\rm ej}$. 

Alpha particles form at a temperature $T_{\alpha}$ determined by nuclear statistical equilibrium (NSE; e.g., \citealt{Meyer94, Chen&Beloborodov07}, see Eq.~13 in the latter). Defining $T_{\alpha}$ as the temperature below which the mass-fraction of alpha particles rises to $\gtrsim 50\%$, and using the relationship $s \propto T^{3}/\rho$ in the first line of Eq.~\eqref{eq:Sshmin}, we find (using the full {\it Helmholtz} EOS described in the main text),
\be
k_{\rm B} T_{\alpha}(s) \approx 0.8\,{\rm MeV}\left(\frac{s}{50k_{\rm B}\rm baryon^{-1}}\right)^{-0.1}
\label{eq:kTalpha}
\ee
in the entropy range $s \sim 10-500 k_{\rm B}{\rm baryon^{-1}}$. Because entropy $s \propto T^{3}/\rho$ is roughly conserved during the optically-thick expansion phase, the $\alpha$ formation density can be estimated as
\be \label{eq:rho_alpha}
\rho_{\alpha} \approx \rho_{\rm sh}\left(\frac{T_{\alpha}}{T_{\rm sh}}\right)^{3} \approx 2.1\times 10^{7}\,{\rm g\,cm^{-3}} \left( \frac{m}{M_{\rm ej}} \right)^{0.9} P_{\rm ext,30}^{0.33} R_{12}^{1.3} M_{1.4}^{-1.3}
\ee

A final key quantity for nucleosynthesis is the expansion time of matter through the temperature range $k_{\rm B}T_{\alpha}$ where $\alpha$-particles form.  Assuming the layer undergoes expansion initially in one-dimension, we estimate the distance from the NS surface at which $\alpha$-formation occurs, $\Delta R_\alpha$, from $m \approx 4\pi R_{\rm ns}^{2}\Delta R_{\alpha}\rho_{\alpha}$. As long as $\Delta R_{\alpha} \lesssim R_{\rm ns},$ the approximation of 1D expansion is reasonable. The corresponding expansion timescale through the $\alpha$-formation region can thus be estimated as:
\be
t_{\alpha} \approx \frac{\Delta R_{\alpha}}{v_{\rm exp}} =  \frac{m}{4 \pi R_{\rm ns}^{2}} \sqrt{\frac{\rho_{\rm sh}}{8 P_{\rm ext}}} \rho_\alpha^{-1} \approx t_{\alpha}(m) \approx 4\times 10^{-3}\,{\rm ms} \left( \frac{m}{ M_{\rm ej}} \right)^{1.35} \rho_{\alpha, 8}^{-1} P_{\rm ext, 30}^{1.43} R_{12}^{3.93} M_{1.4}^{-2.93},
\label{eq:app_t_alpha}
\ee
where we estimate the immediate post shock expansion velocity $v_{\rm exp}$ (distinct from the asymptotic velocity (Eq.~\eqref{eq:v}) of the ejecta during the homologous expansion phase) from the enthalpy, $v_{\rm exp}^2/2 \approx h_{\rm sh}$, and $\rho_{\alpha,8} \equiv \rho_\alpha/(10^8\,\rm g\,cm^{-3})$. Note, this expression is completely general so long as the conditions for 1D expansion are satisfied; the assumption of an ideal ultra-relativistic photon and $e^{\pm}$ gas enters only through $\rho_\alpha$ (Eq.~\eqref{eq:rho_alpha}); however $\rho_\alpha$ can also be calculated without making these assumptions, as done for the models presented in the main text (Sec.~\ref{sec:sky_calcs}).

An approximation condition for achieving a sufficiently high ratio of neutrons to seed nuclei to achieve an $r$-process which extends up to the third $r$-process peak at $A \sim 195$, can be written (e.g., \citealt{Hoffman+97})
\be
\zeta \gtrsim \zeta_{\rm crit} \approx 8\times 10^{9},
\label{eq:zeta_app}
\ee
where
\be
    \zeta \approx 3\times 10^{9}\left(\frac{Y_{e}}{0.45}\right)^{-3}P_{\rm ext,30}^{-1.86}M_{1.4}^{4.93}R_{12}^{-5.93}\left(\frac{m}{M_{\rm ej}}\right)^{-2.55},
\ee
where $s$ is in units of $k_{\rm B}$ baryon$^{-1}$ and $t_{\alpha}$ in seconds. This estimate shows that $\zeta \gtrsim \zeta_{\rm crit}$ is in principle possible for the outer ejecta $m \lesssim M_{\rm ej}/3$ for fiducial parameters, though exhibiting sensitive dependence on the parameters of the problem. The steep dependence $\zeta \propto m^{-2.55}$ is consistent with Fig.~8 of \citetalias{Cehula+24}.

\section{Dependence on Density Evolution} \label{sec:tD_test}

\begin{figure}
    \centering
    \includegraphics[width=1.0\textwidth]{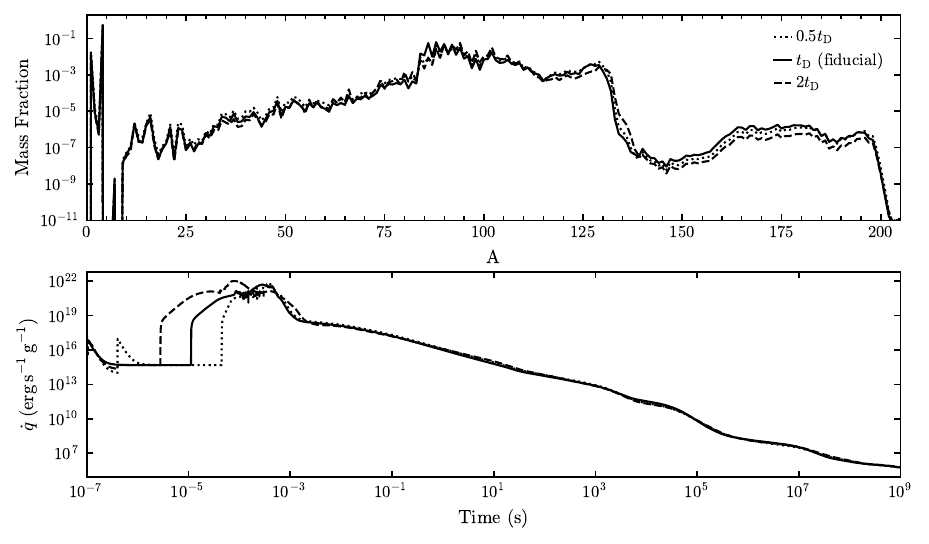}
\caption{Total nucleosynthetic yields and specific radioactive heating for variations from the fiducial transition time $t_{\rm D}$ in the density evolution (Eq~\ref{eq:rhoprofile}).}
    \label{fig:td_test}
\end{figure}

Fiducially we take the transition from approximate one-dimensional planar expansion to homologous expansion to occur at time $t_{\rm D} \equiv 2R_{\rm NS}/v$ (Eq.~\ref{eq:rhoprofile}). Fig.~\ref{fig:td_test} shows that the effects on the nucleosynthesis of varying the transition time to $0.5t_{\rm D}$ or $2t_{\rm D}$ (representing different durations for which the ejecta remains collimated due to e.g. magnetic fields), are negligible.

\section{Convergence Test} \label{sec:tests}
\begin{figure}
    \centering
    \includegraphics[width=1.0\textwidth]{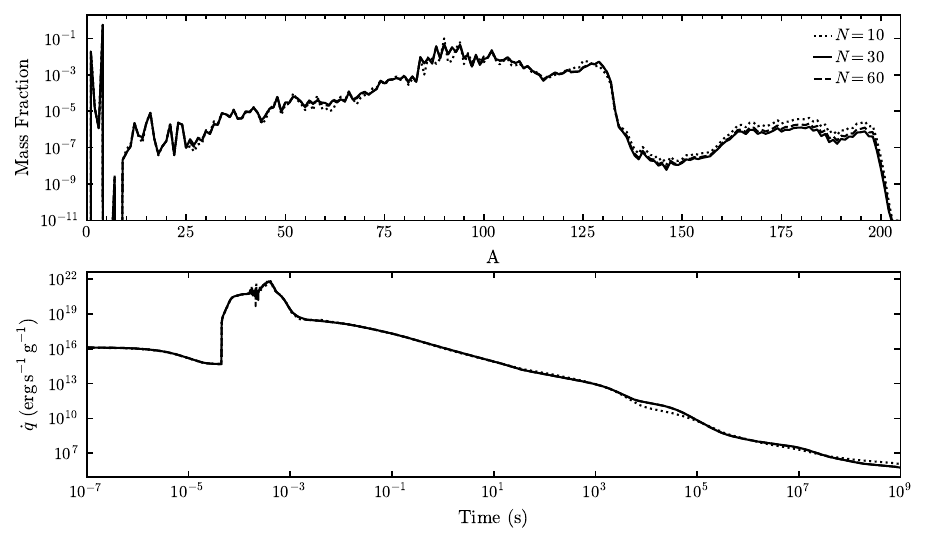}
\caption{Total nucleosynthetic yields and specific radioactive heating for the fiducial model with $N = 10, 30, 60$ ejecta layers.}
    \label{fig:convergence}
\end{figure}

We show in Fig.~\ref{fig:convergence} convergence is achieved for $N \geq 30$ ejecta layers. The difference in the abundance pattern is negligible and the radioactive heating almost identical for the $N=30$ and $N=60$ models.

\section{Power From Neutron Decay} \label{sec:n_decay}
{\it SkyNet} does not calculate the fractional heating contributions from individual nuclei or from specific decay channels. Here we determine the specific heating rate due to free neutron decay by post-processing the {\it SkyNet} output. This is required to estimate the thermalization (Eq.~\eqref{eq:fth}). Because all free protons at the end of a trajectory are produced through $\beta^-$ decay of free neutrons, we have $Y_{n, \rm inital}^{*} = Y_{p, \rm final}$ where $Y_{n, \rm inital}^{*}$ is the fraction of neutrons present at $t = 0$ that will not be captured onto seed nuclei and subsequently decay to free protons. The time dependent fraction of free protons produced from these decaying neutrons is then,
\be
Y_p^*(t) = Y_{p, \rm final} \left( 1 - e^{-t/\tau_n} \right),
\ee
where the mean neutron lifetime, $\tau_n \approx 900$s. The specific heating rate from free neutron decay is then,  
\be
\dot{q}_n(t) = \frac{1}{\Delta t} N_{\rm A} \Delta Y_p^* (\mathcal{M}_n - \mathcal{M}_p)c^2 ,
\label{eq:n_heating}
\ee
where $N_{\rm A}$ is Avogadro's constant and $\mathcal{M}_i = m_i - A_im_u$ is the mass excess of species $i$, consistent with the convention used in {\it SkyNet}. The heating from all other channels is $\dot{q}_{r} = \dot{q} - \dot{q}_n$, which we denote with the subscript ``$r$" since we expect the dominant contributions to be from $r$-process nuclei.

\bibliographystyle{aasjournal}
\bibliography{refs,example,refs2}

\newcommand{\noop}[1]{}
\begin{thebibliography}{}
\expandafter\ifx\csname natexlab\endcsname\relax\def\natexlab#1{#1}\fi
\providecommand{\url}[1]{\href{#1}{#1}}
\providecommand{\dodoi}[1]{doi:~\href{http://doi.org/#1}{\nolinkurl{#1}}}
\providecommand{\doeprint}[1]{\href{http://ascl.net/#1}{\nolinkurl{http://ascl.net/#1}}}
\providecommand{\doarXiv}[1]{\href{https://arxiv.org/abs/#1}{\nolinkurl{https://arxiv.org/abs/#1}}}

\bibitem[{{Abbott} {et~al.}(2017{\natexlab{a}}){Abbott}, {Abbott}, {Abbott}, {et~al.}}]{Abbott+17}
{Abbott}, B.~P., {Abbott}, R., {Abbott}, T.~D., {et~al.} 2017{\natexlab{a}}, \prl, 119, 161101, \dodoi{10.1103/PhysRevLett.119.161101}

\bibitem[{{Abbott} {et~al.}(2017{\natexlab{b}}){Abbott}, {Abbott}, {Abbott}, {Acernese}, {Ackley}, {Adams}, {Adams}, {Addesso}, {Adhikari}, {Adya}, {Affeldt}, {Afrough}, {Agarwal}, {Agathos}, {Agatsuma}, {Aggarwal}, {Aguiar}, {Aiello}, {Ain}, {Ajith}, {Allen}, {Allen}, {Allocca}, {Altin}, {Amato}, {Ananyeva}, {Anderson}, {Anderson}, {Angelova}, {Antier}, {Appert}, {Arai}, {Araya}, {Areeda}, {Arnaud}, {Arun}, {Ascenzi}, {Ashton}, {Ast}, {Aston}, {Astone}, {Atallah}, {Aufmuth}, {Aulbert}, {AultONeal}, {Austin}, {Avila-Alvarez}, {Babak}, {Bacon}, {Bader}, {Bae}, {Baker}, {Baldaccini}, {Ballardin}, {Ballmer}, {Banagiri}, {Barayoga}, {Barclay}, {Barish}, {Barker}, {Barkett}, {Barone}, {Barr}, {Barsotti}, {Barsuglia}, {Barta}, {Barthelmy}, {Bartlett}, {Bartos}, {Bassiri}, {Basti}, {Batch}, {Bawaj}, {Bayley}, {Bazzan}, {B{\'e}csy}, {Beer}, {Bejger}, {Belahcene}, {Bell}, {Berger}, {Bergmann}, {Bero}, {Berry}, {Bersanetti}, {Bertolini}, {Betzwieser}, {Bhagwat}, {Bhandare}, {Bilenko}, {Billingsley}, {Billman}, {Birch},
  {Birney}, {Birnholtz}, {Biscans}, {Biscoveanu}, {Bisht}, {Bitossi}, {Biwer}, {Bizouard}, {Blackburn}, {Blackman}, {Blair}, {Blair}, {Blair}, {Bloemen}, {Bock}, {Bode}, {Boer}, {Bogaert}, {Bohe}, {Bondu}, {Bonilla}, {Bonnand}, {Boom}, {Bork}, {Boschi}, {Bose}, {Bossie}, {Bouffanais}, {Bozzi}, {Bradaschia}, {Brady}, {Branchesi}, {Brau}, {Briant}, {Brillet}, {Brinkmann}, {Brisson}, {Brockill}, {Broida}, {Brooks}, {Brown}, {Brown}, {Brunett}, {Buchanan}, {Buikema}, {Bulik}, {Bulten}, {Buonanno}, {Buskulic}, {Buy}, {Byer}, {Cabero}, {Cadonati}, {Cagnoli}, {Cahillane}, {Calder{\'o}n Bustillo}, {Callister}, {Calloni}, {Camp}, {Canepa}, {Canizares}, {Cannon}, {Cao}, {Cao}, {Capano}, {Capocasa}, {Carbognani}, {Caride}, {Carney}, {Casanueva Diaz}, {Casentini}, {Caudill}, {Cavagli{\`a}}, {Cavalier}, {Cavalieri}, {Cella}, {Cepeda}, {Cerd{\'a}-Dur{\'a}n}, {Cerretani}, {Cesarini}, {Chamberlin}, {Chan}, {Chao}, {Charlton}, {Chase}, {Chassande-Mottin}, {Chatterjee}, {Chatziioannou}, {Cheeseboro}, {Chen}, {Chen}, {Chen},
  {Cheng}, {Chia}, {Chincarini}, {Chiummo}, {Chmiel}, {Cho}, {Cho}, {Chow}, {Christensen}, {Chu}, {Chua}, {Chua}, {Chung}, {Chung}, {Ciani}, {Ciolfi}, {Cirelli}, {Cirone}, {Clara}, {Clark}, {Clearwater}, {Cleva}, {Cocchieri}, {Coccia}, {Cohadon}, {Cohen}, {Colla}, {Collette}, {Cominsky}, {Constancio}, {Conti}, {Cooper}, {Corban}, {Corbitt}, {Cordero-Carri{\'o}n}, {Corley}, {Cornish}, {Corsi}, {Cortese}, {Costa}, {Coughlin}, {Coughlin}, {Coulon}, {Countryman}, {Couvares}, {Covas}, {Cowan}, {Coward}, {Cowart}, {Coyne}, {Coyne}, {Creighton}, {Creighton}, {Cripe}, {Crowder}, {Cullen}, {Cumming}, {Cunningham}, {Cuoco}, {Dal Canton}, {D{\'a}lya}, {Danilishin}, {D'Antonio}, {Danzmann}, {Dasgupta}, {Da Silva Costa}, {Dattilo}, {Dave}, {Davier}, {Davis}, {Daw}, {Day}, {De}, {DeBra}, {Degallaix}, {De Laurentis}, {Del{\'e}glise}, {Del Pozzo}, {Demos}, {Denker}, {Dent}, {De Pietri}, {Dergachev}, {De Rosa}, {DeRosa}, {De Rossi}, {DeSalvo}, {de Varona}, {Devenson}, {Dhurandhar}, {D{\'\i}az}, {Di Fiore}, {Di Giovanni}, {Di
  Girolamo}, {Di Lieto}, {Di Pace}, {Di Palma}, {Di Renzo}, {Doctor}, {Dolique}, {Donovan}, {Dooley}, {Doravari}, {Dorrington}, {Douglas}, {Dovale {\'A}lvarez}, {Downes}, {Drago}, {Dreissigacker}, {Driggers}, {Du}, {Ducrot}, {Dupej}, {Dwyer}, {Edo}, {Edwards}, {Effler}, {Ehrens}, {Eichholz}, {Eikenberry}, {Eisenstein}, {Essick}, {Estevez}, {Etienne}, {Etzel}, {Evans}, {Evans}, {Factourovich}, {Fafone}, {Fair}, {Fairhurst}, {Fan}, {Farinon}, {Farr}, {Farr}, {Fauchon-Jones}, {Favata}, {Fays}, {Fee}, {Fehrmann}, {Feicht}, {Fejer}, {Fernandez-Galiana}, {Ferrante}, {Ferreira}, {Ferrini}, {Fidecaro}, {Finstad}, {Fiori}, {Fiorucci}, {Fishbach}, {Fisher}, {Fitz-Axen}, {Flaminio}, {Fletcher}, {Fong}, {Font}, {Forsyth}, {Forsyth}, {Fournier}, {Frasca}, {Frasconi}, {Frei}, {Freise}, {Frey}, {Frey}, {Fries}, {Fritschel}, {Frolov}, {Fulda}, {Fyffe}, {Gabbard}, {Gadre}, {Gaebel}, {Gair}, {Gammaitoni}, {Ganija}, {Gaonkar}, {Garcia-Quiros}, {Garufi}, {Gateley}, {Gaudio}, {Gaur}, {Gayathri}, {Gehrels}, {Gemme}, {Genin},
  {Gennai}, {George}, {George}, {Gergely}, {Germain}, {Ghonge}, {Ghosh}, {Ghosh}, {Ghosh}, {Giaime}, {Giardina}, {Giazotto}, {Gill}, {Glover}, {Goetz}, {Goetz}, {Gomes}, {Goncharov}, {Gonz{\'a}lez}, {Gonzalez Castro}, {Gopakumar}, {Gorodetsky}, {Gossan}, {Gosselin}, {Gouaty}, {Grado}, {Graef}, {Granata}, {Grant}, {Gras}, {Gray}, {Greco}, {Green}, {Gretarsson}, {Griswold}, {Groot}, {Grote}, {Grunewald}, {Gruning}, {Guidi}, {Guo}, {Gupta}, {Gupta}, {Gushwa}, {Gustafson}, {Gustafson}, {Halim}, {Hall}, {Hall}, {Hamilton}, {Hammond}, {Haney}, {Hanke}, {Hanks}, {Hanna}, {Hannam}, {Hannuksela}, {Hanson}, {Hardwick}, {Harms}, {Harry}, {Harry}, {Hart}, {Haster}, {Haughian}, {Healy}, {Heidmann}, {Heintze}, {Heitmann}, {Hello}, {Hemming}, {Hendry}, {Heng}, {Hennig}, {Heptonstall}, {Heurs}, {Hild}, {Hinderer}, {Hoak}, {Hofman}, {Holt}, {Holz}, {Hopkins}, {Horst}, {Hough}, {Houston}, {Howell}, {Hreibi}, {Hu}, {Huerta}, {Huet}, {Hughey}, {Husa}, {Huttner}, {Huynh-Dinh}, {Indik}, {Inta}, {Intini}, {Isa}, {Isac}, {Isi},
  {Iyer}, {Izumi}, {Jacqmin}, {Jani}, {Jaranowski}, {Jawahar}, {Jim{\'e}nez-Forteza}, {Johnson}, {Jones}, {Jones}, {Jonker}, {Ju}, {Junker}, {Kalaghatgi}, {Kalogera}, {Kamai}, {Kandhasamy}, {Kang}, {Kanner}, {Kapadia}, {Karki}, {Karvinen}, {Kasprzack}, {Katolik}, {Katsavounidis}, {Katzman}, {Kaufer}, {Kawabe}, {K{\'e}f{\'e}lian}, {Keitel}, {Kemball}, {Kennedy}, {Kent}, {Key}, {Khalili}, {Khan}, {Khan}, {Khan}, {Khazanov}, {Kijbunchoo}, {Kim}, {Kim}, {Kim}, {Kim}, {Kim}, {Kim}, {Kimbrell}, {King}, {King}, {Kinley-Hanlon}, {Kirchhoff}, {Kissel}, {Kleybolte}, {Klimenko}, {Knowles}, {Koch}, {Koehlenbeck}, {Koley}, {Kondrashov}, {Kontos}, {Korobko}, {Korth}, {Kowalska}, {Kozak}, {Kr{\"a}mer}, {Kringel}, {Krishnan}, {Kr{\'o}lak}, {Kuehn}, {Kumar}, {Kumar}, {Kumar}, {Kuo}, {Kutynia}, {Kwang}, {Lackey}, {Lai}, {Landry}, {Lang}, {Lange}, {Lantz}, {Lanza}, {Larson}, {Lartaux-Vollard}, {Lasky}, {Laxen}, {Lazzarini}, {Lazzaro}, {Leaci}, {Leavey}, {Lee}, {Lee}, {Lee}, {Lee}, {Lee}, {Lehmann}, {Lenon}, {Leonardi}, {Leroy},
  {Letendre}, {Levin}, {Li}, {Linker}, {Littenberg}, {Liu}, {Lo}, {Lockerbie}, {London}, {Lord}, {Lorenzini}, {Loriette}, {Lormand}, {Losurdo}, {Lough}, {Lousto}, {Lovelace}, {L{\"u}ck}, {Lumaca}, {Lundgren}, {Lynch}, {Ma}, {Macas}, {Macfoy}, {Machenschalk}, {MacInnis}, {Macleod}, {Maga{\~n}a Hernandez}, {Maga{\~n}a-Sandoval}, {Maga{\~n}a Zertuche}, {Magee}, {Majorana}, {Maksimovic}, {Man}, {Mandic}, {Mangano}, {Mansell}, {Manske}, {Mantovani}, {Marchesoni}, {Marion}, {M{\'a}rka}, {M{\'a}rka}, {Markakis}, {Markosyan}, {Markowitz}, {Maros}, {Marquina}, {Marsh}, {Martelli}, {Martellini}, {Martin}, {Martin}, {Martynov}, {Mason}, {Massera}, {Masserot}, {Massinger}, {Masso-Reid}, {Mastrogiovanni}, {Matas}, {Matichard}, {Matone}, {Mavalvala}, {Mazumder}, {McCarthy}, {McClelland}, {McCormick}, {McCuller}, {McGuire}, {McIntyre}, {McIver}, {McManus}, {McNeill}, {McRae}, {McWilliams}, {Meacher}, {Meadors}, {Mehmet}, {Meidam}, {Mejuto-Villa}, {Melatos}, {Mendell}, {Mercer}, {Merilh}, {Merzougui}, {Meshkov}, {Messenger},
  {Messick}, {Metzdorff}, {Meyers}, {Miao}, {Michel}, {Middleton}, {Mikhailov}, {Milano}, {Miller}, {Miller}, {Miller}, {Millhouse}, {Milovich-Goff}, {Minazzoli}, {Minenkov}, {Ming}, {Mishra}, {Mitra}, {Mitrofanov}, {Mitselmakher}, {Mittleman}, {Moffa}, {Moggi}, {Mogushi}, {Mohan}, {Mohapatra}, {Montani}, {Moore}, {Moraru}, {Moreno}, {Morriss}, {Mours}, {Mow-Lowry}, {Mueller}, {Muir}, {Mukherjee}, {Mukherjee}, {Mukherjee}, {Mukund}, {Mullavey}, {Munch}, {Mu{\~n}iz}, {Muratore}, {Murray}, {Napier}, {Nardecchia}, {Naticchioni}, {Nayak}, {Neilson}, {Nelemans}, {Nelson}, {Nery}, {Neunzert}, {Nevin}, {Newport}, {Newton}, {Ng}, {Nguyen}, {Nguyen}, {Nichols}, {Nielsen}, {Nissanke}, {Nitz}, {Noack}, {Nocera}, {Nolting}, {North}, {Nuttall}, {Oberling}, {O'Dea}, {Ogin}, {Oh}, {Oh}, {Ohme}, {Okada}, {Oliver}, {Oppermann}, {Oram}, {O'Reilly}, {Ormiston}, {Ortega}, {O'Shaughnessy}, {Ossokine}, {Ottaway}, {Overmier}, {Owen}, {Pace}, {Page}, {Page}, {Pai}, {Pai}, {Palamos}, {Palashov}, {Palomba}, {Pal-Singh}, {Pan}, {Pan},
  {Pang}, {Pang}, {Pankow}, {Pannarale}, {Pant}, {Paoletti}, {Paoli}, {Papa}, {Parida}, {Parker}, {Pascucci}, {Pasqualetti}, {Passaquieti}, {Passuello}, {Patil}, {Patricelli}, {Pearlstone}, {Pedraza}, {Pedurand}, {Pekowsky}, {Pele}, {Penn}, {Perez}, {Perreca}, {Perri}, {Pfeiffer}, {Phelps}, {Piccinni}, {Pichot}, {Piergiovanni}, {Pierro}, {Pillant}, {Pinard}, {Pinto}, {Pirello}, {Pitkin}, {Poe}, {Poggiani}, {Popolizio}, {Porter}, {Post}, {Powell}, {Prasad}, {Pratt}, {Pratten}, {Predoi}, {Prestegard}, {Price}, {Prijatelj}, {Principe}, {Privitera}, {Prodi}, {Prokhorov}, {Puncken}, {Punturo}, {Puppo}, {P{\"u}rrer}, {Qi}, {Quetschke}, {Quintero}, {Quitzow-James}, {Raab}, {Rabeling}, {Radkins}, {Raffai}, {Raja}, {Rajan}, {Rajbhandari}, {Rakhmanov}, {Ramirez}, {Ramos-Buades}, {Rapagnani}, {Raymond}, {Razzano}, {Read}, {Regimbau}, {Rei}, {Reid}, {Reitze}, {Ren}, {Reyes}, {Ricci}, {Ricker}, {Rieger}, {Riles}, {Rizzo}, {Robertson}, {Robie}, {Robinet}, {Rocchi}, {Rolland}, {Rollins}, {Roma}, {Romano}, {Romel}, {Romie},
  {Rosi{\'n}ska}, {Ross}, {Rowan}, {R{\"u}diger}, {Ruggi}, {Rutins}, {Ryan}, {Sachdev}, {Sadecki}, {Sadeghian}, {Sakellariadou}, {Salconi}, {Saleem}, {Salemi}, {Samajdar}, {Sammut}, {Sampson}, {Sanchez}, {Sanchez}, {Sanchis-Gual}, {Sandberg}, {Sanders}, {Sassolas}, {Sathyaprakash}, {Saulson}, {Sauter}, {Savage}, {Sawadsky}, {Schale}, {Scheel}, {Scheuer}, {Schmidt}, {Schmidt}, {Schnabel}, {Schofield}, {Sch{\"o}nbeck}, {Schreiber}, {Schuette}, {Schulte}, {Schutz}, {Schwalbe}, {Scott}, {Scott}, {Seidel}, {Sellers}, {Sengupta}, {Sentenac}, {Sequino}, {Sergeev}, {Shaddock}, {Shaffer}, {Shah}, {Shahriar}, {Shaner}, {Shao}, {Shapiro}, {Shawhan}, {Sheperd}, {Shoemaker}, {Shoemaker}, {Siellez}, {Siemens}, {Sieniawska}, {Sigg}, {Silva}, {Singer}, {Singh}, {Singhal}, {Sintes}, {Slagmolen}, {Smith}, {Smith}, {Smith}, {Somala}, {Son}, {Sonnenberg}, {Sorazu}, {Sorrentino}, {Souradeep}, {Spencer}, {Srivastava}, {Staats}, {Staley}, {Steinke}, {Steinlechner}, {Steinlechner}, {Steinmeyer}, {Stevenson}, {Stone}, {Stops},
  {Strain}, {Stratta}, {Strigin}, {Strunk}, {Sturani}, {Stuver}, {Summerscales}, {Sun}, {Sunil}, {Suresh}, {Sutton}, {Swinkels}, {Szczepa{\'n}czyk}, {Tacca}, {Tait}, {Talbot}, {Talukder}, {Tanner}, {T{\'a}pai}, {Taracchini}, {Tasson}, {Taylor}, {Taylor}, {Tewari}, {Theeg}, {Thies}, {Thomas}, {Thomas}, {Thomas}, {Thorne}, {Thorne}, {Thrane}, {Tiwari}, {Tiwari}, {Tokmakov}, {Toland}, {Tonelli}, {Tornasi}, {Torres-Forn{\'e}}, {Torrie}, {T{\"o}yr{\"a}}, {Travasso}, {Traylor}, {Trinastic}, {Tringali}, {Trozzo}, {Tsang}, {Tse}, {Tso}, {Tsukada}, {Tsuna}, {Tuyenbayev}, {Ueno}, {Ugolini}, {Unnikrishnan}, {Urban}, {Usman}, {Vahlbruch}, {Vajente}, {Valdes}, {van Bakel}, {van Beuzekom}, {van den Brand}, {Van Den Broeck}, {Vander-Hyde}, {van der Schaaf}, {van Heijningen}, {van Veggel}, {Vardaro}, {Varma}, {Vass}, {Vas{\'u}th}, {Vecchio}, {Vedovato}, {Veitch}, {Veitch}, {Venkateswara}, {Venugopalan}, {Verkindt}, {Vetrano}, {Vicer{\'e}}, {Viets}, {Vinciguerra}, {Vine}, {Vinet}, {Vitale}, {Vo}, {Vocca}, {Vorvick},
  {Vyatchanin}, {Wade}, {Wade}, {Wade}, {Walet}, {Walker}, {Wallace}, {Walsh}, {Wang}, {Wang}, {Wang}, {Wang}, {Wang}, {Ward}, {Warner}, {Was}, {Watchi}, {Weaver}, {Wei}, {Weinert}, {Weinstein}, {Weiss}, {Wen}, {Wessel}, {Wessels}, {Westerweck}, {Westphal}, {Wette}, {Whelan}, {Whitcomb}, {Whiting}, {Whittle}, {Wilken}, {Williams}, {Williams}, {Williamson}, {Willis}, {Willke}, {Wimmer}, {Winkler}, {Wipf}, {Wittel}, {Woan}, {Woehler}, {Wofford}, {Wong}, {Worden}, {Wright}, {Wu}, {Wysocki}, {Xiao}, {Yamamoto}, {Yancey}, {Yang}, {Yap}, {Yazback}, {Yu}, {Yu}, {Yvert}, {Zadro{\.z}ny}, {Zanolin}, {Zelenova}, {Zendri}, {Zevin}, {Zhang}, {Zhang}, {Zhang}, {Zhang}, {Zhao}, {Zhou}, {Zhou}, {Zhu}, {Zhu}, {Zimmerman}, {Zucker}, {Zweizig}, {LIGO Scientific Collaboration}, {Virgo Collaboration}, {Wilson-Hodge}, {Bissaldi}, {Blackburn}, {Briggs}, {Burns}, {Cleveland}, {Connaughton}, {Gibby}, {Giles}, {Goldstein}, {Hamburg}, {Jenke}, {Hui}, {Kippen}, {Kocevski}, {McBreen}, {Meegan}, {Paciesas}, {Poolakkil}, {Preece},
  {Racusin}, {Roberts}, {Stanbro}, {Veres}, {von Kienlin}, {GBM}, {Savchenko}, {Ferrigno}, {Kuulkers}, {Bazzano}, {Bozzo}, {Brandt}, {Chenevez}, {Courvoisier}, {Diehl}, {Domingo}, {Hanlon}, {Jourdain}, {Laurent}, {Lebrun}, {Lutovinov}, {Martin-Carrillo}, {Mereghetti}, {Natalucci}, {Rodi}, {Roques}, {Sunyaev}, {Ubertini}, {INTEGRAL}, {Aartsen}, {Ackermann}, {Adams}, {Aguilar}, {Ahlers}, {Ahrens}, {Samarai}, {Altmann}, {Andeen}, {Anderson}, {Ansseau}, {Anton}, {Arg{\"u}elles}, {Auffenberg}, {Axani}, {Bagherpour}, {Bai}, {Barron}, {Barwick}, {Baum}, {Bay}, {Beatty}, {Becker Tjus}, {Bernardini}, {Besson}, {Binder}, {Bindig}, {Blaufuss}, {Blot}, {Bohm}, {B{\"o}rner}, {Bos}, {Bose}, {B{\"o}ser}, {Botner}, {Bourbeau}, {Bourbeau}, {Bradascio}, {Braun}, {Brayeur}, {Brenzke}, {Bretz}, {Bron}, {Brostean-Kaiser}, {Burgman}, {Carver}, {Casey}, {Casier}, {Cheung}, {Chirkin}, {Christov}, {Clark}, {Classen}, {Coenders}, {Collin}, {Conrad}, {Cowen}, {Cross}, {Day}, {de Andr{\'e}}, {De Clercq}, {DeLaunay}, {Dembinski}, {De
  Ridder}, {Desiati}, {de Vries}, {de Wasseige}, {de With}, {DeYoung}, {D{\'\i}az-V{\'e}lez}, {di Lorenzo}, {Dujmovic}, {Dumm}, {Dunkman}, {Dvorak}, {Eberhardt}, {Ehrhardt}, {Eichmann}, {Eller}, {Evenson}, {Fahey}, {Fazely}, {Felde}, {Filimonov}, {Finley}, {Flis}, {Franckowiak}, {Friedman}, {Fuchs}, {Gaisser}, {Gallagher}, {Gerhardt}, {Ghorbani}, {Giang}, {Glauch}, {Gl{\"u}senkamp}, {Goldschmidt}, {Gonzalez}, {Grant}, {Griffith}, {Haack}, {Hallgren}, {Halzen}, {Hanson}, {Hebecker}, {Heereman}, {Helbing}, {Hellauer}, {Hickford}, {Hignight}, {Hill}, {Hoffman}, {Hoffmann}, {Hokanson-Fasig}, {Hoshina}, {Huang}, {Huber}, {Hultqvist}, {H{\"u}nnefeld}, {In}, {Ishihara}, {Jacobi}, {Japaridze}, {Jeong}, {Jero}, {Jones}, {Kalaczynski}, {Kang}, {Kappes}, {Karg}, {Karle}, {Kauer}, {Keivani}, {Kelley}, {Kheirandish}, {Kim}, {Kim}, {Kintscher}, {Kiryluk}, {Kittler}, {Klein}, {Kohnen}, {Koirala}, {Kolanoski}, {K{\"o}pke}, {Kopper}, {Kopper}, {Koschinsky}, {Koskinen}, {Kowalski}, {Krings}, {Kroll}, {Kr{\"u}ckl}, {Kunnen},
  {Kunwar}, {Kurahashi}, {Kuwabara}, {Kyriacou}, {Labare}, {Lanfranchi}, {Larson}, {Lauber}, {Lesiak-Bzdak}, {Leuermann}, {Liu}, {Lu}, {L{\"u}nemann}, {Luszczak}, {Madsen}, {Maggi}, {Mahn}, {Mancina}, {Maruyama}, {Mase}, {Maunu}, {McNally}, {Meagher}, {Medici}, {Meier}, {Menne}, {Merino}, {Meures}, {Miarecki}, {Micallef}, {Moment{\'e}}, {Montaruli}, {Moore}, {Moulai}, {Nahnhauer}, {Nakarmi}, {Naumann}, {Neer}, {Niederhausen}, {Nowicki}, {Nygren}, {Obertacke Pollmann}, {Olivas}, {O'Murchadha}, {Palczewski}, {Pandya}, {Pankova}, {Peiffer}, {Pepper}, {P{\'e}rez de los Heros}, {Pieloth}, {Pinat}, {Price}, {Przybylski}, {Raab}, {R{\"a}del}, {Rameez}, {Rawlins}, {Rea}, {Reimann}, {Relethford}, {Relich}, {Resconi}, {Rhode}, {Richman}, {Robertson}, {Rongen}, {Rott}, {Ruhe}, {Ryckbosch}, {Rysewyk}, {S{\"a}lzer}, {Sanchez Herrera}, {Sandrock}, {Sandroos}, {Santander}, {Sarkar}, {Sarkar}, {Satalecka}, {Schlunder}, {Schmidt}, {Schneider}, {Schoenen}, {Sch{\"o}neberg}, {Schumacher}, {Seckel}, {Seunarine}, {Soedingrekso},
  {Soldin}, {Song}, {Spiczak}, {Spiering}, {Stachurska}, {Stamatikos}, {Stanev}, {Stasik}, {Stettner}, {Steuer}, {Stezelberger}, {Stokstad}, {St{\"o}ssl}, {Strotjohann}, {Stuttard}, {Sullivan}, {Sutherland}, {Taboada}, {Tatar}, {Tenholt}, {Ter-Antonyan}, {Terliuk}, {Te{\v{s}}i{\'c}}, {Tilav}, {Toale}, {Tobin}, {Toscano}, {Tosi}, {Tselengidou}, {Tung}, {Turcati}, {Turley}, {Ty}, {Unger}, {Usner}, {Vandenbroucke}, {Van Driessche}, {van Eijndhoven}, {Vanheule}, {van Santen}, {Vehring}, {Vogel}, {Vraeghe}, {Walck}, {Wallace}, {Wallraff}, {Wandler}, {Wandkowsky}, {Waza}, {Weaver}, {Weiss}, {Wendt}, {Werthebach}, {Whelan}, {Wiebe}, {Wiebusch}, {Wille}, {Williams}, {Wills}, {Wolf}, {Wood}, {Woolsey}, {Woschnagg}, {Xu}, {Xu}, {Xu}, {Yanez}, {Yodh}, {Yoshida}, {Yuan}, {Zoll}, {IceCube Collaboration}, {Balasubramanian}, {Mate}, {Bhalerao}, {Bhattacharya}, {Vibhute}, {Dewangan}, {Rao}, {Vadawale}, {AstroSat Cadmium Zinc Telluride Imager Team}, {Svinkin}, {Hurley}, {Aptekar}, {Frederiks}, {Golenetskii}, {Kozlova},
  {Lysenko}, {Oleynik}, {Tsvetkova}, {Ulanov}, {Cline}, {IPN Collaboration}, {Li}, {Xiong}, {Zhang}, {Lu}, {Song}, {Cao}, {Chang}, {Chen}, {Chen}, {Chen}, {Chen}, {Chen}, {Chen}, {Cui}, {Cui}, {Deng}, {Dong}, {Du}, {Fu}, {Gao}, {Gao}, {Gao}, {Ge}, {Gu}, {Guan}, {Guo}, {Han}, {Hu}, {Huang}, {Huo}, {Jia}, {Jiang}, {Jiang}, {Jin}, {Jin}, {Li}, {Li}, {Li}, {Li}, {Li}, {Li}, {Li}, {Li}, {Li}, {Li}, {Li}, {Liang}, {Liao}, {Liu}, {Liu}, {Liu}, {Liu}, {Liu}, {Liu}, {Liu}, {Lu}, {Lu}, {Luo}, {Ma}, {Meng}, {Nang}, {Nie}, {Ou}, {Qu}, {Sai}, {Sun}, {Tan}, {Tao}, {Tao}, {Tuo}, {Wang}, {Wang}, {Wang}, {Wang}, {Wang}, {Wen}, {Wu}, {Wu}, {Xiao}, {Xu}, {Xu}, {Yan}, {Yang}, {Yang}, {Yang}, {Zhang}, {Zhang}, {Zhang}, {Zhang}, {Zhang}, {Zhang}, {Zhang}, {Zhang}, {Zhang}, {Zhang}, {Zhang}, {Zhang}, {Zhang}, {Zhang}, {Zhang}, {Zhang}, {Zhang}, {Zhang}, {Zhao}, {Zhao}, {Zhao}, {Zheng}, {Zhu}, {Zhu}, {Zou}, {Insight-HXMT Collaboration}, {Albert}, {Andr{\'e}}, {Anghinolfi}, {Ardid}, {Aubert}, {Aublin}, {Avgitas}, {Baret},
  {Barrios-Mart{\'\i}}, {Basa}, {Belhorma}, {Bertin}, {Biagi}, {Bormuth}, {Bourret}, {Bouwhuis}, {Br{\^a}nza{\c{s}}}, {Bruijn}, {Brunner}, {Busto}, {Capone}, {Caramete}, {Carr}, {Celli}, {Cherkaoui El Moursli}, {Chiarusi}, {Circella}, {Coelho}, {Coleiro}, {Coniglione}, {Costantini}, {Coyle}, {Creusot}, {D{\'\i}az}, {Deschamps}, {De Bonis}, {Distefano}, {Di Palma}, {Domi}, {Donzaud}, {Dornic}, {Drouhin}, {Eberl}, {El Bojaddaini}, {El Khayati}, {Els{\"a}sser}, {Enzenh{\"o}fer}, {Ettahiri}, {Fassi}, {Felis}, {Fusco}, {Gay}, {Giordano}, {Glotin}, {Gr{\'e}goire}, {Ruiz}, {Graf}, {Hallmann}, {van Haren}, {Heijboer}, {Hello}, {Hern{\'a}ndez-Rey}, {H{\"o}ssl}, {Hofest{\"a}dt}, {Hugon}, {Illuminati}, {James}, {de Jong}, {Jongen}, {Kadler}, {Kalekin}, {Katz}, {Kiessling}, {Kouchner}, {Kreter}, {Kreykenbohm}, {Kulikovskiy}, {Lachaud}, {Lahmann}, {Lef{\`e}vre}, {Leonora}, {Lotze}, {Loucatos}, {Marcelin}, {Margiotta}, {Marinelli}, {Mart{\'\i}nez-Mora}, {Mele}, {Melis}, {Michael}, {Migliozzi}, {Moussa}, {Navas}, {Nezri},
  {Organokov}, {P{\u{a}}v{\u{a}}la{\c{s}}}, {Pellegrino}, {Perrina}, {Piattelli}, {Popa}, {Pradier}, {Quinn}, {Racca}, {Riccobene}, {S{\'a}nchez-Losa}, {Salda{\~n}a}, {Salvadori}, {Samtleben}, {Sanguineti}, {Sapienza}, {Sieger}, {Spurio}, {Stolarczyk}, {Taiuti}, {Tayalati}, {Trovato}, {Turpin}, {T{\"o}nnis}, {Vallage}, {Van Elewyck}, {Versari}, {Vivolo}, {Vizzoca}, {Wilms}, {Zornoza}, {Z{\'u}{\~n}iga}, {ANTARES Collaboration}, {Beardmore}, {Breeveld}, {Burrows}, {Cenko}, {Cusumano}, {D'A{\`\i}}, {de Pasquale}, {Emery}, {Evans}, {Giommi}, {Gronwall}, {Kennea}, {Krimm}, {Kuin}, {Lien}, {Marshall}, {Melandri}, {Nousek}, {Oates}, {Osborne}, {Pagani}, {Page}, {Palmer}, {Perri}, {Siegel}, {Sbarufatti}, {Tagliaferri}, {Tohuvavohu}, {Swift Collaboration}, {Tavani}, {Verrecchia}, {Bulgarelli}, {Evangelista}, {Pacciani}, {Feroci}, {Pittori}, {Giuliani}, {Del Monte}, {Donnarumma}, {Argan}, {Trois}, {Ursi}, {Cardillo}, {Piano}, {Longo}, {Lucarelli}, {Munar-Adrover}, {Fuschino}, {Labanti}, {Marisaldi}, {Minervini},
  {Fioretti}, {Parmiggiani}, {Gianotti}, {Trifoglio}, {Di Persio}, {Antonelli}, {Barbiellini}, {Caraveo}, {Cattaneo}, {Costa}, {Colafrancesco}, {D'Amico}, {Ferrari}, {Morselli}, {Paoletti}, {Picozza}, {Pilia}, {Rappoldi}, {Soffitta}, {Vercellone}, {AGILE Team}, {Foley}, {Coulter}, {Kilpatrick}, {Drout}, {Piro}, {Shappee}, {Siebert}, {Simon}, {Ulloa}, {Kasen}, {Madore}, {Murguia-Berthier}, {Pan}, {Prochaska}, {Ramirez-Ruiz}, {Rest}, {Rojas-Bravo}, {1M2H Team}, {Berger}, {Soares-Santos}, {Annis}, {Alexander}, {Allam}, {Balbinot}, {Blanchard}, {Brout}, {Butler}, {Chornock}, {Cook}, {Cowperthwaite}, {Diehl}, {Drlica-Wagner}, {Drout}, {Durret}, {Eftekhari}, {Finley}, {Fong}, {Frieman}, {Fryer}, {Garc{\'\i}a-Bellido}, {Gruendl}, {Hartley}, {Herner}, {Kessler}, {Lin}, {Lopes}, {Louren{\c{c}}o}, {Margutti}, {Marshall}, {Matheson}, {Medina}, {Metzger}, {Mu{\~n}oz}, {Muir}, {Nicholl}, {Nugent}, {Palmese}, {Paz-Chinch{\'o}n}, {Quataert}, {Sako}, {Sauseda}, {Schlegel}, {Scolnic}, {Secco}, {Smith}, {Sobreira}, {Villar},
  {Vivas}, {Wester}, {Williams}, {Yanny}, {Zenteno}, {Zhang}, {Abbott}, {Banerji}, {Bechtol}, {Benoit-L{\'e}vy}, {Bertin}, {Brooks}, {Buckley-Geer}, {Burke}, {Capozzi}, {Carnero Rosell}, {Carrasco Kind}, {Castander}, {Crocce}, {Cunha}, {D'Andrea}, {da Costa}, {Davis}, {DePoy}, {Desai}, {Dietrich}, {Eifler}, {Fernandez}, {Flaugher}, {Fosalba}, {Gaztanaga}, {Gerdes}, {Giannantonio}, {Goldstein}, {Gruen}, {Gschwend}, {Gutierrez}, {Honscheid}, {James}, {Jeltema}, {Johnson}, {Johnson}, {Kent}, {Krause}, {Kron}, {Kuehn}, {Lahav}, {Lima}, {Maia}, {March}, {Martini}, {McMahon}, {Menanteau}, {Miller}, {Miquel}, {Mohr}, {Nichol}, {Ogando}, {Plazas}, {Romer}, {Roodman}, {Rykoff}, {Sanchez}, {Scarpine}, {Schindler}, {Schubnell}, {Sevilla-Noarbe}, {Sheldon}, {Smith}, {Smith}, {Stebbins}, {Suchyta}, {Swanson}, {Tarle}, {Thomas}, {Troxel}, {Tucker}, {Vikram}, {Walker}, {Wechsler}, {Weller}, {Carlin}, {Gill}, {Li}, {Marriner}, {Neilsen}, {Dark Energy Camera GW-EM Collaboration}, {DES Collaboration}, {Haislip}, {Kouprianov},
  {Reichart}, {Sand}, {Tartaglia}, {Valenti}, {Yang}, {DLT40 Collaboration}, {Benetti}, {Brocato}, {Campana}, {Cappellaro}, {Covino}, {D'Avanzo}, {D'Elia}, {Getman}, {Ghirlanda}, {Ghisellini}, {Limatola}, {Nicastro}, {Palazzi}, {Pian}, {Piranomonte}, {Possenti}, {Rossi}, {Salafia}, {Tomasella}, {Amati}, {Antonelli}, {Bernardini}, {Bufano}, {Capaccioli}, {Casella}, {Dadina}, {De Cesare}, {Di Paola}, {Giuffrida}, {Giunta}, {Israel}, {Lisi}, {Maiorano}, {Mapelli}, {Masetti}, {Pescalli}, {Pulone}, {Salvaterra}, {Schipani}, {Spera}, {Stamerra}, {Stella}, {Testa}, {Turatto}, {Vergani}, {Aresu}, {Bachetti}, {Buffa}, {Burgay}, {Buttu}, {Caria}, {Carretti}, {Casasola}, {Castangia}, {Carboni}, {Casu}, {Concu}, {Corongiu}, {Deiana}, {Egron}, {Fara}, {Gaudiomonte}, {Gusai}, {Ladu}, {Loru}, {Leurini}, {Marongiu}, {Melis}, {Melis}, {Migoni}, {Milia}, {Navarrini}, {Orlati}, {Ortu}, {Palmas}, {Pellizzoni}, {Perrodin}, {Pisanu}, {Poppi}, {Righini}, {Saba}, {Serra}, {Serrau}, {Stagni}, {Surcis}, {Vacca}, {Vargiu}, {Hunt},
  {Jin}, {Klose}, {Kouveliotou}, {Mazzali}, {M{\o}ller}, {Nava}, {Piran}, {Selsing}, {Vergani}, {Wiersema}, {Toma}, {Higgins}, {Mundell}, {di Serego Alighieri}, {G{\'o}tz}, {Gao}, {Gomboc}, {Kaper}, {Kobayashi}, {Kopac}, {Mao}, {Starling}, {Steele}, {van der Horst}, {GRAWITA: GRAvitational Wave Inaf TeAm}, {Acero}, {Atwood}, {Baldini}, {Barbiellini}, {Bastieri}, {Berenji}, {Bellazzini}, {Bissaldi}, {Blandford}, {Bloom}, {Bonino}, {Bottacini}, {Bregeon}, {Buehler}, {Buson}, {Cameron}, {Caputo}, {Caraveo}, {Cavazzuti}, {Chekhtman}, {Cheung}, {Chiang}, {Ciprini}, {Cohen-Tanugi}, {Cominsky}, {Costantin}, {Cuoco}, {D'Ammando}, {de Palma}, {Digel}, {Di Lalla}, {Di Mauro}, {Di Venere}, {Dubois}, {Fegan}, {Focke}, {Franckowiak}, {Fukazawa}, {Funk}, {Fusco}, {Gargano}, {Gasparrini}, {Giglietto}, {Giordano}, {Giroletti}, {Glanzman}, {Green}, {Grondin}, {Guillemot}, {Guiriec}, {Harding}, {Horan}, {J{\'o}hannesson}, {Kamae}, {Kensei}, {Kuss}, {La Mura}, {Latronico}, {Lemoine-Goumard}, {Longo}, {Loparco}, {Lovellette},
  {Lubrano}, {Magill}, {Maldera}, {Manfreda}, {Mazziotta}, {McEnery}, {Meyer}, {Michelson}, {Mirabal}, {Monzani}, {Moretti}, {Morselli}, {Moskalenko}, {Negro}, {Nuss}, {Ojha}, {Omodei}, {Orienti}, {Orlando}, {Palatiello}, {Paliya}, {Paneque}, {Pesce-Rollins}, {Piron}, {Porter}, {Principe}, {Rain{\`o}}, {Rando}, {Razzano}, {Razzaque}, {Reimer}, {Reimer}, {Reposeur}, {Rochester}, {Saz Parkinson}, {Sgr{\`o}}, {Siskind}, {Spada}, {Spandre}, {Suson}, {Takahashi}, {Tanaka}, {Thayer}, {Thayer}, {Thompson}, {Tibaldo}, {Torres}, {Torresi}, {Troja}, {Venters}, {Vianello}, {Zaharijas}, {Fermi Large Area Telescope Collaboration}, {Allison}, {Bannister}, {Dobie}, {Kaplan}, {Lenc}, {Lynch}, {Murphy}, {Sadler}, {Australia Telescope Compact Array}, {Hotan}, {James}, {Oslowski}, {Raja}, {Shannon}, {Whiting}, {Australian SKA Pathfinder}, {Arcavi}, {Howell}, {McCully}, {Hosseinzadeh}, {Hiramatsu}, {Poznanski}, {Barnes}, {Zaltzman}, {Vasylyev}, {Maoz}, {Las Cumbres Observatory Group}, {Cooke}, {Bailes}, {Wolf}, {Deller},
  {Lidman}, {Wang}, {Gendre}, {Andreoni}, {Ackley}, {Pritchard}, {Bessell}, {Chang}, {M{\"o}ller}, {Onken}, {Scalzo}, {Ridden-Harper}, {Sharp}, {Tucker}, {Farrell}, {Elmer}, {Johnston}, {Venkatraman Krishnan}, {Keane}, {Green}, {Jameson}, {Hu}, {Ma}, {Sun}, {Wu}, {Wang}, {Shang}, {Hu}, {Ashley}, {Yuan}, {Li}, {Tao}, {Zhu}, {Zhang}, {Suntzeff}, {Zhou}, {Yang}, {Orange}, {Morris}, {Cucchiara}, {Giblin}, {Klotz}, {Staff}, {Thierry}, {Schmidt}, {OzGrav}, {(Deeper}, {Wider}, {program}, {AST3}, {CAASTRO Collaborations}, {Tanvir}, {Levan}, {Cano}, {de Ugarte-Postigo}, {Gonz{\'a}lez-Fern{\'a}ndez}, {Greiner}, {Hjorth}, {Irwin}, {Kr{\"u}hler}, {Mandel}, {Milvang-Jensen}, {O'Brien}, {Rol}, {Rosetti}, {Rosswog}, {Rowlinson}, {Steeghs}, {Th{\"o}ne}, {Ulaczyk}, {Watson}, {Bruun}, {Cutter}, {Figuera Jaimes}, {Fujii}, {Fruchter}, {Gompertz}, {Jakobsson}, {Hodosan}, {J{\`e}rgensen}, {Kangas}, {Kann}, {Rabus}, {Schr{\o}der}, {Stanway}, {Wijers}, {VINROUGE Collaboration}, {Lipunov}, {Gorbovskoy}, {Kornilov}, {Tyurina},
  {Balanutsa}, {Kuznetsov}, {Vlasenko}, {Podesta}, {Lopez}, {Podesta}, {Levato}, {Saffe}, {Mallamaci}, {Budnev}, {Gress}, {Kuvshinov}, {Gorbunov}, {Vladimirov}, {Zimnukhov}, {Gabovich}, {Yurkov}, {Sergienko}, {Rebolo}, {Serra-Ricart}, {Tlatov}, {Ishmuhametova}, {MASTER Collaboration}, {Abe}, {Aoki}, {Aoki}, {Asakura}, {Baar}, {Barway}, {Bond}, {Doi}, {Finet}, {Fujiyoshi}, {Furusawa}, {Honda}, {Itoh}, {Kanda}, {Kawabata}, {Kawabata}, {Kim}, {Koshida}, {Kuroda}, {Lee}, {Liu}, {Matsubayashi}, {Miyazaki}, {Morihana}, {Morokuma}, {Motohara}, {Murata}, {Nagai}, {Nagashima}, {Nagayama}, {Nakaoka}, {Nakata}, {Ohsawa}, {Ohshima}, {Ohta}, {Okita}, {Saito}, {Saito}, {Sako}, {Sekiguchi}, {Sumi}, {Tajitsu}, {Takahashi}, {Takayama}, {Tamura}, {Tanaka}, {Tanaka}, {Terai}, {Tominaga}, {Tristram}, {Uemura}, {Utsumi}, {Yamaguchi}, {Yasuda}, {Yoshida}, {Zenko}, {J-GEM}, {Adams}, {Anupama}, {Bally}, {Barway}, {Bellm}, {Blagorodnova}, {Cannella}, {Chandra}, {Chatterjee}, {Clarke}, {Cobb}, {Cook}, {Copperwheat}, {De}, {Emery},
  {Feindt}, {Foster}, {Fox}, {Frail}, {Fremling}, {Frohmaier}, {Garcia}, {Ghosh}, {Giacintucci}, {Goobar}, {Gottlieb}, {Grefenstette}, {Hallinan}, {Harrison}, {Heida}, {Helou}, {Ho}, {Horesh}, {Hotokezaka}, {Ip}, {Itoh}, {Jacobs}, {Jencson}, {Kasen}, {Kasliwal}, {Kassim}, {Kim}, {Kiran}, {Kuin}, {Kulkarni}, {Kupfer}, {Lau}, {Madsen}, {Mazzali}, {Miller}, {Miyasaka}, {Mooley}, {Myers}, {Nakar}, {Ngeow}, {Nugent}, {Ofek}, {Palliyaguru}, {Pavana}, {Perley}, {Peters}, {Pike}, {Piran}, {Qi}, {Quimby}, {Rana}, {Rosswog}, {Rusu}, {Sadler}, {Van Sistine}, {Sollerman}, {Xu}, {Yan}, {Yatsu}, {Yu}, {Zhang}, {Zhao}, {GROWTH}, {JAGWAR}, {Caltech-NRAO}, {TTU-NRAO}, {NuSTAR Collaborations}, {Chambers}, {Huber}, {Schultz}, {Bulger}, {Flewelling}, {Magnier}, {Lowe}, {Wainscoat}, {Waters}, {Willman}, {Pan-STARRS}, {Ebisawa}, {Hanyu}, {Harita}, {Hashimoto}, {Hidaka}, {Hori}, {Ishikawa}, {Isobe}, {Iwakiri}, {Kawai}, {Kawai}, {Kawamuro}, {Kawase}, {Kitaoka}, {Makishima}, {Matsuoka}, {Mihara}, {Morita}, {Morita}, {Nakahira},
  {Nakajima}, {Nakamura}, {Negoro}, {Oda}, {Sakamaki}, {Sasaki}, {Serino}, {Shidatsu}, {Shimomukai}, {Sugawara}, {Sugita}, {Sugizaki}, {Tachibana}, {Takao}, {Tanimoto}, {Tomida}, {Tsuboi}, {Tsunemi}, {Ueda}, {Ueno}, {Yamada}, {Yamaoka}, {Yamauchi}, {Yatabe}, {Yoneyama}, {Yoshii}, {MAXI Team}, {Coward}, {Crisp}, {Macpherson}, {Andreoni}, {Laugier}, {Noysena}, {Klotz}, {Gendre}, {Thierry}, {Turpin}, {Consortium}, {Im}, {Choi}, {Kim}, {Yoon}, {Lim}, {Lee}, {Lee}, {Kim}, {Ko}, {Joe}, {Kwon}, {Kim}, {Lim}, {Choi}, {KU Collaboration}, {Fynbo}, {Malesani}, {Xu}, {Optical Telescope}, {Smartt}, {Jerkstrand}, {Kankare}, {Sim}, {Fraser}, {Inserra}, {Maguire}, {Leloudas}, {Magee}, {Shingles}, {Smith}, {Young}, {Kotak}, {Gal-Yam}, {Lyman}, {Homan}, {Agliozzo}, {Anderson}, {Angus}, {Ashall}, {Barbarino}, {Bauer}, {Berton}, {Botticella}, {Bulla}, {Cannizzaro}, {Cartier}, {Cikota}, {Clark}, {De Cia}, {Della Valle}, {Dennefeld}, {Dessart}, {Dimitriadis}, {Elias-Rosa}, {Firth}, {Fl{\"o}rs}, {Frohmaier}, {Galbany},
  {Gonz{\'a}lez-Gait{\'a}n}, {Gromadzki}, {Guti{\'e}rrez}, {Hamanowicz}, {Harmanen}, {Heintz}, {Hernandez}, {Hodgkin}, {Hook}, {Izzo}, {James}, {Jonker}, {Kerzendorf}, {Kostrzewa-Rutkowska}, {Kromer}, {Kuncarayakti}, {Lawrence}, {Manulis}, {Mattila}, {McBrien}, {M{\"u}ller}, {Nordin}, {O'Neill}, {Onori}, {Palmerio}, {Pastorello}, {Patat}, {Pignata}, {Podsiadlowski}, {Razza}, {Reynolds}, {Roy}, {Ruiter}, {Rybicki}, {Salmon}, {Pumo}, {Prentice}, {Seitenzahl}, {Smith}, {Sollerman}, {Sullivan}, {Szegedi}, {Taddia}, {Taubenberger}, {Terreran}, {Van Soelen}, {Vos}, {Walton}, {Wright}, {Wyrzykowski}, {Yaron}, {pre=''(''>ePESSTO}, {Chen}, {Kr{\"u}hler}, {Schady}, {Wiseman}, {Greiner}, {Rau}, {Schweyer}, {Klose}, {Nicuesa Guelbenzu}, {GROND}, {Palliyaguru}, {Tech University}, {Shara}, {Williams}, {Vaisanen}, {Potter}, {Romero Colmenero}, {Crawford}, {Buckley}, {Mao}, {SALT Group}, {D{\'\i}az}, {Macri}, {Garc{\'\i}a Lambas}, {Mendes de Oliveira}, {Nilo Castell{\'o}n}, {Ribeiro}, {S{\'a}nchez}, {Schoenell}, {Abramo},
  {Akras}, {Alcaniz}, {Artola}, {Beroiz}, {Bonoli}, {Cabral}, {Camuccio}, {Chavushyan}, {Coelho}, {Colazo}, {Costa-Duarte}, {Cuevas Larenas}, {Dom{\'\i}nguez Romero}, {Dultzin}, {Fern{\'a}ndez}, {Garc{\'\i}a}, {Girardini}, {Gon{\c{c}}alves}, {Gon{\c{c}}alves}, {Gurovich}, {Jim{\'e}nez-Teja}, {Kanaan}, {Lares}, {Lopes de Oliveira}, {L{\'o}pez-Cruz}, {Melia}, {Molino}, {Padilla}, {Pe{\~n}uela}, {Placco}, {Qui{\~n}ones}, {Ram{\'\i}rez Rivera}, {Renzi}, {Riguccini}, {R{\'\i}os-L{\'o}pez}, {Rodriguez}, {Sampedro}, {Schneiter}, {Sodr{\'e}}, {Starck}, {Torres-Flores}, {Tornatore}, {Zadro{\.z}ny}, {Castillo}, {TOROS: Transient Robotic Observatory of South Collaboration}, {Castro-Tirado}, {Tello}, {Hu}, {Zhang}, {Cunniffe}, {Castell{\'o}n}, {Hiriart}, {Caballero-Garc{\'\i}a}, {Jel{\'\i}nek}, {Kub{\'a}nek}, {P{\'e}rez del Pulgar}, {Park}, {Jeong}, {Castro Cer{\'o}n}, {Pandey}, {Yock}, {Querel}, {Fan}, {Wang}, {BOOTES Collaboration}, {Beardsley}, {Brown}, {Crosse}, {Emrich}, {Franzen}, {Gaensler}, {Horsley},
  {Johnston-Hollitt}, {Kenney}, {Morales}, {Pallot}, {Sokolowski}, {Steele}, {Tingay}, {Trott}, {Walker}, {Wayth}, {Williams}, {Wu}, {Murchison Widefield Array}, {Yoshida}, {Sakamoto}, {Kawakubo}, {Yamaoka}, {Takahashi}, {Asaoka}, {Ozawa}, {Torii}, {Shimizu}, {Tamura}, {Ishizaki}, {Cherry}, {Ricciarini}, {Penacchioni}, {Marrocchesi}, {CALET Collaboration}, {Pozanenko}, {Volnova}, {Mazaeva}, {Minaev}, {Krugov}, {Kusakin}, {Reva}, {Moskvitin}, {Rumyantsev}, {Inasaridze}, {Klunko}, {Tungalag}, {Schmalz}, {Burhonov}, {IKI-GW Follow-up Collaboration}, {Abdalla}, {Abramowski}, {Aharonian}, {Ait Benkhali}, {Ang{\"u}ner}, {Arakawa}, {Arrieta}, {Aubert}, {Backes}, {Balzer}, {Barnard}, {Becherini}, {Becker Tjus}, {Berge}, {Bernhard}, {Bernl{\"o}hr}, {Blackwell}, {B{\"o}ttcher}, {Boisson}, {Bolmont}, {Bonnefoy}, {Bordas}, {Bregeon}, {Brun}, {Brun}, {Bryan}, {B{\"u}chele}, {Bulik}, {Capasso}, {Caroff}, {Carosi}, {Casanova}, {Cerruti}, {Chakraborty}, {Chaves}, {Chen}, {Chevalier}, {Colafrancesco}, {Condon}, {Conrad},
  {Davids}, {Decock}, {Deil}, {Devin}, {deWilt}, {Dirson}, {Djannati-Ata{\"\i}}, {Donath}, {O'C. Drury}, {Dutson}, {Dyks}, {Edwards}, {Egberts}, {Emery}, {Ernenwein}, {Eschbach}, {Farnier}, {Fegan}, {Fernandes}, {Fiasson}, {Fontaine}, {Funk}, {F{\"u}ssling}, {Gabici}, {Gallant}, {Garrigoux}, {Gat{\'e}}, {Giavitto}, {Giebels}, {Glawion}, {Glicenstein}, {Gottschall}, {Grondin}, {Hahn}, {Haupt}, {Hawkes}, {Heinzelmann}, {Henri}, {Hermann}, {Hinton}, {Hofmann}, {Hoischen}, {Holch}, {Holler}, {Horns}, {Ivascenko}, {Iwasaki}, {Jacholkowska}, {Jamrozy}, {Jankowsky}, {Jankowsky}, {Jingo}, {Jouvin}, {Jung-Richardt}, {Kastendieck}, {Katarzy{\'n}ski}, {Katsuragawa}, {Kerszberg}, {Khangulyan}, {Kh{\'e}lifi}, {King}, {Klepser}, {Klochkov}, {Klu{\'z}niak}, {Komin}, {Kosack}, {Krakau}, {Kraus}, {Kr{\"u}ger}, {Laffon}, {Lamanna}, {Lau}, {Lees}, {Lefaucheur}, {Lemi{\`e}re}, {Lemoine-Goumard}, {Lenain}, {Leser}, {Lohse}, {Lorentz}, {Liu}, {Lypova}, {Malyshev}, {Marandon}, {Marcowith}, {Mariaud}, {Marx}, {Maurin}, {Maxted},
  {Mayer}, {Meintjes}, {Meyer}, {Mitchell}, {Moderski}, {Mohamed}, {Mohrmann}, {Mor{\r{a}}}, {Moulin}, {Murach}, {Nakashima}, {de Naurois}, {Ndiyavala}, {Niederwanger}, {Niemiec}, {Oakes}, {O'Brien}, {Odaka}, {Ohm}, {Ostrowski}, {Oya}, {Padovani}, {Panter}, {Parsons}, {Pekeur}, {Pelletier}, {Perennes}, {Petrucci}, {Peyaud}, {Piel}, {Pita}, {Poireau}, {Poon}, {Prokhorov}, {Prokoph}, {P{\"u}hlhofer}, {Punch}, {Quirrenbach}, {Raab}, {Rauth}, {Reimer}, {Reimer}, {Renaud}, {de los Reyes}, {Rieger}, {Rinchiuso}, {Romoli}, {Rowell}, {Rudak}, {Rulten}, {Sahakian}, {Saito}, {Sanchez}, {Santangelo}, {Sasaki}, {Schlickeiser}, {Sch{\"u}ssler}, {Schulz}, {Schwanke}, {Schwemmer}, {Seglar-Arroyo}, {Settimo}, {Seyffert}, {Shafi}, {Shilon}, {Shiningayamwe}, {Simoni}, {Sol}, {Spanier}, {Spir-Jacob}, {Stawarz}, {Steenkamp}, {Stegmann}, {Steppa}, {Sushch}, {Takahashi}, {Tavernet}, {Tavernier}, {Taylor}, {Terrier}, {Tibaldo}, {Tiziani}, {Tluczykont}, {Trichard}, {Tsirou}, {Tsuji}, {Tuffs}, {Uchiyama}, {van der Walt}, {van Eldik},
  {van Rensburg}, {van Soelen}, {Vasileiadis}, {Veh}, {Venter}, {Viana}, {Vincent}, {Vink}, {Voisin}, {V{\"o}lk}, {Vuillaume}, {Wadiasingh}, {Wagner}, {Wagner}, {Wagner}, {White}, {Wierzcholska}, {Willmann}, {W{\"o}rnlein}, {Wouters}, {Yang}, {Zaborov}, {Zacharias}, {Zanin}, {Zdziarski}, {Zech}, {Zefi}, {Ziegler}, {Zorn}, {{\.Z}ywucka}, {H.~E.~S.~S. Collaboration}, {Fender}, {Broderick}, {Rowlinson}, {Wijers}, {Stewart}, {ter Veen}, {Shulevski}, {LOFAR Collaboration}, {Kavic}, {Simonetti}, {League}, {Tsai}, {Obenberger}, {Nathaniel}, {Taylor}, {Dowell}, {Liebling}, {Estes}, {Lippert}, {Sharma}, {Vincent}, {Farella}, {Wavelength Array}, {Abeysekara}, {Albert}, {Alfaro}, {Alvarez}, {Arceo}, {Arteaga-Vel{\'a}zquez}, {Avila Rojas}, {Ayala Solares}, {Barber}, {Becerra Gonzalez}, {Becerril}, {Belmont-Moreno}, {BenZvi}, {Berley}, {Bernal}, {Braun}, {Brisbois}, {Caballero-Mora}, {Capistr{\'a}n}, {Carrami{\~n}ana}, {Casanova}, {Castillo}, {Cotti}, {Cotzomi}, {Couti{\~n}o de Le{\'o}n}, {De Le{\'o}n}, {De la Fuente},
  {Diaz Hernandez}, {Dichiara}, {Dingus}, {DuVernois}, {D{\'\i}az-V{\'e}lez}, {Ellsworth}, {Engel}, {Enr{\'\i}quez-Rivera}, {Fiorino}, {Fleischhack}, {Fraija}, {Garc{\'\i}a-Gonz{\'a}lez}, {Garfias}, {Gerhardt}, {Gonz{\~o}lez Mu{\~n}oz}, {Gonz{\'a}lez}, {Goodman}, {Hampel-Arias}, {Harding}, {Hernandez}, {Hernandez-Almada}, {Hona}, {H{\"u}ntemeyer}, {Iriarte}, {Jardin-Blicq}, {Joshi}, {Kaufmann}, {Kieda}, {Lara}, {Lauer}, {Lennarz}, {Le{\'o}n Vargas}, {Linnemann}, {Longinotti}, {Raya}, {Luna-Garc{\'\i}a}, {L{\'o}pez-Coto}, {Malone}, {Marinelli}, {Martinez}, {Martinez-Castellanos}, {Mart{\'\i}nez-Castro}, {Mart{\'\i}nez-Huerta}, {Matthews}, {Miranda-Romagnoli}, {Moreno}, {Mostaf{\'a}}, {Nellen}, {Newbold}, {Nisa}, {Noriega-Papaqui}, {Pelayo}, {Pretz}, {P{\'e}rez-P{\'e}rez}, {Ren}, {Rho}, {Rivi{\`e}re}, {Rosa-Gonz{\'a}lez}, {Rosenberg}, {Ruiz-Velasco}, {Salazar}, {Salesa Greus}, {Sandoval}, {Schneider}, {Schoorlemmer}, {Sinnis}, {Smith}, {Springer}, {Surajbali}, {Tibolla}, {Tollefson}, {Torres}, {Ukwatta},
  {Weisgarber}, {Westerhoff}, {Wisher}, {Wood}, {Yapici}, {Yodh}, {Younk}, {Zhou}, {{\'A}lvarez}, {HAWC Collaboration}, {Aab}, {Abreu}, {Aglietta}, {Albuquerque}, {Albury}, {Allekotte}, {Almela}, {Alvarez Castillo}, {Alvarez-Mu{\~n}iz}, {Anastasi}, {Anchordoqui}, {Andrada}, {Andringa}, {Aramo}, {Arsene}, {Asorey}, {Assis}, {Avila}, {Badescu}, {Balaceanu}, {Barbato}, {Barreira Luz}, {Becker}, {Bellido}, {Berat}, {Bertaina}, {Bertou}, {Biermann}, {Biteau}, {Blaess}, {Blanco}, {Blazek}, {Bleve}, {Boh{\'a}{\v{c}}ov{\'a}}, {Bonifazi}, {Borodai}, {Botti}, {Brack}, {Brancus}, {Bretz}, {Bridgeman}, {Briechle}, {Buchholz}, {Bueno}, {Buitink}, {Buscemi}, {Caballero-Mora}, {Caccianiga}, {Cancio}, {Canfora}, {Caruso}, {Castellina}, {Catalani}, {Cataldi}, {Cazon}, {Chavez}, {Chinellato}, {Chudoba}, {Clay}, {Cobos Cerutti}, {Colalillo}, {Coleman}, {Collica}, {Coluccia}, {Concei{\c{c}}{\~a}o}, {Consolati}, {Contreras}, {Cooper}, {Coutu}, {Covault}, {Cronin}, {D'Amico}, {Daniel}, {Dasso}, {Daumiller}, {Dawson}, {Day}, {de
  Almeida}, {de Jong}, {De Mauro}, {de Mello Neto}, {De Mitri}, {de Oliveira}, {de Souza}, {Debatin}, {Deligny}, {D{\'\i}az Castro}, {Diogo}, {Dobrigkeit}, {D'Olivo}, {Dorosti}, {Dos Anjos}, {Dova}, {Dundovic}, {Ebr}, {Engel}, {Erdmann}, {Erfani}, {Escobar}, {Espadanal}, {Etchegoyen}, {Falcke}, {Farmer}, {Farrar}, {Fauth}, {Fazzini}, {Feldbusch}, {Fenu}, {Fick}, {Figueira}, {Filip{\v{c}}i{\v{c}}}, {Freire}, {Fujii}, {Fuster}, {Ga{\"\i}or}, {Garc{\'\i}a}, {Gat{\'e}}, {Gemmeke}, {Gherghel-Lascu}, {Ghia}, {Giaccari}, {Giammarchi}, {Giller}, {G{\l}as}, {Glaser}, {Golup}, {G{\'o}mez Berisso}, {G{\'o}mez Vitale}, {Gonz{\'a}lez}, {Gorgi}, {Gottowik}, {Grillo}, {Grubb}, {Guarino}, {Guedes}, {Halliday}, {Hampel}, {Hansen}, {Harari}, {Harrison}, {Harvey}, {Haungs}, {Hebbeker}, {Heck}, {Heimann}, {Herve}, {Hill}, {Hojvat}, {Holt}, {Homola}, {H{\"o}randel}, {Horvath}, {Hrabovsk{\'y}}, {Huege}, {Hulsman}, {Insolia}, {Isar}, {Jandt}, {Johnsen}, {Josebachuili}, {Jurysek}, {K{\"a}{\"a}p{\"a}}, {Kampert}, {Keilhauer},
  {Kemmerich}, {Kemp}, {Kieckhafer}, {Klages}, {Kleifges}, {Kleinfeller}, {Krause}, {Krohm}, {Kuempel}, {Kukec Mezek}, {Kunka}, {Kuotb Awad}, {Lago}, {LaHurd}, {Lang}, {Lauscher}, {Legumina}, {Leigui de Oliveira}, {Letessier-Selvon}, {Lhenry-Yvon}, {Link}, {Lo Presti}, {Lopes}, {L{\'o}pez}, {L{\'o}pez Casado}, {Lorek}, {Luce}, {Lucero}, {Malacari}, {Mallamaci}, {Mandat}, {Mantsch}, {Mariazzi}, {Maris}, {Marsella}, {Martello}, {Martinez}, {Mart{\'\i}nez Bravo}, {Mas{\'\i}as Meza}, {Mathes}, {Mathys}, {Matthews}, {Matthiae}, {Mayotte}, {Mazur}, {Medina}, {Medina-Tanco}, {Melo}, {Menshikov}, {Merenda}, {Michal}, {Micheletti}, {Middendorf}, {Miramonti}, {Mitrica}, {Mockler}, {Mollerach}, {Montanet}, {Morello}, {Morlino}, {M{\"u}ller}, {M{\"u}ller}, {Muller}, {M{\"u}ller}, {Mussa}, {Naranjo}, {Nguyen}, {Niculescu-Oglinzanu}, {Niechciol}, {Niemietz}, {Niggemann}, {Nitz}, {Nosek}, {Novotny}, {No{\v{z}}ka}, {N{\'u}{\~n}ez}, {Oikonomou}, {Olinto}, {Palatka}, {Pallotta}, {Papenbreer}, {Parente}, {Parra}, {Paul},
  {Pech}, {Pedreira}, {P{\c{e}}kala}, {Pe{\~n}a-Rodriguez}, {Pereira}, {Perlin}, {Perrone}, {Peters}, {Petrera}, {Phuntsok}, {Pierog}, {Pimenta}, {Pirronello}, {Platino}, {Plum}, {Poh}, {Porowski}, {Prado}, {Privitera}, {Prouza}, {Quel}, {Querchfeld}, {Quinn}, {Ramos-Pollan}, {Rautenberg}, {Ravignani}, {Ridky}, {Riehn}, {Risse}, {Ristori}, {Rizi}, {Rodrigues de Carvalho}, {Rodriguez Fernandez}, {Rodriguez Rojo}, {Roncoroni}, {Roth}, {Roulet}, {Rovero}, {Ruehl}, {Saffi}, {Saftoiu}, {Salamida}, {Salazar}, {Saleh}, {Salina}, {S{\'a}nchez}, {Sanchez-Lucas}, {Santos}, {Santos}, {Sarazin}, {Sarmento}, {Sarmiento-Cano}, {Sato}, {Schauer}, {Scherini}, {Schieler}, {Schimp}, {Schmidt}, {Scholten}, {Schov{\'a}nek}, {Schr{\"o}der}, {Schr{\"o}der}, {Schulz}, {Schumacher}, {Sciutto}, {Segreto}, {Shadkam}, {Shellard}, {Sigl}, {Silli}, {{\v{S}}m{\'\i}da}, {Snow}, {Sommers}, {Sonntag}, {Soriano}, {Squartini}, {Stanca}, {Stani{\v{c}}}, {Stasielak}, {Stassi}, {Stolpovskiy}, {Strafella}, {Streich}, {Suarez}, {Suarez-Dur{\'a}n},
  {Sudholz}, {Suomij{\"a}rvi}, {Supanitsky}, {{\v{S}}up{\'\i}k}, {Swain}, {Szadkowski}, {Taboada}, {Taborda}, {Timmermans}, {Todero Peixoto}, {Tomankova}, {Tom{\'e}}, {Torralba Elipe}, {Travnicek}, {Trini}, {Tueros}, {Ulrich}, {Unger}, {Urban}, {Vald{\'e}s Galicia}, {Vali{\~n}o}, {Valore}, {van Aar}, {van Bodegom}, {van den Berg}, {van Vliet}, {Varela}, {Vargas C{\'a}rdenas}, {V{\'a}zquez}, {Veberi{\v{c}}}, {Ventura}, {Vergara Quispe}, {Verzi}, {Vicha}, {Villase{\~n}or}, {Vorobiov}, {Wahlberg}, {Wainberg}, {Walz}, {Watson}, {Weber}, {Weindl}, {Wiede{\'n}ski}, {Wiencke}, {Wilczy{\'n}ski}, {Wirtz}, {Wittkowski}, {Wundheiler}, {Yang}, {Yushkov}, {Zas}, {Zavrtanik}, {Zavrtanik}, {Zepeda}, {Zimmermann}, {Ziolkowski}, {Zong}, {Zuccarello}, {Pierre Auger Collaboration}, {Kim}, {Schulze}, {Bauer}, {Corral-Santana}, {de Gregorio-Monsalvo}, {Gonz{\'a}lez-L{\'o}pez}, {Hartmann}, {Ishwara-Chandra}, {Mart{\'\i}n}, {Mehner}, {Misra}, {Micha{\l}owski}, {Resmi}, {ALMA Collaboration}, {Paragi}, {Agudo}, {An}, {Beswick},
  {Casadio}, {Frey}, {Jonker}, {Kettenis}, {Marcote}, {Moldon}, {Szomoru}, {van Langevelde}, {Yang}, {Euro VLBI Team}, {Cwiek}, {Cwiok}, {Czyrkowski}, {Dabrowski}, {Kasprowicz}, {Mankiewicz}, {Nawrocki}, {Opiela}, {Piotrowski}, {Wrochna}, {Zaremba}, {{\.Z}arnecki}, {Pi of Sky Collaboration}, {Haggard}, {Nynka}, {Ruan}, {Chandra Team at McGill University}, {Bland}, {Booler}, {Devillepoix}, {de Gois}, {Hancock}, {Howie}, {Paxman}, {Sansom}, {Towner}, {Desert Fireball Network}, {Tonry}, {Coughlin}, {Stubbs}, {Denneau}, {Heinze}, {Stalder}, {Weiland}, {ATLAS}, {Eatough}, {Kramer}, {Kraus}, {Time Resolution Universe Survey}, {Troja}, {Piro}, {Becerra Gonz{\'a}lez}, {Butler}, {Fox}, {Khandrika}, {Kutyrev}, {Lee}, {Ricci}, {Ryan}, {S{\'a}nchez-Ram{\'\i}rez}, {Veilleux}, {Watson}, {Wieringa}, {Burgess}, {van Eerten}, {Fontes}, {Fryer}, {Korobkin}, {Wollaeger}, {RIMAS}, {RATIR}, {Camilo}, {Foley}, {Goedhart}, {Makhathini}, {Oozeer}, {Smirnov}, {Fender}, {Woudt}, \& {South Africa/MeerKAT}}]{Abbot+2017_EM}
---. 2017{\natexlab{b}}, \apjl, 848, L12, \dodoi{10.3847/2041-8213/aa91c9}

\bibitem[{{Adams} {et~al.}(2013){Adams}, {Kochanek}, {Beacom}, {Vagins}, \& {Stanek}}]{Adams+13}
{Adams}, S.~M., {Kochanek}, C.~S., {Beacom}, J.~F., {Vagins}, M.~R., \& {Stanek}, K.~Z. 2013, \apj, 778, 164, \dodoi{10.1088/0004-637X/778/2/164}

\bibitem[{{Arcavi}(2018)}]{Arcavi18}
{Arcavi}, I. 2018, \apjl, 855, L23, \dodoi{10.3847/2041-8213/aab267}

\bibitem[{{Arcavi} {et~al.}(2017){Arcavi}, {Hosseinzadeh}, {Howell}, {McCully}, {Poznanski}, {Kasen}, {Barnes}, {Zaltzman}, {Vasylyev}, {Maoz}, \& {Valenti}}]{Arcavi+2017}
{Arcavi}, I., {Hosseinzadeh}, G., {Howell}, D.~A., {et~al.} 2017, \nat, 551, 64, \dodoi{10.1038/nature24291}

\bibitem[{{Arnett}(1980)}]{Arnett80}
{Arnett}, W.~D. 1980, \apj, 237, 541, \dodoi{10.1086/157898}

\bibitem[{{Barnes} {et~al.}(2016){Barnes}, {Kasen}, {Wu}, \& {Mart{\'\i}nez-Pinedo}}]{Barnes+16}
{Barnes}, J., {Kasen}, D., {Wu}, M.-R., \& {Mart{\'\i}nez-Pinedo}, G. 2016, \apj, 829, 110, \dodoi{10.3847/0004-637X/829/2/110}

\bibitem[{{Barnes} \& {Metzger}(2022)}]{Barnes&Metzger22}
{Barnes}, J., \& {Metzger}, B.~D. 2022, \apjl, 939, L29, \dodoi{10.3847/2041-8213/ac9b41}

\bibitem[{{Beniamini} {et~al.}(2019){Beniamini}, {Hotokezaka}, {van der Horst}, \& {Kouveliotou}}]{Beniamini2019}
{Beniamini}, P., {Hotokezaka}, K., {van der Horst}, A., \& {Kouveliotou}, C. 2019, \mnras, 487, 1426, \dodoi{10.1093/mnras/stz1391}

\bibitem[{{Beniamini} {et~al.}(2024){Beniamini}, {Wadiasingh}, {Trigg}, {Chirenti}, {Burns}, {Younes}, {Negro}, \& {Granot}}]{Beniamini+2024}
{Beniamini}, P., {Wadiasingh}, Z., {Trigg}, A., {et~al.} 2024, arXiv e-prints, arXiv:2411.16846, \dodoi{10.48550/arXiv.2411.16846}

\bibitem[{{Bochenek} {et~al.}(2020){Bochenek}, {Ravi}, {Belov}, {Hallinan}, {Kocz}, {Kulkarni}, \& {McKenna}}]{Bochenek+2020}
{Bochenek}, C.~D., {Ravi}, V., {Belov}, K.~V., {et~al.} 2020, \nat, 587, 59, \dodoi{10.1038/s41586-020-2872-x}

\bibitem[{{Boggs} {et~al.}(2007){Boggs}, {Zoglauer}, {Bellm}, {Hurley}, {Lin}, {Smith}, {Wigger}, \& {Hajdas}}]{Boggs+2007}
{Boggs}, S.~E., {Zoglauer}, A., {Bellm}, E., {et~al.} 2007, \apj, 661, 458, \dodoi{10.1086/516732}

\bibitem[{{Burbidge} {et~al.}(1957){Burbidge}, {Burbidge}, {Fowler}, \& {Hoyle}}]{Burbidge+57}
{Burbidge}, E.~M., {Burbidge}, G.~R., {Fowler}, W.~A., \& {Hoyle}, F. 1957, Reviews of Modern Physics, 29, 547, \dodoi{10.1103/RevModPhys.29.547}

\bibitem[{{Burns} {et~al.}(2021){Burns}, {Svinkin}, {Hurley}, {Wadiasingh}, {Negro}, {Younes}, {Hamburg}, {Ridnaia}, {Cook}, {Cenko}, {Aloisi}, {Ashton}, {Baring}, {Briggs}, {Christensen}, {Frederiks}, {Goldstein}, {Hui}, {Kaplan}, {Kasliwal}, {Kocevski}, {Roberts}, {Savchenko}, {Tohuvavohu}, {Veres}, \& {Wilson-Hodge}}]{Burns+21}
{Burns}, E., {Svinkin}, D., {Hurley}, K., {et~al.} 2021, \apjl, 907, L28, \dodoi{10.3847/2041-8213/abd8c8}

\bibitem[{{Cameron}(1957)}]{Cameron57}
{Cameron}, A.~G.~W. 1957, \aj, 62, 9, \dodoi{10.1086/107435}

\bibitem[{{Cameron} {et~al.}(2005){Cameron}, {Chandra}, {Ray}, {Kulkarni}, {Frail}, {Wieringa}, {Nakar}, {Phinney}, {Miyazaki}, {Tsuboi}, {Okumura}, {Kawai}, {Menten}, \& {Bertoldi}}]{Cameron+2005}
{Cameron}, P.~B., {Chandra}, P., {Ray}, A., {et~al.} 2005, \nat, 434, 1112, \dodoi{10.1038/nature03605}

\bibitem[{{Cehula} {et~al.}(2024){Cehula}, {Thompson}, \& {Metzger}}]{Cehula+24}
{Cehula}, J., {Thompson}, T.~A., \& {Metzger}, B.~D. 2024, \mnras, 528, 5323, \dodoi{10.1093/mnras/stae358}

\bibitem[{{Chen} \& {Beloborodov}(2007)}]{Chen&Beloborodov07}
{Chen}, W.-X., \& {Beloborodov}, A.~M. 2007, \apj, 657, 383, \dodoi{10.1086/508923}

\bibitem[{{CHIME/FRB Collaboration} {et~al.}(2020){CHIME/FRB Collaboration}, {Andersen}, {Bandura}, {Bhardwaj}, {Bij}, {Boyce}, {Boyle}, {Brar}, {Cassanelli}, {Chawla}, {Chen}, {Cliche}, {Cook}, {Cubranic}, {Curtin}, {Denman}, {Dobbs}, {Dong}, {Fandino}, {Fonseca}, {Gaensler}, {Giri}, {Good}, {Halpern}, {Hill}, {Hinshaw}, {H{\"o}fer}, {Josephy}, {Kania}, {Kaspi}, {Landecker}, {Leung}, {Li}, {Lin}, {Masui}, {McKinven}, {Mena-Parra}, {Merryfield}, {Meyers}, {Michilli}, {Milutinovic}, {Mirhosseini}, {M{\"u}nchmeyer}, {Naidu}, {Newburgh}, {Ng}, {Patel}, {Pen}, {Pinsonneault-Marotte}, {Pleunis}, {Quine}, {Rafiei-Ravandi}, {Rahman}, {Ransom}, {Renard}, {Sanghavi}, {Scholz}, {Shaw}, {Shin}, {Siegel}, {Singh}, {Smegal}, {Smith}, {Stairs}, {Tan}, {Tendulkar}, {Tretyakov}, {Vanderlinde}, {Wang}, {Wulf}, \& {Zwaniga}}]{CHIME+20}
{CHIME/FRB Collaboration}, {Andersen}, B.~C., {Bandura}, K.~M., {et~al.} 2020, \nat, 587, 54, \dodoi{10.1038/s41586-020-2863-y}

\bibitem[{{Combi} \& {Siegel}(2023)}]{Combi&Siegel23}
{Combi}, L., \& {Siegel}, D.~M. 2023, \prl, 131, 231402, \dodoi{10.1103/PhysRevLett.131.231402}

\bibitem[{{Corbett} {et~al.}(2023){Corbett}, {Carney}, {Gonzalez}, {Fors}, {Galliher}, {Glazier}, {Howard}, {Law}, {Quimby}, {Ratzloff}, \& {Soto}}]{Corbett+23}
{Corbett}, H., {Carney}, J., {Gonzalez}, R., {et~al.} 2023, \apjs, 265, 63, \dodoi{10.3847/1538-4365/acbd41}

\bibitem[{{C{\^o}t{\'e}} {et~al.}(2019){C{\^o}t{\'e}}, {Eichler}, {Arcones}, {Hansen}, {Simonetti}, {Frebel}, {Fryer}, {Pignatari}, {Reichert}, {Belczynski}, \& {Matteucci}}]{Cote+19}
{C{\^o}t{\'e}}, B., {Eichler}, M., {Arcones}, A., {et~al.} 2019, \apj, 875, 106, \dodoi{10.3847/1538-4357/ab10db}

\bibitem[{{Coulter} {et~al.}(2017){Coulter}, {Foley}, {Kilpatrick}, {Drout}, {Piro}, {Shappee}, {Siebert}, {Simon}, {Ulloa}, {Kasen}, {Madore}, {Murguia-Berthier}, {Pan}, {Prochaska}, {Ramirez-Ruiz}, {Rest}, \& {Rojas-Bravo}}]{Coulter+2017}
{Coulter}, D.~A., {Foley}, R.~J., {Kilpatrick}, C.~D., {et~al.} 2017, Science, 358, 1556, \dodoi{10.1126/science.aap9811}

\bibitem[{Cyburt {et~al.}(2010)Cyburt, Amthor, Ferguson, Meisel, Smith, Warren, Heger, Hoffman, Rauscher, Sakharuk, Schatz, Thielemann, \& Wiescher}]{cyburt:10}
Cyburt, R.~H., Amthor, A.~M., Ferguson, R., {et~al.} 2010, \apjs, 189, 240, \dodoi{10.1088/0067-0049/189/1/240}

\bibitem[{{Demidov} \& {Lyubarsky}(2023)}]{Demidov2023}
{Demidov}, I., \& {Lyubarsky}, Y. 2023, \mnras, 518, 810, \dodoi{10.1093/mnras/stac3120}

\bibitem[{{Desai} {et~al.}(2022){Desai}, {Siegel}, \& {Metzger}}]{Desai+2022}
{Desai}, D., {Siegel}, D.~M., \& {Metzger}, B.~D. 2022, \apj, 931, 104, \dodoi{10.3847/1538-4357/ac69da}

\bibitem[{{Desai} {et~al.}(2023){Desai}, {Siegel}, \& {Metzger}}]{Desai+2023}
{Desai}, D.~K., {Siegel}, D.~M., \& {Metzger}, B.~D. 2023, \apj, 954, 192, \dodoi{10.3847/1538-4357/acea83}

\bibitem[{{Duncan} \& {Thompson}(1992)}]{Duncan&Thompson92}
{Duncan}, R.~C., \& {Thompson}, C. 1992, \apjl, 392, L9, \dodoi{10.1086/186413}

\bibitem[{{Evans} \& {Mathews}(1988)}]{Evans&Mathews1988}
{Evans}, C.~R., \& {Mathews}, G.~J. 1988, in Origin and Distribution of the Elements, ed. G.~J. {Mathews}, 619

\bibitem[{{Evans} {et~al.}(1980){Evans}, {Klebesadel}, {Laros}, {Cline}, {Desai}, {Teegarden}, {Pizzichini}, {Hurley}, {Niel}, \& {Vedrenne}}]{Evans+1980}
{Evans}, W.~D., {Klebesadel}, R.~W., {Laros}, J.~G., {et~al.} 1980, \apjl, 237, L7, \dodoi{10.1086/183222}

\bibitem[{{Farouqi} {et~al.}(2022){Farouqi}, {Thielemann}, {Rosswog}, \& {Kratz}}]{Farouqi+2022}
{Farouqi}, K., {Thielemann}, F.~K., {Rosswog}, S., \& {Kratz}, K.~L. 2022, \aap, 663, A70, \dodoi{10.1051/0004-6361/202141038}

\bibitem[{{Feroci} {et~al.}(2001){Feroci}, {Hurley}, {Duncan}, \& {Thompson}}]{Feroci+01}
{Feroci}, M., {Hurley}, K., {Duncan}, R.~C., \& {Thompson}, C. 2001, \apj, 549, 1021, \dodoi{10.1086/319441}

\bibitem[{{Fischer} {et~al.}(2020){Fischer}, {Wu}, {Wehmeyer}, {Bastian}, {Mart{\'\i}nez-Pinedo}, \& {Thielemann}}]{Fischer+2020}
{Fischer}, T., {Wu}, M.-R., {Wehmeyer}, B., {et~al.} 2020, \apj, 894, 9, \dodoi{10.3847/1538-4357/ab86b0}

\bibitem[{{Frail} {et~al.}(1999){Frail}, {Kulkarni}, \& {Bloom}}]{Frail+1999}
{Frail}, D.~A., {Kulkarni}, S.~R., \& {Bloom}, J.~S. 1999, \nat, 398, 127, \dodoi{10.1038/18163}

\bibitem[{{Frankel} \& {Metropolis}(1947)}]{frankel:47}
{Frankel}, S., \& {Metropolis}, N. 1947, Physical Review, 72, 914, \dodoi{10.1103/PhysRev.72.914}

\bibitem[{{Frederiks} {et~al.}(2007){Frederiks}, {Golenetskii}, {Palshin}, {Aptekar}, {Ilyinskii}, {Oleinik}, {Mazets}, \& {Cline}}]{Frederiks+07}
{Frederiks}, D.~D., {Golenetskii}, S.~V., {Palshin}, V.~D., {et~al.} 2007, Astronomy Letters, 33, 1, \dodoi{10.1134/S106377370701001X}

\bibitem[{{Fuller} {et~al.}(1982){Fuller}, {Fowler}, \& {Newman}}]{fuller:82}
{Fuller}, G.~M., {Fowler}, W.~A., \& {Newman}, M.~J. 1982, \apjs, 48, 279, \dodoi{10.1086/190779}

\bibitem[{{Gaensler} {et~al.}(2005){Gaensler}, {Kouveliotou}, {Gelfand}, {Taylor}, {Eichler}, {Wijers}, {Granot}, {Ramirez-Ruiz}, {Lyubarsky}, {Hunstead}, {Campbell-Wilson}, {van der Horst}, {McLaughlin}, {Fender}, {Garrett}, {Newton-McGee}, {Palmer}, {Gehrels}, \& {Woods}}]{Gaensler+2005}
{Gaensler}, B.~M., {Kouveliotou}, C., {Gelfand}, J.~D., {et~al.} 2005, \nat, 434, 1104, \dodoi{10.1038/nature03498}

\bibitem[{{Gelfand} {et~al.}(2005){Gelfand}, {Lyubarsky}, {Eichler}, {Gaensler}, {Taylor}, {Granot}, {Newton-McGee}, {Ramirez-Ruiz}, {Kouveliotou}, \& {Wijers}}]{Gelfand+2005}
{Gelfand}, J.~D., {Lyubarsky}, Y.~E., {Eichler}, D., {et~al.} 2005, \apjl, 634, L89, \dodoi{10.1086/498643}

\bibitem[{{Gill} \& {Heyl}(2007)}]{Gill&Heyl2007}
{Gill}, R., \& {Heyl}, J. 2007, \mnras, 381, 52, \dodoi{10.1111/j.1365-2966.2007.12254.x}

\bibitem[{{Gill} \& {Heyl}(2010)}]{Gill+Heyl2010}
{Gill}, R., \& {Heyl}, J.~S. 2010, \mnras, 407, 1926, \dodoi{10.1111/j.1365-2966.2010.17038.x}

\bibitem[{{Goriely} {et~al.}(2014){Goriely}, {Bauswein}, {Janka}, {Sida}, {Lema{\^\i}tre}, {Panebianco}, {Dubray}, \& {Hilaire}}]{Goriely+14}
{Goriely}, S., {Bauswein}, A., {Janka}, H.~T., {et~al.} 2014, in American Institute of Physics Conference Series, Vol. 1594, Origin of Matter and Evolution of Galaxies 2013, ed. S.~{Jeong}, N.~{Imai}, H.~{Miyatake}, \& T.~{Kajino} (AIP), 357--364, \dodoi{10.1063/1.4874095}

\bibitem[{{Gottlieb} \& {Loeb}(2020)}]{Gottlieb&Loeb20}
{Gottlieb}, O., \& {Loeb}, A. 2020, \mnras, 493, 1753, \dodoi{10.1093/mnras/staa363}

\bibitem[{{Granot} {et~al.}(2006){Granot}, {Ramirez-Ruiz}, {Taylor}, {Eichler}, {Lyubarsky}, {Wijers}, {Gaensler}, {Gelfand}, \& {Kouveliotou}}]{Granot+2006}
{Granot}, J., {Ramirez-Ruiz}, E., {Taylor}, G.~B., {et~al.} 2006, \apj, 638, 391, \dodoi{10.1086/497680}

\bibitem[{{Grichener} {et~al.}(2022){Grichener}, {Kobayashi}, \& {Soker}}]{Grichener+22}
{Grichener}, A., {Kobayashi}, C., \& {Soker}, N. 2022, \apjl, 926, L9, \dodoi{10.3847/2041-8213/ac4f68}

\bibitem[{{Hoffman} {et~al.}(1997){Hoffman}, {Woosley}, \& {Qian}}]{Hoffman+97}
{Hoffman}, R.~D., {Woosley}, S.~E., \& {Qian}, Y.-Z. 1997, \apj, 482, 951, \dodoi{10.1086/304181}

\bibitem[{{Hotokezaka} {et~al.}(2018){Hotokezaka}, {Beniamini}, \& {Piran}}]{Hotokezaka+2018}
{Hotokezaka}, K., {Beniamini}, P., \& {Piran}, T. 2018, International Journal of Modern Physics D, 27, 1842005, \dodoi{10.1142/S0218271818420051}

\bibitem[{{Hotokezaka} {et~al.}(2016){Hotokezaka}, {Wanajo}, {Tanaka}, {Bamba}, {Terada}, \& {Piran}}]{Hotokezaka+16}
{Hotokezaka}, K., {Wanajo}, S., {Tanaka}, M., {et~al.} 2016, \mnras, 459, 35, \dodoi{10.1093/mnras/stw404}

\bibitem[{{Hurley} {et~al.}(1999){Hurley}, {Cline}, {Mazets}, {Barthelmy}, {Butterworth}, {Marshall}, {Palmer}, {Aptekar}, {Golenetskii}, {Il'Inskii}, {Frederiks}, {McTiernan}, {Gold}, \& {Trombka}}]{Hurley+1999}
{Hurley}, K., {Cline}, T., {Mazets}, E., {et~al.} 1999, \nat, 397, 41, \dodoi{10.1038/16199}

\bibitem[{{Hurley} {et~al.}(2005){Hurley}, {Boggs}, {Smith}, {Duncan}, {Lin}, {Zoglauer}, {Krucker}, {Hurford}, {Hudson}, {Wigger}, {Hajdas}, {Thompson}, {Mitrofanov}, {Sanin}, {Boynton}, {Fellows}, {von Kienlin}, {Lichti}, {Rau}, \& {Cline}}]{Hurley+05}
{Hurley}, K., {Boggs}, S.~E., {Smith}, D.~M., {et~al.} 2005, \nat, 434, 1098, \dodoi{10.1038/nature03519}

\bibitem[{Ioka(2001)}]{Ioka2001}
Ioka, K. 2001, Monthly Notices of the Royal Astronomical Society, 327, 639, \dodoi{10.1046/j.1365-8711.2001.04756.x}

\bibitem[{{Issa} {et~al.}(2024){Issa}, {Gottlieb}, {Metzger}, {Jacquemin-Ide}, {Liska}, {Foucart}, {Halevi}, \& {Tchekhovskoy}}]{Issa+24}
{Issa}, D., {Gottlieb}, O., {Metzger}, B., {et~al.} 2024, arXiv e-prints, arXiv:2410.02852, \dodoi{10.48550/arXiv.2410.02852}

\bibitem[{{Kallman} {et~al.}(2021){Kallman}, {Bautista}, {Deprince}, {Garc{\'\i}a}, {Mendoza}, {Ogorzalek}, {Palmeri}, \& {Quinet}}]{Kallman+21}
{Kallman}, T., {Bautista}, M., {Deprince}, J., {et~al.} 2021, \apj, 908, 94, \dodoi{10.3847/1538-4357/abccd6}

\bibitem[{{Kasen} {et~al.}(2013){Kasen}, {Badnell}, \& {Barnes}}]{Kasen+13}
{Kasen}, D., {Badnell}, N.~R., \& {Barnes}, J. 2013, \apj, 774, 25, \dodoi{10.1088/0004-637X/774/1/25}

\bibitem[{{Kaspi} \& {Beloborodov}(2017)}]{Kaspi2017}
{Kaspi}, V.~M., \& {Beloborodov}, A.~M. 2017, \araa, 55, 261, \dodoi{10.1146/annurev-astro-081915-023329}

\bibitem[{{Khatami} \& {Kasen}(2019)}]{Khatami&Kasen19}
{Khatami}, D.~K., \& {Kasen}, D.~N. 2019, \apj, 878, 56, \dodoi{10.3847/1538-4357/ab1f09}

\bibitem[{{Kouveliotou} {et~al.}(1998){Kouveliotou}, {Dieters}, {Strohmayer}, {van Paradijs}, {Fishman}, {Meegan}, {Hurley}, {Kommers}, {Smith}, {Frail}, \& {Murakami}}]{Kouveliotou+98}
{Kouveliotou}, C., {Dieters}, S., {Strohmayer}, T., {et~al.} 1998, \nat, 393, 235, \dodoi{10.1038/30410}

\bibitem[{{Kulkarni}(2005)}]{Kulkarni05}
{Kulkarni}, S.~R. 2005, arXiv e-prints, astro, \dodoi{10.48550/arXiv.astro-ph/0510256}

\bibitem[{{Kulkarni} {et~al.}(2021){Kulkarni}, {Harrison}, {Grefenstette}, {Earnshaw}, {Andreoni}, {et~al.}}]{Kulkarni+21}
{Kulkarni}, S.~R., {Harrison}, F.~A., {Grefenstette}, B.~W., {et~al.} 2021, arXiv e-prints, arXiv:2111.15608, \dodoi{10.48550/arXiv.2111.15608}

\bibitem[{{Langanke} \& {Mart{\'{\i}}nez-Pinedo}(2000)}]{langanke:00}
{Langanke}, K., \& {Mart{\'{\i}}nez-Pinedo}, G. 2000, Nucl. Phys. A, 673, 481, \dodoi{10.1016/S0375-9474(00)00131-7}

\bibitem[{{Lattimer} {et~al.}(1977){Lattimer}, {Mackie}, {Ravenhall}, \& {Schramm}}]{Lattimer+1977}
{Lattimer}, J.~M., {Mackie}, F., {Ravenhall}, D.~G., \& {Schramm}, D.~N. 1977, \apj, 213, 225, \dodoi{10.1086/155148}

\bibitem[{{Lattimer} \& {Schramm}(1974)}]{Lattimer&Schramm74}
{Lattimer}, J.~M., \& {Schramm}, D.~N. 1974, \apjl, 192, L145, \dodoi{10.1086/181612}

\bibitem[{{Lattimer} \& {Schramm}(1976)}]{Lattimer+Schramm76}
---. 1976, \apj, 210, 549, \dodoi{10.1086/154860}

\bibitem[{{Li} \& {Paczy{\'n}ski}(1998)}]{Li&Paczynski98}
{Li}, L.-X., \& {Paczy{\'n}ski}, B. 1998, Astrophys. J. Lett., 507, L59, \dodoi{10.1086/311680}

\bibitem[{{Lippuner} \& {Roberts}(2015)}]{Lippuner&Roberts15}
{Lippuner}, J., \& {Roberts}, L.~F. 2015, \apj, 815, 82, \dodoi{10.1088/0004-637X/815/2/82}

\bibitem[{{Lippuner} \& {Roberts}(2017)}]{Lippuner&Roberts17}
---. 2017, \apjs, 233, 18, \dodoi{10.3847/1538-4365/aa94cb}

\bibitem[{Lodders(2020)}]{Lodders+2020}
Lodders, K. 2020, Solar Elemental Abundances,  Oxford University Press, \dodoi{10.1093/acrefore/9780190647926.013.145}

\bibitem[{{Lorimer} {et~al.}(2007){Lorimer}, {Bailes}, {McLaughlin}, {Narkevic}, \& {Crawford}}]{Lorimer+07}
{Lorimer}, D.~R., {Bailes}, M., {McLaughlin}, M.~A., {Narkevic}, D.~J., \& {Crawford}, F. 2007, Science, 318, 777, \dodoi{10.1126/science.1147532}

\bibitem[{{Lugaro} {et~al.}(2014){Lugaro}, {Heger}, {Osrin}, {Goriely}, {Zuber}, {Karakas}, {Gibson}, {Doherty}, {Lattanzio}, \& {Ott}}]{Lugaro+2014}
{Lugaro}, M., {Heger}, A., {Osrin}, D., {et~al.} 2014, Science, 345, 650, \dodoi{10.1126/science.1253338}

\bibitem[{{Lyutikov}(2003)}]{Lyutikov2003}
{Lyutikov}, M. 2003, \mnras, 339, 623, \dodoi{10.1046/j.1365-8711.2003.06141.x}

\bibitem[{Lyutikov(2006)}]{Lyutikov2006}
Lyutikov, M. 2006, Monthly Notices of the Royal Astronomical Society, 367, 1594, \dodoi{10.1111/j.1365-2966.2006.10069.x}

\bibitem[{{Mamdouh} {et~al.}(2001){Mamdouh}, {Pearson}, {Rayet}, \& {Tondeur}}]{mamdouh:01}
{Mamdouh}, A., {Pearson}, J.~M., {Rayet}, M., \& {Tondeur}, F. 2001, Nuc. Phys. A, 679, 337, \dodoi{10.1016/S0375-9474(00)00358-4}

\bibitem[{{Margalit} {et~al.}(2020){Margalit}, {Beniamini}, {Sridhar}, \& {Metzger}}]{Margalit+2020}
{Margalit}, B., {Beniamini}, P., {Sridhar}, N., \& {Metzger}, B.~D. 2020, \apjl, 899, L27, \dodoi{10.3847/2041-8213/abac57}

\bibitem[{{Margalit} {et~al.}(2019){Margalit}, {Berger}, \& {Metzger}}]{Margalit+19}
{Margalit}, B., {Berger}, E., \& {Metzger}, B.~D. 2019, \apj, 886, 110, \dodoi{10.3847/1538-4357/ab4c31}

\bibitem[{{Margalit} \& {Metzger}(2018)}]{Margalit&Metzger18}
{Margalit}, B., \& {Metzger}, B.~D. 2018, \apjl, 868, L4, \dodoi{10.3847/2041-8213/aaedad}

\bibitem[{{Margalit} \& {Metzger}(2019)}]{Margalit&Metzger19}
---. 2019, \apjl, 880, L15, \dodoi{10.3847/2041-8213/ab2ae2}

\bibitem[{{Mazets} {et~al.}(1979){Mazets}, {Golentskii}, {Ilinskii}, {Aptekar}, \& {Guryan}}]{Mazets+1979}
{Mazets}, E.~P., {Golentskii}, S.~V., {Ilinskii}, V.~N., {Aptekar}, R.~L., \& {Guryan}, I.~A. 1979, \nat, 282, 587, \dodoi{10.1038/282587a0}

\bibitem[{{Mereghetti} {et~al.}(2005){Mereghetti}, {G{\"o}tz}, {von Kienlin}, {Rau}, {Lichti}, {Weidenspointner}, \& {Jean}}]{Mereghetti+05}
{Mereghetti}, S., {G{\"o}tz}, D., {von Kienlin}, A., {et~al.} 2005, \apjl, 624, L105, \dodoi{10.1086/430669}

\bibitem[{{Mereghetti} {et~al.}(2015){Mereghetti}, {Pons}, \& {Melatos}}]{Mereghetti+2015}
{Mereghetti}, S., {Pons}, J.~A., \& {Melatos}, A. 2015, \ssr, 191, 315, \dodoi{10.1007/s11214-015-0146-y}

\bibitem[{{Metzger}(2019)}]{Metzger19}
{Metzger}, B.~D. 2019, Living Reviews in Relativity, 23, 1, \dodoi{10.1007/s41114-019-0024-0}

\bibitem[{{Metzger} {et~al.}(2010){Metzger}, {Arcones}, {Quataert}, \& {Mart{\'{\i}}nez-Pinedo}}]{Metzger+10}
{Metzger}, B.~D., {Arcones}, A., {Quataert}, E., \& {Mart{\'{\i}}nez-Pinedo}, G. 2010, \mnras, 402, 2771, \dodoi{10.1111/j.1365-2966.2009.16107.x}

\bibitem[{{Metzger} {et~al.}(2015){Metzger}, {Bauswein}, {Goriely}, \& {Kasen}}]{Metzger+15_n_precursor}
{Metzger}, B.~D., {Bauswein}, A., {Goriely}, S., \& {Kasen}, D. 2015, \mnras, 446, 1115, \dodoi{10.1093/mnras/stu2225}

\bibitem[{{Metzger} {et~al.}(2007){Metzger}, {Thompson}, \& {Quataert}}]{Metzger+07}
{Metzger}, B.~D., {Thompson}, T.~A., \& {Quataert}, E. 2007, \apj, 659, 561, \dodoi{10.1086/512059}

\bibitem[{{Metzger} {et~al.}(2018){Metzger}, {Thompson}, \& {Quataert}}]{Metzger+18}
---. 2018, \apj, 856, 101, \dodoi{10.3847/1538-4357/aab095}

\bibitem[{{Meyer}(1989)}]{Meyer1989}
{Meyer}, B.~S. 1989, \apj, 343, 254, \dodoi{10.1086/167702}

\bibitem[{{Meyer}(1994)}]{Meyer94}
---. 1994, \araa, 32, 153, \dodoi{10.1146/annurev.aa.32.090194.001101}

\bibitem[{{Meyer} \& {Brown}(1997)}]{Meyer&Brown97}
{Meyer}, B.~S., \& {Brown}, J.~S. 1997, \apjs, 112, 199, \dodoi{10.1086/313032}

\bibitem[{{Meyer} {et~al.}(1992){Meyer}, {Mathews}, {Howard}, {Woosley}, \& {Hoffman}}]{Meyer+92}
{Meyer}, B.~S., {Mathews}, G.~J., {Howard}, W.~M., {Woosley}, S.~E., \& {Hoffman}, R.~D. 1992, \apj, 399, 656, \dodoi{10.1086/171957}

\bibitem[{{Michilli} {et~al.}(2018){Michilli}, {Seymour}, {Hessels}, {Spitler}, {Gajjar}, {Archibald}, {Bower}, {Chatterjee}, {Cordes}, {Gourdji}, {Heald}, {Kaspi}, {Law}, {Sobey}, {Adams}, {Bassa}, {Bogdanov}, {Brinkman}, {Demorest}, {Fernandez}, {Hellbourg}, {Lazio}, {Lynch}, {Maddox}, {Marcote}, {McLaughlin}, {Paragi}, {Ransom}, {Scholz}, {Siemion}, {Tendulkar}, {van Rooy}, {Wharton}, \& {Whitlow}}]{Michilli+18}
{Michilli}, D., {Seymour}, A., {Hessels}, J.~W.~T., {et~al.} 2018, \nat, 553, 182, \dodoi{10.1038/nature25149}

\bibitem[{{Miller} {et~al.}(2020){Miller}, {Sprouse}, {Fryer}, {Ryan}, {Dolence}, {Mumpower}, \& {Surman}}]{Miller+20}
{Miller}, J.~M., {Sprouse}, T.~M., {Fryer}, C.~L., {et~al.} 2020, \apj, 902, 66, \dodoi{10.3847/1538-4357/abb4e3}

\bibitem[{{M{\"o}ller} {et~al.}(2016){M{\"o}ller}, {Sierk}, {Ichikawa}, \& {Sagawa}}]{Moller+2016}
{M{\"o}ller}, P., {Sierk}, A.~J., {Ichikawa}, T., \& {Sagawa}, H. 2016, Atomic Data and Nuclear Data Tables, 109, 1, \dodoi{10.1016/j.adt.2015.10.002}

\bibitem[{{M{\"o}sta} {et~al.}(2014){M{\"o}sta}, {Richers}, {Ott}, {Haas}, {Piro}, {Boydstun}, {Abdikamalov}, {Reisswig}, \& {Schnetter}}]{Mosta+14}
{M{\"o}sta}, P., {Richers}, S., {Ott}, C.~D., {et~al.} 2014, \apjl, 785, L29, \dodoi{10.1088/2041-8205/785/2/L29}

\bibitem[{{Mumpower} {et~al.}(2012){Mumpower}, {McLaughlin}, \& {Surman}}]{Mumpower+12}
{Mumpower}, M.~R., {McLaughlin}, G.~C., \& {Surman}, R. 2012, \prc, 86, 035803, \dodoi{10.1103/PhysRevC.86.035803}

\bibitem[{{Nevins} \& {Roberts}(2023)}]{Nevins&Roberts23}
{Nevins}, B., \& {Roberts}, L.~F. 2023, \mnras, 520, 3986, \dodoi{10.1093/mnras/stad372}

\bibitem[{{Norman} \& {Schramm}(1979)}]{Norman+Schramm1979}
{Norman}, E.~B., \& {Schramm}, D.~N. 1979, \apj, 228, 881, \dodoi{10.1086/156914}

\bibitem[{{Oda} {et~al.}(1994){Oda}, {Hino}, {Muto}, {Takahara}, \& {Sato}}]{oda:94}
{Oda}, T., {Hino}, M., {Muto}, K., {Takahara}, M., \& {Sato}, K. 1994, Atomic Data and Nuclear Data Tables, 56, 231, \dodoi{10.1006/adnd.1994.1007}

\bibitem[{{Ott} \& {Kratz}(2008)}]{Ott+Kratz2008}
{Ott}, U., \& {Kratz}, K.-L. 2008, \nar, 52, 396, \dodoi{10.1016/j.newar.2008.05.001}

\bibitem[{{Ou} {et~al.}(2024){Ou}, {Ji}, {Frebel}, {Naidu}, \& {Limberg}}]{Ou+2024}
{Ou}, X., {Ji}, A.~P., {Frebel}, A., {Naidu}, R.~P., \& {Limberg}, G. 2024, \apj, 974, 232, \dodoi{10.3847/1538-4357/ad6f9b}

\bibitem[{{Palmer} {et~al.}(2005){Palmer}, {Barthelmy}, {Gehrels}, {Kippen}, {Cayton}, {Kouveliotou}, {Eichler}, {Wijers}, {Woods}, {Granot}, {Lyubarsky}, {Ramirez-Ruiz}, {Barbier}, {Chester}, {Cummings}, {Fenimore}, {Finger}, {Gaensler}, {Hullinger}, {Krimm}, {Markwardt}, {Nousek}, {Parsons}, {Patel}, {Sakamoto}, {Sato}, {Suzuki}, \& {Tueller}}]{Palmer+2005}
{Palmer}, D.~M., {Barthelmy}, S., {Gehrels}, N., {et~al.} 2005, \nat, 434, 1107, \dodoi{10.1038/nature03525}

\bibitem[{{Panov} {et~al.}(2010){Panov}, {Korneev}, {Rauscher}, {Mart{\'{\i}}nez-Pinedo}, {Keli{\'c}-Heil}, {Zinner}, \& {Thielemann}}]{panov:10}
{Panov}, I.~V., {Korneev}, I.~Y., {Rauscher}, T., {et~al.} 2010, \aap, 513, A61, \dodoi{10.1051/0004-6361/200911967}

\bibitem[{Parfrey {et~al.}(2013)Parfrey, Beloborodov, \& Hui}]{Parfrey+2013}
Parfrey, K., Beloborodov, A.~M., \& Hui, L. 2013, The Astrophysical Journal, 774, 92, \dodoi{10.1088/0004-637X/774/2/92}

\bibitem[{{Patel} {et~al.}(2024){Patel}, {Goldberg}, {Renzo}, \& {Metzger}}]{Patel+24}
{Patel}, A., {Goldberg}, J.~A., {Renzo}, M., \& {Metzger}, B.~D. 2024, arXiv e-prints, arXiv:2401.13035, \dodoi{10.48550/arXiv.2401.13035}

\bibitem[{{Patel} {et~al.}(2025){Patel}, {Metzger}, {Cehula}, {Burns}, {Goldberg}, \& {Thompson}}]{Patel+2025}
{Patel}, A., {Metzger}, B.~D., {Cehula}, J., {et~al.} 2025, arXiv e-prints, arXiv:2501.09181, \dodoi{10.48550/arXiv.2501.09181}

\bibitem[{{Prasanna} {et~al.}(2023){Prasanna}, {Coleman}, {Raives}, \& {Thompson}}]{Prasanna+2023}
{Prasanna}, T., {Coleman}, M. S.~B., {Raives}, M.~J., \& {Thompson}, T.~A. 2023, \mnras, 526, 3141, \dodoi{10.1093/mnras/stad2948}

\bibitem[{{Prasanna} {et~al.}(2024){Prasanna}, {Coleman}, \& {Thompson}}]{Prasanna+2024}
{Prasanna}, T., {Coleman}, M. S.~B., \& {Thompson}, T.~A. 2024, \apj, 973, 91, \dodoi{10.3847/1538-4357/ad4d90}

\bibitem[{{Qian} \& {Woosley}(1996)}]{Qian&Woosley96}
{Qian}, Y., \& {Woosley}, S.~E. 1996, \apj, 471, 331, \dodoi{10.1086/177973}

\bibitem[{{Qian} \& {Wasserburg}(2007)}]{Qian&Wasserburg07}
{Qian}, Y.-Z., \& {Wasserburg}, G.~J. 2007, \physrep, 442, 237, \dodoi{10.1016/j.physrep.2007.02.006}

\bibitem[{{Rodi} {et~al.}(2025){Rodi}, {Pacholski}, {Mereghetti}, {Arrigoni}, {Bazzano}, {Natalucci}, {Salvaterra}, \& {Ubertini}}]{Rodi+2025}
{Rodi}, J.~C., {Pacholski}, D.~P., {Mereghetti}, S., {et~al.} 2025, \apjl, 979, L25, \dodoi{10.3847/2041-8213/ada6b7}

\bibitem[{{Sagiv} {et~al.}(2014){Sagiv}, {Gal-Yam}, {Ofek}, {Waxman}, {Aharonson}, {Kulkarni}, {Nakar}, {Maoz}, {Trakhtenbrot}, {Phinney}, {Topaz}, {Beichman}, {Murthy}, \& {Worden}}]{Sagiv+14}
{Sagiv}, I., {Gal-Yam}, A., {Ofek}, E.~O., {et~al.} 2014, \aj, 147, 79, \dodoi{10.1088/0004-6256/147/4/79}

\bibitem[{{Sato}(1974)}]{Sato1974}
{Sato}, K. 1974, Progress of Theoretical Physics, 51, 726, \dodoi{10.1143/PTP.51.726}

\bibitem[{{Sautron} {et~al.}(2025){Sautron}, {McEwen}, {Younes}, {P{\'e}tri}, {Beniamini}, \& {Huppenkothen}}]{Sautron+2025}
{Sautron}, M., {McEwen}, A.~E., {Younes}, G., {et~al.} 2025, arXiv e-prints, arXiv:2503.11875.
\newblock \doarXiv{2503.11875}

\bibitem[{{Shibata} \& {Hotokezaka}(2019)}]{Shibata&Hotokezaka19}
{Shibata}, M., \& {Hotokezaka}, K. 2019, Annual Review of Nuclear and Particle Science, 69, 41, \dodoi{10.1146/annurev-nucl-101918-023625}

\bibitem[{{Siegel}(2019)}]{Siegel19}
{Siegel}, D.~M. 2019, European Physical Journal A, 55, 203, \dodoi{10.1140/epja/i2019-12888-9}

\bibitem[{{Simon} {et~al.}(2023){Simon}, {Brown}, {Mutlu-Pakdil}, {Ji}, {Drlica-Wagner}, {Avila}, {Mart{\'\i}nez-V{\'a}zquez}, {Li}, {Balbinot}, {Bechtol}, {Frebel}, {Geha}, {Hansen}, {James}, {Pace}, {Aguena}, {Alves}, {Andrade-Oliveira}, {Annis}, {Bacon}, {Bertin}, {Brooks}, {Burke}, {Carnero Rosell}, {Carrasco Kind}, {Carretero}, {Costanzi}, {da Costa}, {De Vicente}, {Desai}, {Doel}, {Everett}, {Ferrero}, {Frieman}, {Garc{\'\i}a-Bellido}, {Gatti}, {Gerdes}, {Gruen}, {Gruendl}, {Gschwend}, {Gutierrez}, {Hinton}, {Hollowood}, {Honscheid}, {Kuehn}, {Kuropatkin}, {Marshall}, {Mena-Fern{\'a}ndez}, {Miquel}, {Palmese}, {Paz-Chinch{\'o}n}, {Pereira}, {Pieres}, {Plazas Malag{\'o}n}, {Raveri}, {Rodriguez-Monroy}, {Sanchez}, {Santiago}, {Scarpine}, {Sevilla-Noarbe}, {Smith}, {Suchyta}, {Swanson}, {Tarle}, {To}, {Vincenzi}, {Weaverdyck}, \& {Wilkinson}}]{Simon+2023}
{Simon}, J.~D., {Brown}, T.~M., {Mutlu-Pakdil}, B., {et~al.} 2023, \apj, 944, 43, \dodoi{10.3847/1538-4357/aca9d1}

\bibitem[{{Sneden} {et~al.}(2008){Sneden}, {Cowan}, \& {Gallino}}]{Sneden+08}
{Sneden}, C., {Cowan}, J.~J., \& {Gallino}, R. 2008, \araa, 46, 241, \dodoi{10.1146/annurev.astro.46.060407.145207}

\bibitem[{{Spitler} {et~al.}(2016){Spitler}, {Scholz}, {Hessels}, {Bogdanov}, {Brazier}, {Camilo}, {Chatterjee}, {Cordes}, {Crawford}, {Deneva}, {Ferdman}, {Freire}, {Kaspi}, {Lazarus}, {Lynch}, {Madsen}, {McLaughlin}, {Patel}, {Ransom}, {Seymour}, {Stairs}, {Stappers}, {van Leeuwen}, \& {Zhu}}]{Spitler+16}
{Spitler}, L.~G., {Scholz}, P., {Hessels}, J.~W.~T., {et~al.} 2016, \nat, 531, 202, \dodoi{10.1038/nature17168}

\bibitem[{{Svirski} {et~al.}(2011){Svirski}, {Nakar}, \& {Ofek}}]{Svirski+11}
{Svirski}, G., {Nakar}, E., \& {Ofek}, E.~O. 2011, \mnras, 415, 2485, \dodoi{10.1111/j.1365-2966.2011.18872.x}

\bibitem[{{Symbalisty} \& {Schramm}(1982)}]{Symbalisty&Schramm82}
{Symbalisty}, E., \& {Schramm}, D.~N. 1982, Astrophys. J. Lett., 22, 143

\bibitem[{{Tanaka} {et~al.}(2020){Tanaka}, {Kato}, {Gaigalas}, \& {Kawaguchi}}]{Tanaka+20}
{Tanaka}, M., {Kato}, D., {Gaigalas}, G., \& {Kawaguchi}, K. 2020, \mnras, 496, 1369, \dodoi{10.1093/mnras/staa1576}

\bibitem[{{Taylor} {et~al.}(2005){Taylor}, {Gelfand}, {Gaensler}, {Granot}, {Kouveliotou}, {Fender}, {Ramirez-Ruiz}, {Eichler}, {Lyubarsky}, {Garrett}, \& {Wijers}}]{Taylor+2005}
{Taylor}, G.~B., {Gelfand}, J.~D., {Gaensler}, B.~M., {et~al.} 2005, \apjl, 634, L93, \dodoi{10.1086/491648}

\bibitem[{{Thielemann} {et~al.}(2020){Thielemann}, {Wehmeyer}, \& {Wu}}]{Thielemann+2020}
{Thielemann}, F.-K., {Wehmeyer}, B., \& {Wu}, M.-R. 2020, in Journal of Physics Conference Series, Vol. 1668, Journal of Physics Conference Series (IOP), 012044, \dodoi{10.1088/1742-6596/1668/1/012044}

\bibitem[{{Thompson} \& {Duncan}(1995)}]{Thompson&Duncan1995}
{Thompson}, C., \& {Duncan}, R.~C. 1995, \mnras, 275, 255, \dodoi{10.1093/mnras/275.2.255}

\bibitem[{Thompson \& Duncan(2001)}]{Thompson&Duncan2001}
Thompson, C., \& Duncan, R.~C. 2001, The Astrophysical Journal, 561, 980, \dodoi{10.1086/323256}

\bibitem[{{Thompson} {et~al.}(2001){Thompson}, {Burrows}, \& {Meyer}}]{Thompson+01}
{Thompson}, T.~A., {Burrows}, A., \& {Meyer}, B.~S. 2001, \apj, 562, 887, \dodoi{10.1086/323861}

\bibitem[{{Thompson} {et~al.}(2003){Thompson}, {Burrows}, \& {Pinto}}]{Thompson+03}
{Thompson}, T.~A., {Burrows}, A., \& {Pinto}, P.~A. 2003, \apj, 592, 434, \dodoi{10.1086/375701}

\bibitem[{{Thompson} {et~al.}(2004){Thompson}, {Chang}, \& {Quataert}}]{Thompson+04}
{Thompson}, T.~A., {Chang}, P., \& {Quataert}, E. 2004, \apj, 611, 380, \dodoi{10.1086/421969}

\bibitem[{{Thompson} \& {ud-Doula}(2018)}]{Thompson&udDoula18}
{Thompson}, T.~A., \& {ud-Doula}, A. 2018, \mnras, 476, 5502, \dodoi{10.1093/mnras/sty480}

\bibitem[{{Timmes} \& {Swesty}(2000)}]{Timmes&Swesty00}
{Timmes}, F.~X., \& {Swesty}, F.~D. 2000, \apjs, 126, 501, \dodoi{10.1086/313304}

\bibitem[{{Tsujimoto}(2021)}]{Tsujimoto+2021}
{Tsujimoto}, T. 2021, \apjl, 920, L32, \dodoi{10.3847/2041-8213/ac2c75}

\bibitem[{Turolla {et~al.}(2015)Turolla, Zane, \& Watts}]{Turolla+2015}
Turolla, R., Zane, S., \& Watts, A.~L. 2015, Reports on Progress in Physics, 78, 116901, \dodoi{10.1088/0034-4885/78/11/116901}

\bibitem[{{van de Voort} {et~al.}(2020){van de Voort}, {Pakmor}, {Grand}, {Springel}, {G{\'o}mez}, \& {Marinacci}}]{vandeVoort+20}
{van de Voort}, F., {Pakmor}, R., {Grand}, R. J.~J., {et~al.} 2020, \mnras, 494, 4867, \dodoi{10.1093/mnras/staa754}

\bibitem[{{Vlasov} {et~al.}(2017){Vlasov}, {Metzger}, {Lippuner}, {Roberts}, \& {Thompson}}]{Vlasov+17}
{Vlasov}, A.~D., {Metzger}, B.~D., {Lippuner}, J., {Roberts}, L.~F., \& {Thompson}, T.~A. 2017, \mnras, 468, 1522, \dodoi{10.1093/mnras/stx478}

\bibitem[{{Wahl}(2002)}]{wahl:02}
{Wahl}, A.~C. 2002, Systematics of Fission-Product Yields, Tech. Rep. LA-13928, Los Alamos National Laboratory, Los Alamos, NM, \dodoi{10.2172/809946}

\bibitem[{{Wanajo}(2013)}]{Wanajo13}
{Wanajo}, S. 2013, \apjl, 770, L22, \dodoi{10.1088/2041-8205/770/2/L22}

\bibitem[{{Werner} {et~al.}(2023){Werner}, {{\v{R}}{\'\i}pa}, {Th{\"o}ne}, {M{\"u}nz}, {Kurf{\"u}rst}, {Jel{\'\i}nek}, {Hroch}, {Ben{\'a}{\v{c}}ek}, {Topinka}, {Lukes-Gerakopoulos}, {Zaja{\v{c}}ek}, {Labaj}, {Pri{\v{s}}egen}, {Krti{\v{c}}ka}, {Merc}, {P{\'a}l}, {Pejcha}, {D{\'a}niel}, {Jon}, {{\v{S}}o{\v{s}}ovi{\v{c}}ka}, {Grome{\v{s}}}, {V{\'a}clav{\'\i}k}, {Steiger}, {Segi{\v{n}}{\'a}k}, {Behar}, {Tarem}, {Salh}, {Reich}, {Ben-Ami}, {Barschke}, {Berge}, {Tohuvavohu}, {Sivanandam}, {Bulla}, {Popov}, \& {Chang}}]{Werner+23}
{Werner}, N., {{\v{R}}{\'\i}pa}, J., {Th{\"o}ne}, C., {et~al.} 2023, arXiv e-prints, arXiv:2306.15080, \dodoi{10.48550/arXiv.2306.15080}

\bibitem[{{Winteler} {et~al.}(2012){Winteler}, {K{\"a}ppeli}, {Perego}, {Arcones}, {Vasset}, {Nishimura}, {Liebend{\"o}rfer}, \& {Thielemann}}]{Winteler+12}
{Winteler}, C., {K{\"a}ppeli}, R., {Perego}, A., {et~al.} 2012, \apjl, 750, L22, \dodoi{10.1088/2041-8205/750/1/L22}

\bibitem[{{Woosley} \& {Hoffman}(1992)}]{Woosley&Hoffman92}
{Woosley}, S.~E., \& {Hoffman}, R.~D. 1992, \apj, 395, 202, \dodoi{10.1086/171644}

\bibitem[{{Zevin} {et~al.}(2019){Zevin}, {Kremer}, {Siegel}, {Coughlin}, {Tsang}, {Berry}, \& {Kalogera}}]{Zevin+2019}
{Zevin}, M., {Kremer}, K., {Siegel}, D.~M., {et~al.} 2019, \apj, 886, 4, \dodoi{10.3847/1538-4357/ab498b}

\bibitem[{{Zhang} {et~al.}(2020){Zhang}, {Liu}, {Zhong}, \& {Wang}}]{Hai-Ming+20}
{Zhang}, H.-M., {Liu}, R.-Y., {Zhong}, S.-Q., \& {Wang}, X.-Y. 2020, \apjl, 903, L32, \dodoi{10.3847/2041-8213/abc2c9}

\end{thebibliography}

\end{document}